\documentclass[letterpaper,11pt]{article}
\pdfoutput=1 

\usepackage{jheppub} 
                     
\usepackage[T1]{fontenc} 
\usepackage{amsmath}
\usepackage{amssymb}
\usepackage{mathtools}
\usepackage{slashed}
\usepackage{hyperref}
\usepackage[dvipsnames]{xcolor}
\usepackage{relsize}
\usepackage{empheq}
\usepackage{xspace}
\usepackage{scalerel}
\usepackage{soul}

\newcommand\tab[1][1cm]{\hspace*{#1}}

\newcommand{\eq}[1]{Eq.~\eqref{#1}}
\newcommand{\eqs}[2]{Eqs.~\eqref{#1} and \eqref{#2}}

\newcommand{\Fig}[1]{Fig.~\ref{fig:#1}}

\DeclarePairedDelimiter{\ceil}{\lceil}{\rceil}

\newcommand{\bn}{{\bar n}}
\newcommand{\as}{{\alpha_{s}}}
\newcommand{\T}{{\tau}}
\newcommand{\e}{{\epsilon}}
\newcommand{\al}{{\alpha}}
\newcommand{\bt}{{\beta}}

\newcommand{\nn}{\nonumber} 
\newcommand{\bnspace}{\!\!\!\!\!\!\!\!\!\!\!}
\newcommand{\ns}{\!\!\!\!\!}
\newcommand{\V}[1]{{\vec{#1}_{\perp}^{\, 2}}}
\newcommand{\vt}[1]{{\vec{#1}_{\perp}}}
\newcommand{\MV}[1]{{\vert \vec{#1}_{\perp} \vert}}
\newcommand{\Eventtwo}{\textsc{Event2}\xspace}
\newcommand{\df}{\mathrm{d}}

\newcommand{\comment}[1]{\ignorespaces}

\preprint{\vbox{\hbox{UWThPh 2019-5}}}

\title{\boldmath One-loop Angularity Distributions with Recoil using Soft-Collinear Effective Theory}

 \author[a]{Ankita Budhraja,}
 \author[a]{Ambar Jain,}
 \author[b,c]{Massimiliano Procura}
 \affiliation[a]{Indian Institute of Science Education and Research,\\Bhopal Bypass Road, Bhauri, Bhopal 462 066, Madhya Pradesh, India}
 \affiliation[b]{Fakult\"at f\"ur Physik, Universit\"at Wien, Boltzmanngasse 5, 1090 Wien, Austria}
 \affiliation[c]{Theoretical Physics Department, CERN, 1 Esplanade des Particules, Geneva 23, Switzerland}

\emailAdd{ankitab@iiserb.ac.in}
\emailAdd{ambarj@iiserb.ac.in}
\emailAdd{mprocura@univie.ac.at}

\abstract{Angularities are event shapes whose sensitivity to the splitting angle of a collinear emission is controlled by a continuous parameter $b$, with $ -1 < b < \infty$. When measured with respect to the thrust axis, this class of QCD observables includes thrust ($b=1$) and jet broadening ($b=0$), the former being insensitive to the recoil of soft against collinear radiation, while the latter being maximally sensitive to it. Presently available analytic results for angularity distributions with $b \neq 0$ can be applied only close to the thrust limit since recoil effects have so far been neglected. As a first step to establish a comprehensive theoretical framework based on Soft-Collinear Effective Theory valid for all recoil-sensitive angularities, we compute for the first time angularity distributions at one-loop order in $\alpha_s$ for {\it all values of $b$} taking into account recoil effects. In the differential cross section, these amount to 
novel sub-leading singular contributions and/or power corrections, where the former are characterized by fractional powers of the angularity and contribute appreciably close to the peak region, also for $b \gtrsim 0.5$. 
Our calculations are checked against various limits known in the literature and agree with the numerical output of the \Eventtwo generator.

}

\begin{document} 
\maketitle
\flushbottom


\section{Introduction}
\label{sec:intro}

Event shapes describe geometrical properties of the energy-momentum flow in QCD events and probe strong interactions at various energy scales. They were among the first observables introduced to test QCD and have been used over the years for tuning parton showers and non-perturbative components of Monte Carlo event generators as well as to gain insight into hadronization effects in QCD  (see {\it e.g.}~\cite{Dasgupta:2003iq} and references therein). Moreover, the comparison between high-precision $e^+ e^-$ collider data against  fixed-order and resummed event shape calculations has provided accurate determinations of the strong coupling $\alpha_s$, as in~\cite{Becher:2008cf,Davison:2008vx,Abbate:2010xh,Gehrmann:2012sc,Hoang:2015hka}. More recently, event shape studies have also been extended to applications at hadron colliders (see {\it e.g.}~\cite{Banfi:2004nk,Banfi:2010xy}), and more exclusive, non-global event shape and jet shape measurements have acquired an important role in jet substructure studies (see {\it e.g.}~\cite{Banfi:2002hw,Dasgupta:2001sh,Hornig:2016ahz,Larkoski:2017jix}).

In this paper we focus on a specific class of global $e^+e^-$ event shapes, called angularities, which are defined as~\cite{Berger:2003iw}
\begin{align}
&\T_b = \frac{1}{Q}\, \sum_{i\, \e X}\, \vert p_{\perp i} \vert \, e^{-b\vert \eta_i \vert} \, , \label{eq:angularities}
\end{align}
where $Q$ is the center-of-mass energy.\footnote{In the most commonly used notation, the angularity exponent $b$ is replaced by $1-a$~\cite{Berger:2003iw, Berger:2004xf, Hornig:2009vb, Ellis:2010rwa}. However, \eq{eq:angularities} is more convenient for our study, where a key role is played by the jet broadening limit $b \to 0$.}
Here the sum runs over all final-state particles $i$, with $p_{\perp i}$ and $\eta_i$ denoting transverse momentum and pseudo-rapidity, respectively, of the $i^{\rm th}$ particle with respect to an appropriately chosen axis. In available data analyses~\cite{Achard:2011zz} this is identified with the thrust axis~\cite{Farhi:1977sg, Brandt:1964sa}, which is defined by the unit vector $\hat{t}$ that maximises the ratio $\sum_i |\vec p_i \cdot \hat{t}|$/$\sum_i  |{\vec p}_i|$. 

For any given kinematic configuration, \eq{eq:angularities} defines a one-parameter family of infrared and collinear safe observables for any real number $b$ in the interval $-1<b<\infty$. Angularities are a generalization of the classic QCD observables thrust and jet broadening, which are given by the special cases\footnote{The conventional definition of total jet broadening~\cite{Rakow:1981qn,Catani:1992jc} amounts to $\T_0/2$.}
\begin{align}
\text {Thrust} : \quad &b = 1, \qquad \T_1 = \frac{1}{Q}\, \sum_{i\, \e X}\, \vert p_{\perp i} \vert \, e^{-\vert \eta_i \vert}\, , \label{eq:thrust} \\
\text {Broadening} : \quad &b = 0, \qquad \T_0 = \frac{1}{Q}\, \sum_{i\, \e X}\, \vert p_{\perp i} \vert \, .\label{eq:broadening}
\end{align}

The possibility to tune the parameter $b$ exposes us to a wealth of information which is not accessible when looking at a single event shape such as thrust or broadening.
Indeed, a continuous change in the angularity exponent leads to a smooth variation of the sensitivity of $\T_b$ to the splitting angle of a collinear emission. Varying $b$ affects the transverse size of the jets that dominate each region of the $\T_b$-distribution, changes
 the proportion of two-jet-like and three-or-more-jet-like events in the peak region as well as non-perturbative corrections, as discussed in detail in Ref.~\cite{Bell:2018gce}.

For angularities with $b \gtrsim 1$, the direction of the thrust axis is insensitive to recoil by soft radiation, but as $b \to 0$, this effect cannot be ignored~\cite{Dokshitzer:1998kz}. One possibility to deal with this is to choose a recoil-insensitive axis~\cite{Larkoski:2014uqa}.\footnote{We refer to~\cite{Procura:2018zpn,Banfi:2018mcq} for recent studies of angularity distributions with respect to the winner-take-all axis up to next-to-next-to-leading logarithmic (NNLL) accuracy with next-to-leading fixed-order (NLO) corrections.} In this paper, instead, we stick to the traditional thrust axis for the following reasons. First of all, the available analytic expressions for angularity distributions with respect to the thrust axis (with the exception of jet broadening) have been applied only for $b \gtrsim  0.5$ because of neglected recoil effects. Here we derive novel complementary results valid also for $b<0.5$, thereby opening up the possibility of comparing against existing LEP data~\cite{Achard:2011zz} to improve the extraction of the strong coupling from angularities, by extending the work of Ref.~\cite{Bell:2018gce}. Secondly, we aim to establish a novel general framework based on factorization for recoil-sensitive angularities with respect to the thrust axis, and this paper is a first step in this direction. Here we focus on fixed-order predictions derived in the framework of Soft-Collinear Effective Theory (SCET)~\cite{Bauer:2000ew, Bauer:2000yr, Bauer:2001ct, Bauer:2001yt, Bauer:2002nz} for recoil-sensitive angularities, which allows us to illustrate interesting features of the interplay between collinear radiation and soft radiation in shaping up the  $\T_b$-distribution when we continuously vary $b$ {\it over its whole range}.  
In a companion paper~\cite{Budhraja:2018}, we will address the resummation of logarithmically enhanced terms and the validation of the logarithmic structure of recoil-sensitive angularity cross sections at ${\cal O}(\alpha_s^2)$ as predicted by a SCET factorization formula.

For $\T \ll 1$, it is known that these angularity cross sections for $b \gtrsim 0.5$ and $b=0$ can be expressed in a factorized form in terms of a hard function multiplying convolutions of jet functions with a soft function, each encoding physics at a single dynamical scale (hard, collinear and soft, respectively), thereby facilitating the resummation of large logarithmic terms. 
Calculations of these angularity distributions with respect to the thrust axis were carried out at NLL accuracy in Refs.~\cite{Berger:2003iw,Berger:2003pk,Hornig:2009vb,Chiu:2012ir,Almeida:2014uva}. Recently these observables have been analyzed up to NNLL including NNLO corrections~\cite{Bell:2018gce} within SCET for the purpose of a precision determination of $\al_s(m_Z)$ from LEP data, thereby providing important complementary information to similar analyses of event shapes like thrust~\cite{Becher:2008cf,Abbate:2010xh} and $C$-parameter~\cite{Hoang:2014wka}. NNLL+NLO accuracy for angularities has also been reached using the semi-numerical method {\tt ARES}~\cite{Banfi:2014sua} in~\cite{Banfi:2018mcq}, while the NNLL resummation of jet broadening in the SCET framework was achieved in~\cite{Becher:2012qc}.

\begin{figure}
\label{fig:modes}
\centering
\includegraphics[width=8cm]{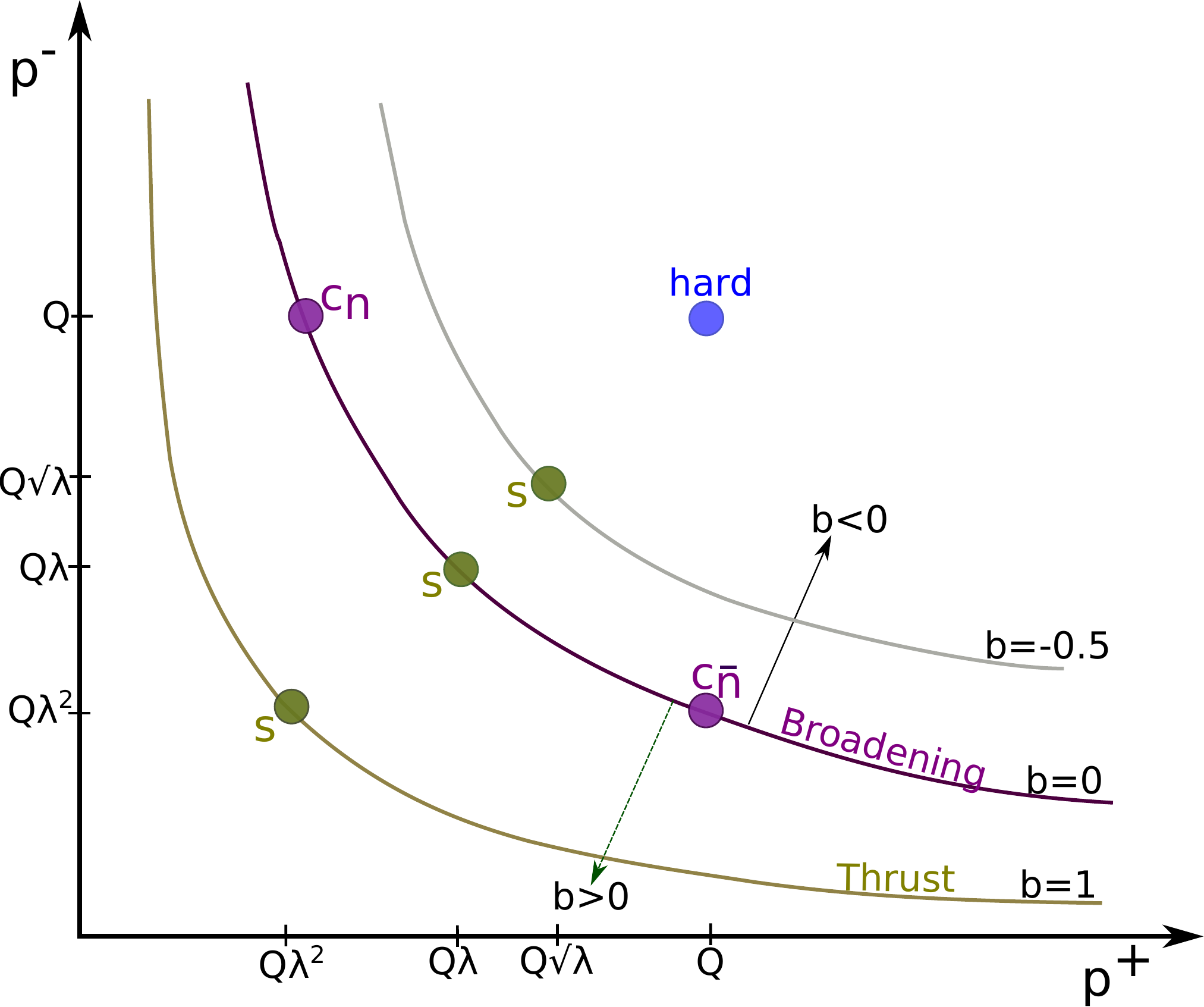}
\caption{Invariant mass-hyperbolae and momentum scaling of collinear ($c_n$, $c_{\bar{n}}$) and soft ($s$) degrees of freedom for angularities with different exponents $b$. The power counting parameter $\lambda$ is defined as $\lambda^{1+b} \sim \tau_b$.}
\end{figure}

By varying the angularity exponent $b$, the components of the momenta carried by the soft degrees of freedom in SCET can either be suppressed or not, compared to the collinear modes, as in the case of thrust and broadening, respectively. 
If one neglects masses of both hadrons$\footnote{Hadron masses will be relevant for calculating non-perturbative power corrections to the event-shape distributions~\cite{Salam:2001bd, Mateu:2012nk}, which is beyond the scope of this work.}$ and quarks, the soft and collinear modes, in the light-cone coordinates, satisfy the condition 
\begin{equation}
\label{eq:onshell}
p^+\,p^-\sim p_{\perp}^2\, ,
\end{equation} 
together with the additional constraint $\vert p_{\perp} \vert \, e^{-b\vert \eta \vert} \sim \T_b$ imposed by the angularity measurement. For the left ($\bn$-collinear) and right ($n$-collinear) hemispheres defined by the plane orthogonal to $\hat{t}$, this constraint amounts to 
\begin{equation}
\begin{split}
\label{eq:scaling}
\text{Right hemisphere ($p^{+} < p^{-}$)} : \quad & \vert p_{\perp} \vert \left(\frac{p^+}{p^-}\right)^{\!\!\frac{b}{2}} = (p^{+})^{\frac{1+b}{2}} (p^{-})^{\frac{1-b}{2}} \sim \T_b \, , \\ 
\text{Left hemisphere ($p^{+} > p^{-}$)} : \quad & \vert p_{\perp} \vert \left(\frac{p^-}{p^+}\right)^{\!\!\frac{b}{2}} = (p^{+})^{\frac{1-b}{2}} (p^{-})^{\frac{1+b}{2}} \sim \T_b \, .  
\end{split}
\end{equation}
In our calculations, we will refer to the angularities measured on the right and left hemisphere as $\tau_R$ and $\tau_L$, respectively. The event-wide angularity is obtained by summing over the two hemispheres, $\tau_b= \tau_L +\tau_R$.

In the expressions above, the momentum $p^{\mu}$ is decomposed into light-cone components as
\begin{align}
\label{eq:lightCone}
p^{\mu} = (p^+,p^-,p_\perp^{\mu}) = \frac{n^{\mu}}{2} p^- + \frac{\bn^{\mu}}{2} p^+ + p_\perp^{\mu} ; \quad p^{+} = n\cdot p\, , \quad p^{-} = \bn \cdot p \, ,
\end{align}
with $n^{\mu} = (1,0,0,1)$, $\bn^{\mu} = (1,0,0,-1)$ and $n \cdot \bn = 2$, where we have assumed that the thrust axis is along the $\hat z$ axis ($\hat n=\hat{t}$). If we define the power counting parameter $\lambda$ as
\begin{equation}
\label{eq:PowerCounting}
\lambda^{1+b}  \sim \T_b \, ,
\end{equation}
then the momenta corresponding to collinear and soft radiation obeying \eqs{eq:onshell}{eq:scaling} have the following parametric scaling in $\lambda$,
\begin{align}
\label{eq:ModeScaling}
p_n^{\mu} \sim Q(\lambda^2,1,\lambda), \quad p_{\bn}^{\mu} \sim Q(1,\lambda^2,\lambda), \quad p_{s}^{\mu} \sim Q(\lambda^{1+b},\lambda^{1+b},\lambda^{1+b}) \, .
\end{align}

For $b=1$ ($b=0$), namely in the thrust (broadening) case, $p_s^\mu$ has so-called ultrasoft (soft) scaling. The relative momentum scalings determined by the angularity exponent $b$ controls the separation of the mass hyperbolae for soft and collinear modes, as illustrated in \Fig{modes}. For angularities with $b<0$ ($b>0$), the soft mode lies on a heavier (lighter) mass hyperbola compared to the collinear modes.\footnote{For simplicity, we will keep the name ``soft'' for both cases.} For values of $b$ near $0$, the soft and collinear modes have similar scalings and recoil effects are particularly important.

If soft and collinear modes have relative scalings such that $p_c^{2} \gg p_s^{2}$, which happens for $b \gtrsim 1$, then jet and soft sectors can be factorized on the basis of their invariant mass scales using the SCET$_{\text{I}}$ framework, according to a {\it thrust-like} factorization theorem as in~\cite{Hornig:2009vb}. 
On the other hand, when the soft and collinear modes have similar invariant masses, $p_s^2\sim p_c^2$, as is realized close to the jet broadening limit, the appropriate effective theory is SCET$_{\text{II}}$, which accounts for the recoil effects of soft against collinear radiation via convolutions in the transverse momenta carried by jet and soft functions, and enables the resummation of rapidity logarithms~\cite{Chiu:2011qc,Chiu:2012ir} corresponding to rapidity divergences.
Eq.~(\ref{eq:ModeScaling}) exhibits the fact that there is a continuous scaling of modes connecting these two regimes.

The main content of this paper can be summarized as follows. Starting from a ${\rm SCET_{II}}$ factorization formula that can be derived in full analogy with the one for jet broadening in Ref.~\cite{Chiu:2012ir}, we derive analytic expressions for the angularity distributions at one-loop for both positive and negative values of $b$ and explicitly show their continuous behavior as $b \to 0$. We demonstrate that our cross section for $b>0$ reduces to the known result from thrust-like factorization  as $b$ approaches 1 augmented by recoil effects which we are able to compute.  
In the differential angularity distributions with $0.5 < b < 1$, recoil effects amount to a sub-leading singular term and power corrections. The former is given by an integrable fractional power of $\tau_b$ whose numerical contribution at one-loop suggests that its all-order resummation might play an important role in precision calculations. Our analytic expressions for the angularity cross sections with $b<0$ are the first derived within a SCET-based approach.\footnote{For an analysis based on the {\tt ARES} method, we refer to~\cite{Banfi:2018mcq}.}  For both positive and negative values of $b$, we show how to recover the known results for the jet broadening distribution as well as the broadening jet and soft functions as $b \to 0$. Finally, we successfully compare our analytical expressions for $b \leq 0.5$ against \Eventtwo and find that including sub-leading terms from our calculations can significantly improve the agreement with \Eventtwo for the accessible range of values of $\tau$. 

The authors of Ref.~\cite{Larkoski:2014uqa} succeeded in defining a universal factorization framework for angularities measured with respect to a recoil-insensitive axis, like the winner-take-all or the broadening axis. This choice greatly simplifies the form of the factorization theorem by avoiding transverse momentum convolutions. Rapidity logarithms are important only for the special case of $b=0$, while angularity cross sections for any other value of $b$ get factorized and resummed in ${\rm SCET_I}$. Such a comprehensive picture has been lacking so far for recoil-sensitive angularities measured with respect to the thrust axis. In this paper we make the first step to fill this gap and propose ${\rm SCET_{II}}$ factorization as a universal setup to calculate angularity distributions over the whole range of exponents $b$. Our forthcoming analysis of resummation~\cite{Budhraja:2018} will provide important complementary information to validate  
this approach.

The plan of the paper is as follows. In Sec.~\ref{sec:Factorization} we present a {\it broadening-like} factorization theorem for generic angularities with respect to the thrust axis, define the relevant jet and soft functions and compare our theoretical framework based on ${\rm SCET_{II}}$ against {\it thrust-like} factorization. In Secs.~\ref{sec:OneLoopCalculations} and~\ref{sec:SoftFunc} we calculate bare angularity jet and soft functions to one-loop order in $\as$, for arbitrary values of $b$, discuss the origin and nature of divergences and how these cancel in the cross section. In Secs.~\ref{JetBroadLimit} and~\ref{subsec:SoftBroadLimit}, it is shown how these functions reduce to the known broadening jet and soft functions in the limits $b \rightarrow 0^{\pm}$. Sec.~\ref{sec:AngularityCrossSection} is devoted to the calculation of the one-loop hemisphere double-differential and single-differential angularity cross sections and 
to an analysis of
the small-$\tau$ small-$b$ limit, which requires special care and is shown to smoothly connect with the jet broadening limit.
In Sec.~\ref{sec:AngularityCrossSection}, we also derive the thrust limit ($b \to 1$) of our expressions and provide a comparison with the thrust-like factorization results from Ref.~\cite{Hornig:2009vb} for $b \geq 1$. There we will show the additional contributions (sub-leading singular terms and power corrections) that our SCET$_{\text{II}}$ approach allows us to work out. The most technical details of our calculations are left to the appendices. 
The comparison of our results against \Eventtwo is the subject of Sec.~\ref{sec:NumericalResults}. We present our conclusions in Sec.~\ref{sec:Conclusions}.


\section{Factorization theorems for angularities with respect to the thrust axis}
\label{sec:Factorization}

The focus of our analysis is on the effects of recoil of soft against collinear radiation in angularity distributions with respect to the thrust axis, for the whole range of angularity exponents. Jet broadening is the prototype of a recoil-sensitive angularity and its factorization theorem accounts for these effects. Starting from the generalization of the jet broadening factorization theorem to the case of angularity exponents away from $b=0$, we will calculate recoil effects and show how (and by how much) they get suppressed if one approaches the thrust limit ($b=1$) or takes $b>1$. Furthermore, using this approach we will explore also the regime of negative values of $b$.

The factorization theorem for jet broadening was thoroughly discussed in Refs.~\cite{Becher:2011pf,Chiu:2012ir}. In the following, we will use the notation of Ref.~\cite{Chiu:2012ir}, where the factorization formula was obtained using the ${\rm SCET_{II}}$ formalism.
For angularities away from $b=0$, one can repeat the same derivation with two modifications.
First, the definition of the observable obviously changes. Secondly, the scaling of the gluon fields in the soft Wilson line becomes $A_s \sim Q(\lambda^{1+b}, \lambda^{1+b}, \lambda^{1+b})$. When $b$ is close to zero, soft gluons throw collinear quarks off-shell and get encoded into a soft Wilson line $S_{n,\bar{n}}$ \cite{Bauer:2001yt}, as in the case of jet broadening. When $b \geq 1$, the gluons approach ultrasoft scaling and decouple from collinear quarks via a field redefinition yielding an ultrasoft Wilson line $Y_{n,\bn}$ {\cite{Bauer:2001yt}}, as in the case of thrust factorization. In either case, (ultra)soft gluons decouple from collinear quarks/antiquarks to give (ultra)soft Wilson line factors, thereby leading to a factorization theorem of the same form, up to transverse-momentum dependence accounting for recoil of soft against collinear modes. In this paper, we will denote by  $S_{n,\bar{n}}$ both of these Wilson lines in the fundamental representation entering the definition of the soft function. 

Following and adapting the derivation in Ref.~\cite{Chiu:2012ir}, the {\it broadening-like} factorization theorem for the double-differential hemisphere angularity distribution with generic exponent $b$, is given by
\begin{equation}
\label{eq:BroadeningfactorizeRestated}
\begin{split}
\frac{1}{\sigma_0}\frac{\df\sigma}{\df\T_{L} \df\T_{R}} = H(Q;\mu)\! \int\! \df\T_n\, \df\T_{\bn}\, \df\T_n^s\, \df\T_{\bn}^s\, \delta(\T_{R}-\T_n-\T_n^s)\, \delta(\T_{L}-\T_{\bn}-\T_{\bn}^s) \int\! \df\V{p} \, \df\V{k} \\ 
{\cal J}(\T_n,\V{p};\mu,\nu/Q)\, {\cal J}(\T_{\bn},\V{k};\mu,\nu/Q)\, {\cal S}(\T_n^s,\T_{\bn}^s,\V{p},\V{k};\mu,\nu)\, ,
\end{split}
\end{equation}
%
where the contributions to the hemisphere angularities from soft and collinear sectors are denoted by  $\T_n^s$,  $\T_{\bn}^s$, $\T_n$ and $\T_{\bn}$, respectively. 
The definitions of the jet and the soft functions are more general than for the jet broadening case in Ref.~\cite{Chiu:2012ir} since the scaling of kinematical quantities, the fields and the measurement operator are all $b$-dependent. For notational convenience, this dependence is not written explicitly.
The bare angularity jet function for a jet in the $\hat n$-hemisphere, in $d$ space-time dimensions (with $d=4-2 \e$) is
\begin{equation}
\label{eq:JetDefine}
{\cal J}(\T_n,\V{p}) = \frac{(2\pi)^{3-2\e}}{N_c}{\rm tr}\langle 0\vert \frac{\bar{\slashed{n}}}{2}\chi_n(0)\delta(Q -\bn \cdot {\cal{\hat{P}}})\, \delta(\T_n-\hat{\T}_{\hat{n}})\delta^{(2-2\e)}(\vt{p} - {\cal{\hat{P}}}_{\perp})\bar\chi_{n}(0) \vert 0\rangle \,,
\end{equation}
where the operator $\hat{\T}_{\hat n} $ measures the angularity on the final-state $n$-collinear jet. ${\hat{\cal P}}$ denotes the label momentum operator~\cite{Bauer:2001ct, Bauer:2001yt} and $\chi_{n}$ is the collinear quark field~\cite{Bauer:2001ct, Bauer:2001yt}.\footnote{Since the angularity distribution is gauge invariant, we can always choose a non-singular gauge to carry out our calculations, thereby avoiding the inclusion of T-Wilson lines \cite{Idilbi:2010im} in the definition of SCET matrix elements.} The jet function for the $\hat{\bn}$-hemisphere is simply obtained through the replacement $n \rightarrow \bn$ in~\eq{eq:JetDefine}.

The bare angularity soft function, ${\cal S}(\T_n^s,\T_{\bn}^s,\V{p},\V{k})$ in \eq{eq:BroadeningfactorizeRestated} is given by the following matrix element of Wilson lines $S_{n, \bn}$, 
\begin{align}
\label{eq:SoftDefine}
{\cal S}(\T_n^s,\T_{\bn}^s,\V{p},\V{k}) &= \frac{\pi^{2-2\e}(\V{p})^{-\e}(\V{k})^{-\e}}{N_c\, \Gamma^2(1-\e)}\, {\rm{tr}}\langle 0\vert S_{\bn}^{\dagger}(0)\, S_n(0)\, \delta^{(2-2\e)}(\vt{p}+\mathbb{P}_{n\perp})\, \delta(\T_n^s-\hat{\T}_{\hat n}^s) \nn \\ 
& \quad \delta^{(2-2\e)}(\vt{k}+\mathbb{\bar{P}}_{n\perp})\, \delta(\T_{\bn}^s-\hat{\T}_{\hat \bn}^s)\, S_n^{\dagger}(0)\, S_{\bn}(0) \vert 0\rangle\, ,
\end{align}
which is also a generalization of the broadening soft function in~\cite{Chiu:2012ir} in the sense that the soft Wilson line is now obeying the scaling suitable for a  generic angularity, with $A_s \sim Q(\lambda^{1+b}, \lambda^{1+b}, \lambda^{1+b})$ rather than $A_s \sim Q(\lambda,\lambda,\lambda)$ specific to broadening. Here, $\mathbb{P}_{n\perp}$ and $\mathbb{\bar{P}}_{n\perp}$ are the operators that measure the net transverse momentum of all the particles within a hemisphere~\cite{Chiu:2012ir}. The crucial point is that, whether the gluon field scaling in a Wilson line is more soft- or ultrasoft-like, the transverse momentum convolutions between jet and soft sectors will be kept. This allows us to include the effects of recoil at any intermediate stage of the calculation. 

The hard function $H(Q;\mu)$ in \eq{eq:BroadeningfactorizeRestated} instead, is identical to the cases of thrust and broadening. It encodes virtual corrections in the $q \bar{q}$ production at the hard scale $Q$, and is given by the squared Wilson coefficient of the matching of QCD onto SCET currents~\cite{Manohar:2003vb,Bauer:2003di}. Up to order $\as$,
\begin{equation}
\label{eq:Hard}
H(Q;\mu) = 1 + \frac{\as(\mu) C_F}{\pi} \left(\!-\frac{1}{\e^2} - \frac{1}{\e} \ln\!\frac{\mu^2}{Q^2} - \frac{3}{2\, \e} -4 + \frac{7{\,}\pi^2}{12} - \frac{1}{2} \ln^2\!\frac{\mu^2}{Q^2} - \frac{3}{2} \ln\!\frac{\mu^2}{Q^2}\right) \, ,
\end{equation}
where we have explicitly included the ultraviolet (UV) divergences of the bare hard function calculated using dimensional regularization.

Hard, jet and soft functions depend upon the UV renormalization scale $\mu$ according to the familiar $\mu$-renormalization group equations (RGEs). In addition to $\mu$, the jet and the soft functions also depend upon the renormalization scale $\nu$ tied to rapidity divergences.
In \eq{eq:BroadeningfactorizeRestated}, both $\mu$- and $\nu$-scale dependences cancel out among the individual functions order by order in perturbation theory, up to a residual $\mu$-dependence at higher orders in $\as$.  
The study of the renormalization group evolution of ${\rm SCET_{II}}$ angularity jet and soft functions is beyond the scope of this paper and will be the subject of a future publication~\cite{Budhraja:2018}.

We now discuss how the {\it thrust-like} factorization theorem in~\cite{Hornig:2009vb} follows from \eq{eq:BroadeningfactorizeRestated}. From the definition of the jet function we note that the transverse momenta $\vec p_\perp$ and $\vec k_\perp$ involved in the convolutions have the scaling of label momenta, and are thus expected to be of order $\lambda$. In the case of thrust ($b=1$), the operators $\mathbb{P}_{n\perp}$ and $\mathbb{\bar{P}}_{n\perp}$ in the soft function of \eq{eq:SoftDefine}, measure the transverse momentum of ultrasoft modes, which scales like $\lambda^2$. This is the scaling of residual momenta, which is suppressed compared to $\vec p_\perp$ and $\vec k_\perp$. Dropping correspondingly the operators $\mathbb{P}_{n\perp}$ and $\mathbb{\bar{P}}_{n\perp}$ in the $\delta$-functions of \eq{eq:SoftDefine}, one can factor $\delta^{(2-2\epsilon)}(\vec p_\perp) \delta^{(2-2\epsilon)}(\vec k_\perp)$ out of the soft function and perform the $\vec p_\perp^{\,2}$ and $\vec k_\perp^{\,2}$ integrals in the factorization theorem. This reduces the jet and the soft functions of \eqs{eq:JetDefine}{eq:SoftDefine}, respectively, to
\begin{align}
{\cal J}(\T_n) =& \,\, \frac{(2\pi)^{3-2\e}}{N_c}{\rm tr}\langle 0\vert \frac{\bar{\slashed{n}}}{2}\chi_n(0)\delta(Q-\bn \cdot {\cal{\hat{P}}})\, \delta(\T_n-\hat{\T}_{\hat n})\delta^{(2-2\e)}({\cal{\hat{P}}}_{\perp})\bar\chi_{n}(0) \vert 0\rangle \, , \nonumber   \\
{\cal S}(\T_n^s,\T_{\bn}^s) =& \,\, \frac{1}{N_c}\, {\rm{tr}}\langle 0\vert S_{\bn}^{\dagger}(0)\, S_n(0)\,  \delta(\T_n^s-\hat{\T}_{\hat n}^s) 
  \delta(\T_{\bn}^s-\hat{\T}_{\hat \bn}^s)\, S_n^{\dagger}(0)\, S_{\bn}(0) \vert 0\rangle\, , \label{eq:SoftThrust}
\end{align}
which are the well-known jet and soft functions for thrust. The presence of $\delta^{(2-2\e)}({\cal{\hat{P}}}_{\perp})$ in the thrust jet function ensures that the total transverse momentum of the particles in each hemisphere   
is zero.
 The factorization formula for the double-differential hemisphere thrust distribution in the dijet limit is well-known,
\begin{align}
\label{eq:ThrustFactorize}
\frac{1}{\sigma_0}\frac{\df\sigma}{\df\T_{L} \df\T_{R}} = H(Q;\mu)\! \int\! \df\T_n\, \df\T_{\bn}\, \df\T_n^s\, \df\T_{\bn}^s\,\, \delta(\T_{R}-\T_n-\T_n^s)\, \delta(\T_{L}-\T_{\bn}-\T_{\bn}^s) \nn \\ 
\times\, {\cal J}(\T_n;\mu)\, {\cal J}(\T_{\bn};\mu)\, {\cal S}(\T_n^s,\T_{\bn}^s;\mu)\,  ,
\end{align}
see {\it e.g.}~\cite{Schwartz:2007ib,Becher:2008cf} for a derivation using the ${\rm SCET_{I}}$ formalism. If recoil effects are negligible, one can repeat the argument (adapting momentum scalings, measurement functions as well as the definitions of jet and soft functions) to derive a thrust-like factorization formula for double-differential angularity cross sections~\cite{Berger:2003iw}, which looks exactly like \eq{eq:ThrustFactorize} (see Ref.~\cite{Hornig:2009vb}) in the same way \eq{eq:BroadeningfactorizeRestated} looks like the factorization formula for jet broadening.

In Sec.~\ref{sec:AngularityCrossSection} we will show that when $b > 0$, the angularity distributions derived from the general, broadening-like factorization theorem contain all the singular terms from thrust-like factorization obtained in~\cite{Hornig:2009vb} together with suppressed singular terms and power corrections. In particular, we will show that at one-loop, the single-differential angularity cross section for strictly positive (and not vanishingly small) values of $b$, obtained from \eq{eq:BroadeningfactorizeRestated}, has the form
\begin{equation}
\begin{split}
\label{eq:posbCS}
\bigg[\frac{1}{\sigma_0}\, \frac{\rm \df\sigma^{(+)}}{\df\T}\bigg]^{\rm {\cal O}(\alpha_s)}_{\rm SCET_{II}} 
\ns &= \frac{\as(\mu){\,}C_F}{\pi}  \bigg\{ A^{(+)}(b)\, \delta(\T) \!\!- \frac{3}{1 + b} \bigg[ \frac{1}{\T} \bigg]_{+} \ns- \frac{4}{1+b} \bigg[\frac{\ln\T}{\T}\bigg]_{+}\ns -\frac{4}{1+b} \frac{\ln(1-r)}{\T} \bigg\} , \\
\end{split}
\end{equation}
where $\sigma_0$ is the Born cross section, whose expression can be found {\it e.g.} in App. A of Ref.~\cite{Abbate:2010xh}, $A^{(+)}(b)$ is $b$-dependent constant given in \eq{eq:NLODiffCrossSection} and $r$ is the solution of the equation
\begin{equation}
\frac{r}{(1-r)^{1+b}} = \T^b \, .
\end{equation}
The first three terms in \eq{eq:posbCS} agree with the one-loop singular result from thrust-like factorization~\cite{Hornig:2009vb}, while the last term gives the recoil effects due to ${\rm SCET_{II}}$ factorization theorem. Formally, this term, which we will refer to as {\em recoil correction} henceforth, is an integrable singularity for $0<b<1$, but of significant numerical size, as will be shown in section \ref{sec:NumericalResults}. Interestingly, the recoil correction reduces to a power correction in the thrust limit $(b \rightarrow 1)$ and becomes singular in the broadening limit $(b \rightarrow 0)$, thus smoothly connecting the singular thrust and broadening cross-section formulas at one loop. This will be shown in Sec.~\ref{sec:AngularityCrossSection}. In the small-$\T$ limit, the recoil correction can be expanded in powers of $\T^b$ as 
\begin{equation}
\frac{\ln(1-r)}{\T} = \sum_{n=1}^{\ceil*{1/b}-1} \frac{c_n}{\T^{1-n\,b}}  \, + {\rm{power \; corrections}}\, .
\end{equation}
Here the symbol $\ceil*{x}$ denotes the {\em ceiling function}, making $(\ceil*{1/b}-1)$ the greatest integer strictly smaller than $1/b$. The summation, with coefficients $c_n$  dependent only on $b$, provides singular contributions for $0<b<1$. The singularity $1/\T^{1-n b}$ is less severe than $1/\T$ but more severe than $\ln\T$, and thus we refer to it as a sub-leading singularity. Since this is integrable, it does not require a plus prescription.
The appearance of fractional powers in the one-loop cross section is unique to angularities and is due to the recoil effects. While it is customary in the effective field theory approach to drop power corrections and keep only terms that are homogenous in power counting, here we cannot drop the recoil effect as a power correction, since it is produced by the leading-power ${\rm SCET}_{\rm II}$ Lagrangian and is an artefact of the inhomogenous scaling of recoiling transverse momenta of the jet and soft functions. Indeed, the net transverse momentum of the jet function is equal and opposite to the net transverse momentum of the soft function, making its scaling 
ambiguous for an arbitrary $b$: one can choose it to scale like $\lambda$ or $\lambda^{1+b}$. When $b\sim 1$, a multipole expansion in powers of $p_\perp^2$ reduces the factorization theorem to that of thrust-like angularities and consistently reduces the recoil correction to power correction which is numerically small. For small $b$ the ambiguity is inherent in the problem and leads to inhomogeneity in the power counting. The recoil correction is numerically significant compared to the singular terms, it reduces to a singular term in small-$b$ limit and modulates smoothly between the thrust and broadening limits. Thus one should consider this term as a leading correction due to recoil.

We stress that for $0.5<b<1$, our one-loop angularity distribution contains one new (sub-leading) singular contribution compared to the thrust-like computation that scales like $\T^{1-b}$. 
Since this contribution diverges for $\tau \rightarrow 0$ and is numerically sizable, this term will require a resummation to all orders in perturbation theory, which will be discussed in \cite{Budhraja:2018}.
 
Similarly, for the case of strictly negative (and not vanishingly small) values of $b$, we will show that the one-loop single-differential angularity cross section has the form
\begin{equation}
\begin{split}
\label{eq:negbCS}
\bigg[\frac{1}{\sigma_0}\, \frac{\rm \df\sigma^{(-)}}{\df\T}\bigg]^{\rm {\cal O}(\alpha_s)}_{\rm SCET_{II}} 
\ns &= \frac{\as(\mu){\,}C_F}{\pi}  \bigg\{ A^{(-)}(b)\, \delta(\T)\!\!- \frac{3}{1 + b} \bigg[ \frac{1}{\T} \bigg]_{+}\ns - \frac{4}{(1+b)^2} \bigg[\frac{\ln\T}{\T}\bigg]_{+}\ns - \frac{4}{(1+b)^2} \frac{\ln(1-s)}{\T} \bigg\} , \\
\end{split}
\end{equation}
where the $b$-dependent constant $A^{(-)}(b)$ is given in \eq{eq:NLODiffCrossSectionNegativeBCase} of Sec.~\ref{sec:AngularityCrossSection} and $s$ is given by the solution of the equation
\begin{equation}
\frac{s}{(1-s)^{\frac{1}{1+b}}} = \T^{-\frac{b}{1+b}}\, .
\end{equation}
In the small-$\T$ limit, the last term of \eq{eq:negbCS} can be expanded in powers of $\T^{-\frac{b}{1+b}}$ as
\begin{equation}
\frac{\ln(1-s)}{\T} = \sum_{n=1}^{\ceil*{1/|b|}-2}\!\! \frac{c^{'}_n}{\T^{1+\frac{n\,b}{1+b}}} + {\rm{power \; corrections}}\, , 
\end{equation}
where $(\ceil*{1/|b|}-2)$ gives the greatest integer strictly smaller than $1/|b|-1$. 
This result for angularity exponents $b<0$ (or $a>1$) is similar in structure to \eq{eq:posbCS}, where the terms in the summation are sub-leading integrable singularities for $-1/2<b<0$. Although the sub-leading terms are power-suppressed in comparison to the leading singular terms, they cannot be expanded away as these terms are produced by the leading-order SCET$_{\rm II}$ Lagrangian and capture important recoil contributions. Moreover, the $\ln(1-s)/\T$ term becomes a leading singularity in $b \rightarrow 0$ limit and reproduces the well-known result of the singular broadening cross-section at one loop. Along with \eq{eq:posbCS}, this makes the singular angularity cross section continuous at $b=0$. The numerical significance of the sub-leading singular terms relative to the singular contributions is studied in Sec.~\ref{sec:NumericalResults}, where we show that the sub-leading singular terms progressively become a leading contribution as $b$ increases from $-1/2$ to $0$. Note that 
for $-1<b<-1/2$, all terms in the summation contribute as power corrections. In Sec.~\ref{sec:AngularityCrossSection} we will derive both \eq{eq:posbCS} and \eq{eq:negbCS} and will demonstrate the smoothness of the cross section at $b = 0$. The importance of the resummation of sub-leading singular terms will be addressed in a future publication~\cite{Budhraja:2018}.


\section{Angularity jet function at one loop}
\label{sec:OneLoopCalculations}

In this section, we present our results for the angularity jet function to one-loop  order in $\as$ using the definition given in \eq{eq:JetDefine}, which includes the recoil effects. We cover both cases of positive-$b$ and negative-$b$ angularities for $b>-1$. This calculation is unique in many ways. Unlike the calculation of thrust and broadening jet functions the zero-bin contributions are non-vanishing here. There is an interesting interplay of zero-bin and rapidity divergences, due to which the jet function for positive $b$ has an overall rapidity divergence while for negative $b$, the rapidity divergences cancel between the  unsubtracted and zero-bin contributions to the jet function. We stress that these rapidity divergences are not regulated by the angularity exponent $b$. Moreover, the presence of rapidity divergences and rapidity logarithms in the one-loop jet function makes the calculation for an arbitrary angularity exponent $b$ highly non-trivial.


\subsection{Angularity jet function at one-loop for $b>0$}
\label{subsec:JetNLOPositiveBCase}

Using the definition of the bare quark jet function given in \eq{eq:JetDefine}, the tree-level jet function corresponding to the quark jet in the right hemisphere ($R$) is given by
\begin{equation}
\label{eq:TreeLevelJet}
{\cal J}^{(0)}(\T_{R},\V{p}) = \delta\Bigg(\T_{R}-\bigg(\frac{\MV{p}}{Q}\bigg)^{\!\!\!1+b}\Bigg)\, .
\end{equation}
The other jet function corresponding to the left hemisphere ($L$) is obtained through the replacements $\T_{R}\rightarrow \T_{L}$ and $\V{p} \rightarrow \V{k}$.  

We compute the one-loop jet function only for the case $\vt{p} = 0$
since this suffices for the computation of the cross section both at one-loop and up to NLL accuracy.\footnote{Beyond this order, one would require the one-loop jet function with full $\vt{p}$ dependence, which is beyond the scope of this work.} Indeed, the convolution of the one-loop jet function with the tree-level soft function, which is proportional to $\delta(\V{p})\, \delta(\V{k})$ (see \eq{eq:TreeLevelSoft}), sets the net transverse momentum of the jet equal to zero. 
The results for the NLL resummed angularity distributions from broadening-like factorization will be presented in~\cite{Budhraja:2018}.

\begin{figure}[t]
 \includegraphics[width=15cm]{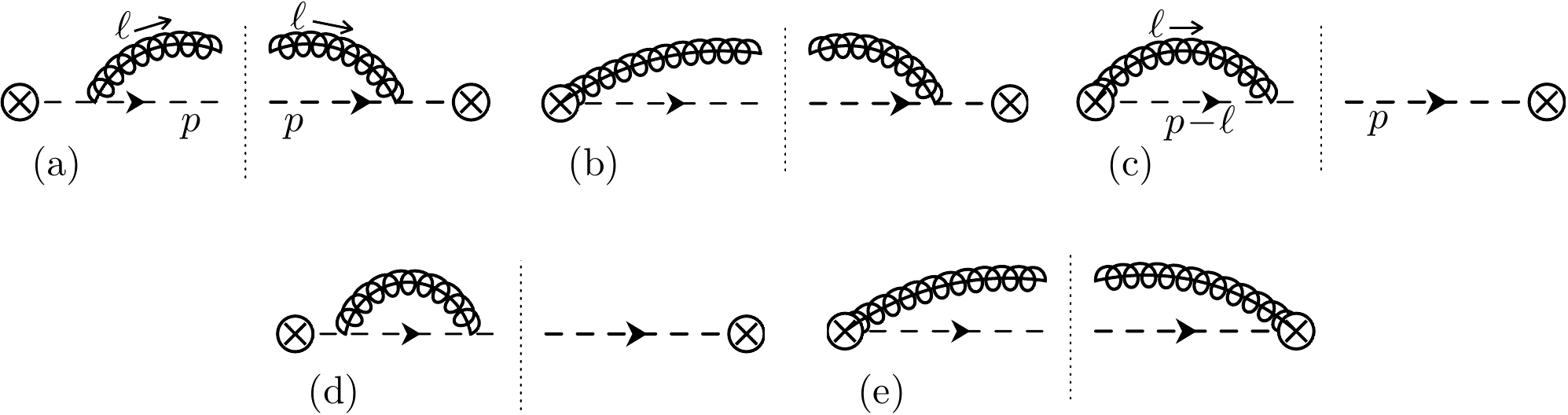}
 \centering
 \caption{\label{fig:JetDiagrams} Diagrams contributing to the angularity jet function at one-loop order. Here, $p^{\mu}$ is the quark momentum and $l^{\mu}$ is the gluon momentum. The mirror image graphs for (b), (c) and (d) are not shown here. Diagram (d) denotes the quark wave-function renormalization contribution.}
 \end{figure}

The relevant one-loop diagrams are those shown in Fig.~\ref{fig:JetDiagrams}, together with mirror images for graphs (b), (c) and (d). The diagrams in Fig.~\ref{fig:JetDiagrams} are similar to the case of thrust, broadening or other similar di-jet observables, however the results computed below cannot be obtained from any of these known calculations because of non-trivial zero-bin and rapidity divergences due to the  dependence on the exponent $b$. 

We use the $\eta$-regulator as proposed in~\cite{Chiu:2011qc, Chiu:2012ir} for regulating rapidity divergences and dimensional regularization for both UV and infrared (IR) divergences. Calculations in the following are performed in Feynman gauge. However, the results we obtained are valid for any choice of covariant gauge owing to collinear gauge invariance of the jet function~\cite{Chiu:2012ir}. Virtual correction diagrams (c) and (d) are given by scaleless integrals and hence vanish in pure dimensional regularization. Diagram (e) is zero in Feynman gauge. 
The bare one-loop jet function is thus given as the sum of the contributions from diagrams (a) and (b) together with corresponding zero-bin subtractions~\cite{Manohar:2006nz}, which remove double counting between jet and soft regions. 

Diagram (a) does not require any rapidity regulator. The zero-bin contribution for this graph is given by a scaleless integral and thus vanishes in pure dimensional regularization. 
We refer to App.~\ref{appsubsec:JetA} for details of our calculation. The bare jet diagram (a) is given by
\begin{align}
\label{eq:JetNaiveA}
{\cal J}^{(1)}_{\rm{a}}(\T_{R}, 0) 
& = \frac{\as(\mu) C_F}{\pi}\, \frac{e^{\e \gamma_E}}{\Gamma(1-\e)} \Big(\frac{\mu}{Q} \Big)^{\!2\e}\! \frac{1}{\T_{R}^{1+\frac{2\e}{1+b}}}\,\,  \frac{1-\e}{1+b} \int_{0}^{1}\! \df x\, x\, [(1 - x)^{-b} + x^{- b}]^{\!\frac{2\e}{1+b}}\, , 
\end{align}
where $x = l^{-}/Q$ is the light-cone momentum fraction of the radiated gluon. The factor $1/\T^{1+\bt}$ should be interpreted as a plus distribution with infinity boundary, satisfying the condition $\displaystyle \int_{0}^{\infty}\! \df\T\, \Big[1/\T^{1+\bt} \Big]_{+}^{\infty}\!\! = 0$. 
The definition of this distribution and related useful identities are collected in App.~\ref{appsec:PlusDistributions}. 
We expand the result in \eq{eq:JetNaiveA} in powers of $\e$
to identify divergent and finite parts,
\begin{eqnarray}
\label{eq:JetDiagramA}
{\cal J}^{(1)}_{\rm{a}}(\T_{R}, 0) = \frac{\as(\mu) C_F}{\pi} \bigg[\!\!-\! \frac{1}{4\, \e}\, \delta(\T_{R})\! +\! \frac{1}{2(1+b)} \frac{1}{m} \!\left[\frac{1}{\T_{R}/m}\right]_{+} \ns + i_a \, \delta(\T_{R})  + \mathcal{O}(\e) \bigg] , 
\end{eqnarray}
where
\begin{equation}
\label{eq:IntegralADefine}
i_a = \frac{1}{4(1+b)} - \frac{1}{1+b}\int_{0}^{1}\! \df x\, x\, \ln\!\Big(1+\Big(\frac{x}{1-x} \Big)^{\! b} \Big)\, ,
\end{equation} 
and $m = (\mu/Q)^{\! 1+b}$.
Here $\left[1/\T\right]_{+}$ denotes the plus distribution with boundary at $\T = +1$, see App.~\ref{appsec:PlusDistributions}. 

For the jet diagram (b), we include a factor of 2 to account for its mirror image. Moreover, as we are going to show, diagram (b) also yields a non-vanishing zero-bin subtraction. The result of the unsubtracted jet diagram (b) is 
\begin{equation}
\label{eq:JetNaiveB}
{\cal J}^{(1)}_{\rm{b, unsub}}(\T_{R}, 0)  
= \frac{2}{1+b}\, \frac{\as(\mu) C_F}{\pi}\, \frac{e^{\e \gamma_E}w^2}{\Gamma(1-\e)}\, \Big(\frac{\mu}{Q} \Big)^{\!2\e}\! \Big(\frac{\nu}{Q} \Big)^{\!\eta}\! \frac{1}{\T_{R}^{1+\frac{2\e}{1+b}}}\! \int_0^1\!\!\! \df x\, [(1 - x)^{- b} + x^{-b}]^{\!\frac{2\e}{1+b}}\, \frac{(1-x)}{x^{1+\eta}} ,
\end{equation}
where $\nu$ is the rapidity renormalization scale and $w$ is a bookkeeping parameter to keep track of the rapidity divergences.$\footnote{$w = 1 + {\cal O}(\eta)$ and will be important for obtaining the renormalization group equations.}$ Details of the calculation are presented in App.~\ref{appsubsubsec:JetBPositive}.

Here again, we expand the integrand in the regulators $\eta$ and $\e$ before evaluating the integral, as in the case of jet diagram (a). We stress that the full $\e$-dependence must be retained in the term containing the rapidity divergence as discussed in Ref.~\cite{Chiu:2012ir} and the proper order in taking the limits must be $\eta \to 0$, then $\e \to 0$ with $\eta/\e^n \to 0$ for all $n>0$.
The unsubtracted result for the jet diagram (b) exhibits a singularity at $x \rightarrow 0$ which is regulated by $\eta$. However, a careful examination of the integral shows that 
$\eta$ is required to regulate the divergence at $x \rightarrow 0$ only when $ b \leq 0$. Hence, the $\eta$-regulator can be safely dropped for positive-$b$ angularities, as the $x \rightarrow 0$ limit in this case is already regulated by $\e$. Thus for $b>0$, we get
\begin{align}
\label{eq:JetNaivePositiveB}
{\cal J}^{(1+)}_{b,\rm{unsub}}(\T_R,0) &= \frac{\al_s(\mu)\, C_F}{\pi}  \bigg[\frac{1+b}{2\,b\,\e^2}\, \delta(\T_R) + \frac{1}{\e} \delta(\T_R) - \frac{1}{b\,\e} \frac{1}{m} \bigg[\frac{1}{\T_R/m}\bigg]_{+} \ns  + \frac{2}{b(1+b)} \frac{1}{m} \bigg[\frac{\ln(\T_R/m)}{\T_R/m}\bigg]_{+} \ns  \nn \\
&\quad  -\frac{2}{1+b}\, \frac{1}{m}\, \bigg[\frac{1}{\T_R/m}\bigg]_{+} \ns  \! -\! \frac{2}{1+b}\,\delta(\T_R)\!\! \int_0^1 \df x\, \frac{(1-x)}{x} \ln\Big(1+\Big(\frac{x}{1-x}\Big)^{\! b}\Big) \nn \\
&\quad - \frac{\pi^2}{24}\, \frac{1+b}{b}\, \delta(\T_R) \! +\frac{2\,b}{1+b}\, \delta(\T_R)  + \mathcal{O}(\e) \bigg] \, ,
\end{align}
$\text{with}, m = (\mu/Q)^{\! 1+b}$ as before.
The zero-bin contribution ${\cal J}^{(1+)}_{\rm{b, 0}}$ for this graph is obtained by taking the gluon momentum to scale as $l^{\mu} \sim Q(\lambda^{1+b},\lambda^{1+b},\lambda^{1+b})$. For the explicit calculation we refer to App.~\ref{appsubsec:ZeroBin}. We obtain
\begin{align}
\label{eq:ZeroBin}
{\cal J}^{(1+)}_{\rm{b, 0}}(\T_R,0) 
&= \frac{2}{1+b} \frac{\as(\mu) C_F}{\pi} \frac{e^{\e \gamma_E}}{\Gamma(1-\e)} w^2\, \Big(\frac{\nu}{Q}\Big)^{\!\eta}\, \Big(\frac{\mu}{Q}\Big)^{\!2\e} \frac{1}{\T_R^{1+\frac{2\e}{1+b}}} \frac{\Gamma(\frac{\eta}{b})\, \Gamma(- \frac{2\e}{1+b} - \frac{\eta}{b})}{b{\,}\Gamma(-\frac{2\e}{1+b})}\, ,
\end{align}
which upon expansion in $\eta$ and then in $\e$ unveils the rapidity divergence,

\begin{align} 
{\cal J}_{b,0}^{(1+)}(\T_{R}, 0) &= \frac{\as(\mu)\, C_F}{\pi} \bigg[\frac{2\, w^2\, e^{\e\, \gamma_E}}{\eta\, (1+b)\, \Gamma(1-\e)}\, \Big(\frac{\mu}{Q}\Big)^{\! 2\e}\! \bigg[\frac{1}{\T_R^{1+\frac{2\e}{1+b}}}\bigg]_{+}^{\infty} \ns+ \frac{1+b}{2\,b\,\e^2}\, \delta(\T_R) - \frac{1}{b\,\e} \frac{1}{m} \bigg[\frac{1}{\T_R/m}\bigg]_{+} \ns  \nn \\
&\quad -\frac{1}{\e} \ln\!\frac{\nu}{Q}\, \delta(\T_R) + \frac{2}{b(1+b)} \frac{1}{m} \bigg[\frac{\ln(\T_R/m)}{\T_R/m}\bigg]_{+} \ns +\frac{2}{1+b}\, \ln\!\frac{\nu}{Q}\, \frac{1}{m} \bigg[\frac{1}{\T_R/m}\bigg]_{+}\ns \nn \\
&\quad -\frac{\pi^2}{24\,b}\, \frac{b^2+2\,b+9}{1+b}\,  \delta(\T_R) + \mathcal{O}(\eta,\e) \bigg] \, .
\end{align} 
The total contribution of diagram (b) including the zero-bin subtraction for $b>0$ is then 
\begin{align} 
\label{eq:JetDiagramB}
{\cal J}_{b}^{(1+)}(\T_{R}, 0) &= {\cal J}^{(1+)}_{b,\rm{unsub}}(\T_{R}, 0) - {\cal J}^{(1+)}_{\rm{b, 0}}(\T_{R}, 0) \nn \\
&= \frac{\as(\mu) C_F}{\pi} \bigg[\!\!- \frac{2{\,}w^2\, e^{\e \gamma_E}}{\eta{\,}(1+b){\,}\Gamma(1-\e)}\, \Big(\frac{\mu}{Q}\Big)^{\!2\e} \bigg[\frac{1}{\T_{R}^{1+\frac{2\e}{1+ b}}} \bigg]_{+}^{\infty} \ns + \frac{1}{\e}\, \ln\!\frac{\nu}{Q}\, \delta(\T_{R}) + \frac{1}{\e}\, \delta(\T_{R})   \nn \\ 
& \quad  - \frac{2}{1+b} \, \ln\!\frac{\nu}{Q}\, \frac{1}{m} \bigg[\frac{1}{\T_{R}/m} \bigg]_{+} \ns - \frac{2}{1+b} \frac{1}{m} \bigg[\frac{1}{\T_{R}/m}\bigg]_{+} \ns - i_b^{(+)}{\,}\delta(\T_{R})  + \mathcal{O}(\eta,\e) \bigg]\, ,
\end{align} 
where
\begin{equation}
\label{eq:IntegralBDefine}
i_b^{(+)} = \frac{2}{1+b} \int_{0}^{1} \df x\, \frac{(1-x)}{x} \ln\!\Big(1\!+\!\Big(\frac{x}{1-x}\Big)^{\! b} \Big) - \frac{2{\,} b}{1+b} - \frac{\pi^2}{3\, b(1+b)}\, .
\end{equation}
It is interesting to note that the zero-bin plays an important role in this calculation. In the case of the broadening jet function in Ref.~\cite{Chiu:2012ir}, the rapidity divergence emerged from the unsubtracted equivalent of diagram (b) and the zero-bin vanished, using the same regulators and gauge as done here. Instead, for positive-$b$ angularities the rapidity divergence emerges from the zero-bin subtraction, which is non-vanishing. 

Adding the results for diagram (a) and (b), we get the bare one-loop right-hemisphere jet function for positive-$b$ angularities,
\begin{align}
\label{eq:JetResult}
&{\cal J}^{(1+)}(\T_{R} , 0) = \frac{\as(\mu) C_F}{\pi} \bigg[\!\!- \frac{2{\,} w^2\, e^{\e \gamma_E}}{\eta{\,}(1+b){\,}\Gamma(1-\e)}\, \Big(\frac{\mu}{Q}\Big)^{\!2\e} \bigg[\frac{1}{\T_{R}^{1+\frac{2\e}{1+ b}}} \bigg]_{+}^{\infty} \ns + \frac{1}{\e}\, \ln\!\frac{\nu}{Q}\, \delta(\T_{R}) + \frac{3}{4\,\e}\, \delta(\T_{R}) \nn \\ 
& \qquad  - \frac{2}{1+b} \, \ln\!\frac{\nu}{Q}\, \frac{1}{m} \bigg[\frac{1}{\T_{R}/m} \bigg]_{+}\ns - \frac{3}{2(1+b)} \frac{1}{m} \bigg[\frac{1}{\T_{R}/m}\bigg]_{+}\ns +(i_a  - i_b^{(+)}){\,}\delta(\T_{R}) + \mathcal{O}(\eta,\e)  \bigg]\, . 
\end{align}
The bare one-loop left-hemisphere jet function has the same form but with $\T_R$ replaced by $\T_L$. We stress that both the angularity jet function and the soft function, as we are going to show, contain factors of $1/b$ which make them singular as $b \to 0$. The situation here is similar to what happens in the formalism of Ref.~\cite{Larkoski:2014uqa} for angularities measured with respect to a recoil-insensitive axis. Their jet and soft functions still yielded a well-defined resummed cross section for small $b$ due to the decreasing range of integration in the RG evolution~\cite{Larkoski:2014uqa}. The same feature will emerge in our case \cite{Budhraja:2018}.


\subsection{Angularity jet function at one-loop for $b<0$}
\label{subsec:JetNLONegativeBCase}

The tree-level jet function remains the same as given in \eq{eq:TreeLevelJet} for negative-$b$ angularities. 
At one loop, the result for the jet diagram (a) also remains unchanged and the integral in \eq{eq:IntegralADefine} is non-singular for both positive and negative-$b$ values. Hence, the same expressions as \eqs{eq:JetDiagramA}{eq:IntegralADefine} hold for all $b > -1$ angularities. However, as noted earlier, the unsubtracted jet integral for diagram (b) given in \eq{eq:JetNaiveB}  exhibits a singularity as $x \rightarrow 0$ and the $\eta$-regulator is essential to deal with this limit when $b \leq 0$. The result for the unsubtracted jet diagram (b) upon expansion in $\eta$ and $\e$, for $-1<b<0$ is given by
\begin{align}
\label{eq:JetNaiveNegativeB}
{\cal J}^{(1-)}_{\rm{b, unsub}}(\T_R,0) &= \frac{\al_s(\mu)\, C_F}{\pi} \bigg[\!\!-\! \frac{2{\,} w^2\, e^{\e \gamma_E}}{\eta{\,}(1+b){\,}\Gamma(1-\e)}\, \Big(\frac{\mu}{Q}\Big)^{\!2\e} \bigg[\frac{1}{\T_{R}^{1+\frac{2\e}{1+ b}}} \bigg]_{+}^{\infty} \ns + \frac{1}{\e} \ln\!\frac{\nu}{Q}\, \delta(\T_R) +\frac{1}{\e}\, \delta(\T_R)  \nn \\
&\quad - \frac{2}{1+b} \ln\!\frac{\nu}{Q}\, \frac{1}{m} \bigg[\frac{1}{\T_R/m}\bigg]_{+} \ns -\frac{2}{1+b} \frac{1}{m} \bigg[\frac{1}{\T_R/m}\bigg]_{+}\ns +\frac{2\,b}{1+b}\,\delta(\T_R) - \frac{\pi^2\,b}{3(1+b)}\, \delta(\T_R)  \nn \\
&\quad -\! \frac{2}{1+b}\,\delta(\T_R) \!\!\int_0^1\!\! \df x\, \frac{(1-x)}{x} \ln\!\Big(1+\!\Big(\frac{x}{1-x}\Big)^{\!-b}\Big) \!+ \mathcal{O}(\eta,\e)\bigg]\, . 
\end{align}
The intermediate steps of the calculation can be found in App.~\ref{appsubsec:JetBNegative}.
The zero-bin diagram flips sign for negative-$b$ (see App.~\ref{appsubsec:ZeroBin}) leading to
\begin{align}
\label{eq:ZeroBinNegativeB}
{\cal J}^{(1-)}_{\rm{(b, 0)}}(\T_{R},0) &= -\frac{2}{1+b} \frac{\as(\mu) C_F}{\pi} \frac{e^{\e \gamma_E}}{\Gamma(1-\e)} w^2\, \Big(\frac{\nu}{Q}\Big)^{\!\eta}\, \Big(\frac{\mu}{Q}\Big)^{\!2\e} \frac{1}{\T_R^{1+\frac{2\e}{1+b}}} \frac{\Gamma(\frac{\eta}{b})\, \Gamma(- \frac{2\e}{1+b} - \frac{\eta}{b})}{b{\,}\Gamma(-\frac{2\e}{1+b})}\, \nn \\
&= \frac{\as(\mu)\, C_F}{\pi} \bigg[\!\!-\frac{2\, w^2\, e^{\e\, \gamma_E}}{\eta\, (1+b)\, \Gamma(1-\e)}\, \Big(\frac{\mu}{Q}\Big)^{\! 2\e}\! \bigg[\frac{1}{\T_R^{1+\frac{2\e}{1+b}}}\bigg]_{+}^{\infty} \ns - \frac{1+b}{2\,b\,\e^2}\, \delta(\T_R) + \frac{1}{b\,\e} \frac{1}{m} \bigg[\frac{1}{\T_R/m}\bigg]_{+} \ns \nn  \\
&\quad +\frac{1}{\e} \ln\!\frac{\nu}{Q}\, \delta(\T_R) - \frac{2}{b(1+b)}\, \frac{1}{m}\, \bigg[\frac{\ln(\T_R/m)}{\T_R/m}\bigg]_{+} \ns -\frac{2}{1+b} \, \ln\!\frac{\nu}{Q}\, \frac{1}{m} \bigg[\frac{1}{\T_R/m}\bigg]_{+}\ns \nn \\
& \quad +\frac{\pi^2}{24\,b}\, \frac{b^2+2\,b+9}{1+b}\, \delta(\T_R) \! + \mathcal{O}(\eta,\e) \bigg] \, . 
\end{align}
The total contribution of diagram (b) for angularities with $b<0$ is then
\begin{align}
\label{eq:JetDiagramBNegativeCase}
{\cal J}_{\rm b}^{(1-)}(\T_{R} , 0) &= \frac{\as(\mu) C_F}{\pi} \bigg[ \frac{1+b}{2\, b\, \e^2}\, \delta(\T_R) - \frac{1}{b\, \e}\, \frac{1}{m} \bigg[\frac{1}{\T_{R}/m} \bigg]_{+} \!\!\!  + \frac{1}{\e} \delta(\T_R) + \frac{2}{b\, (1+b)}\, \frac{1}{m} \bigg[\frac{\ln(\T_R/m)}{\T_{R}/m} \bigg]_{+} \ns  \nn \\ 
& \quad  - \frac{2}{1+b}\, \frac{1}{m} \bigg[\frac{1}{\T_{R}/m}\bigg]_{+}\ns -\frac{\pi^2}{24}\, \frac{1+b}{b} \delta(\T_R) - i^{(-)}_b{\,}\delta(\T_{R}) + \mathcal{O}(\e) \bigg]\, ,
\end{align}
where
\begin{equation}
\label{eq:IntegralBDefineNegativeB}
i^{(-)}_b = \frac{2}{1+b} \int_{0}^{1} \df x\, \frac{(1-x)}{x} \ln\!\Big(1\!+\!\Big(\frac{x}{1-x}\Big)^{\! -b} \Big) - \frac{2{\,} b}{1+b} + \frac{\pi^2}{3(1+b)}\, \Big(\frac{b^2+1}{b}\Big)\, .
\end{equation}
Adding the contributions of diagrams (a) and (b), we obtain the bare jet function for $b<0$,
\begin{align}
\label{eq:JetResultNegativeB}
{\cal J}^{(1-)}(\T_{R} , 0) &= \frac{\as(\mu) C_F}{\pi} \bigg[\frac{1+b}{2\,b\,\e^2}\, \delta(\T_R) - \frac{1}{b\, \e}\, \frac{1}{m} \bigg[\frac{1}{\T_{R}/m} \bigg]_{+} \ns + \frac{3}{4\,\e}\, \delta(\T_{R}) + \frac{2}{b(1+b)} \frac{1}{m} \bigg[\frac{\ln(\T_R/m)}{\T_{R}/m} \bigg]_{+}\ns \nn \\ 
& \quad   - \frac{3}{2(1+b)} \frac{1}{m} \bigg[\frac{1}{\T_{R}/m}\bigg]_{+}\ns - \frac{\pi^2}{24} \frac{1+b}{b}\, \delta(\T_R) +(i_a  - i^{(-)}_b){\,}\delta(\T_{R}) + \mathcal{O}(\eta,\e) \bigg] .
\end{align}
It is interesting to note that the zero-bin graph exhibits a rapidity divergence for both positive-$b$ and negative-$b$ integrals. It is only the unsubtracted jet function graph (b) that does not contain any rapidity divergence for positive-$b$ angularities. 
The negative-$b$ angularity jet function has no net rapidity divergence as both the unsubtracted jet diagram and its zero-bin contain rapidity divergences which cancel in the sum. In addition, it contains $\e$-divergences proportional to $1/b$ which cancel against the soft function divergence (see \eq{eq:SoftExpandedNegativeB} below), as expected since the hard function is universal for any angularity.

\subsection{Jet Broadening Limit of the one-loop jet function}
\label{JetBroadLimit}
We will now show that we are able to reproduce the jet broadening ($b=0$) jet function from our previous results. This provides a cross-check on our calculation. However, it is important to stress that one cannot retrieve the jet broadening limit from the angularity jet function {\it after} expanding in $\eta$ and $\e$, as the limits $\eta,\e \rightarrow 0$ do not commute with $b \rightarrow 0$.

The unexpanded result of \eq{eq:JetNaiveA} for the jet graph (a) in the limit $b \rightarrow 0^{\pm}$ reduces to 
\begin{align}
\label{eq:JetBroadeningA}
\lim_{b \rightarrow 0^{\pm}}{\cal J}^{(1)}_{\rm{a}}(\T_{R}, 0) &= \frac{\as(\mu) C_F}{\pi}\, \frac{e^{\e \gamma_E}}{\Gamma(1-\e)} \Big(\frac{\mu}{Q} \Big)^{\!2\e} \bigg[\frac{1}{\T_{R}^{1+2\e}} \bigg]_{+}^{\infty} (1-\e) \int_{0}^{1}\! \df x\, x\, 2^{2\e}\, \nn \\
& = \frac{\as(\mu) C_F}{2\pi}\, \frac{e^{\e \gamma_E}}{\Gamma(1-\e)} \Big(\frac{2\, \mu}{Q} \Big)^{\!2\e} \bigg[\frac{1}{\T_{R}^{1+2\e}} \bigg]_{+}^{\infty} (1-\e)\, .
\end{align}
The jet diagram (b) of \eq{eq:JetNaiveB} including the zero-bin subtraction (see \eq{appeq:ZeroBinIntegral}), in the limit $b \rightarrow 0^{\pm}$, gives
\begin{align}
\label{eq:JetBroadeningB}
\lim_{b \rightarrow 0^{\pm}}{\cal J}^{(1)}_{\rm{b}}(\T_{R}, 0) &= 2\, \frac{\as(\mu) C_F}{\pi}\, \frac{e^{\e \gamma_E}\, w^2 }{\Gamma(1-\e)} \Big(\frac{\mu}{Q} \Big)^{\!2\e} \Big(\frac{\nu}{Q} \Big)^{\!\eta} \bigg[\frac{1}{\T_{R}^{1+2\e}} \bigg]_{+}^{\infty}\! \bigg[ \int_{0}^{1}\!\! \df x\, 2^{2\e} \frac{(1-x)}{x^{1+\eta}} -\!\! \int_0^{\infty}\!\!\!\! \df x \frac{1}{x^{1+\eta}}\, 2^{2\e}\bigg]\nn \\
&= 2\, \frac{\as(\mu) C_F}{\pi}\, \frac{e^{\e \gamma_E}\, w^2}{\Gamma(1-\e)} \Big(\frac{2\, \mu}{Q} \Big)^{\!2\e} \Big(\frac{\nu}{Q} \Big)^{\!\eta} \bigg[\frac{1}{\T_{R}^{1+2\e}} \bigg]_{+}^{\infty} \bigg[\!-\frac{1}{\eta}-1+ \mathcal{O}(\eta) \bigg]\, . 
\end{align}
Adding the contributions in \eqs{eq:JetBroadeningA}{eq:JetBroadeningB}, we get the one-loop jet function in the limit $b \rightarrow 0$,
\begin{equation}
\label{eq:BroadeningJetFunc}
\lim_{b \rightarrow 0}{\cal J}^{(1)}(\T_{R}, 0) = \frac{\as(\mu) C_F}{2\pi}\, \frac{e^{\e \gamma_E}}{\Gamma(1-\e)} \Big(\frac{2\, \mu}{Q} \Big)^{\!2\e} \bigg[\frac{1}{\T_{R}^{1+2\e}} \bigg]_{+}^{\infty} \bigg[ (1-\e) + 4\, w^2 \Big(\frac{\nu}{Q} \Big)^{\!\eta} \bigg(\!\!-\frac{1}{\eta}-1\!\bigg)\bigg]\, ,
\end{equation}
which agrees with the broadening jet function given in Eq.~(6.38) of Ref.~\cite{Chiu:2012ir}.


\section{Angularity soft function at one loop}
\label{sec:SoftFunc}

In this section, we present our calculation of the bare angularity soft function to one loop, based on the definition given in \eq{eq:SoftDefine}. The one-loop soft function for angularities is a generalization of the one-loop soft function for the jet broadening case.  However, the presence of the exponent $b$, not only makes the calculation significantly more difficult than in the case of jet broadening, but also the choice of distribution variables becomes highly non-trivial, due to the presence of coupled distributions. As in the case of the jet function and consistent with it, we find a net rapidity divergence for the soft function when $b>0$, while no net rapidity divergence is present when $b<0$. We provide novel one-loop results of the soft function for both the positive and the negative value of the angularity exponent.

\subsection{Soft function calculation for $b>0$}
\label{subsec:SoftNLO}

\begin{figure}[t]
 \includegraphics[width=15cm]{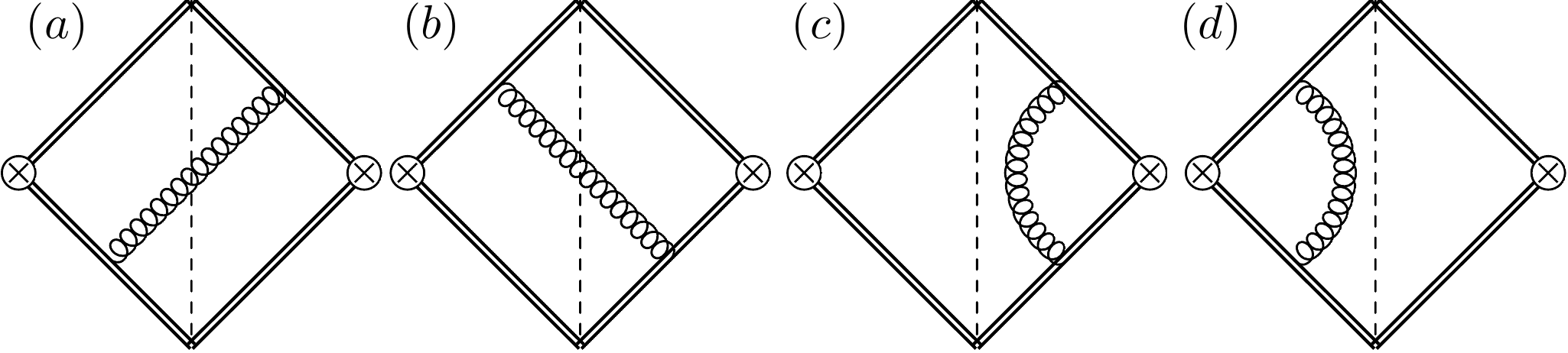}
 \centering
 \caption{\label{fig:SoftDiagrams} Diagrams contributing to the angularity soft function at one-loop order. Diagrams (a) and (b) give identical contribution due to symmetry.}
 \end{figure}

According to \eq{eq:SoftDefine}, the tree-level soft function is simply given by
\begin{equation}
\label{eq:TreeLevelSoft}
{\cal S}^{(0)}(\T_{L},\T_{R},\V{p},\V{k}) = \delta(\T_{L})\, \delta(\T_{R})\, \delta(\V{p})\, \delta(\V{k})\, .
\end{equation}
The relevant Feynman diagrams for the one-loop computation are shown in Fig.~\ref{fig:SoftDiagrams}.
As in the case of the jet function, we will use the $\eta$-regulator for rapidity divergences and dimensional regularization for both UV and IR divergences. Due to our choice of gauge, a non-zero contribution arises only when a gluon is exchanged between $n$-collinear and $\bar{n}$-collinear Wilson lines. Virtual correction diagrams (c) and (d) are given by scaleless integrals and therefore vanish due to our choice of regulators. 
Since (a) and (b) give identical contributions, we only need to compute one diagram at this order. The details of this calculation are presented in App.~\ref{appsec:SoftIntegrals}.
The bare one-loop soft function for all $b>-1$ is given by
\begin{align}
\label{eq:OneLoopSoft}
{\cal S}^{(1)}(\T_{L},\T_{R},\V{p},\V{k})
& = \frac{\as(\mu){\,}C_F}{\pi} \frac{\mu^{2\e}e^{\e \gamma_E}}{\Gamma(1-\e)} \nu^{\eta}\, w^2\, Q\, \delta(\T_{L})\, \delta(\V{k})\, \theta\Big(\Big(\frac{\MV{p}}{Q{\,}\T_{R}}\Big)^{\!1/b} \ns - 1\Big)  \nn \\
& \quad \times (\V{p})^{\,\,-1-\e-\frac{\eta+1}{2}} \frac{\Big\vert 1\!-\Big(\frac{Q{\,}\T_{R}}{\MV{p}}\Big)^{\!2/b}\Big\vert^{-\eta}}{\big\vert b\big\vert{\,}\Big(\frac{Q{\,}\T_{R}}{\MV{p}}\Big)^{\!1-\eta/b}} + \bigg\{ 
  \begin {array} {c} 
  \T_L \leftrightarrow \T_R \\ 
  \V{k} \leftrightarrow \V{p} 
  \end {array} \bigg\}\, .
\end{align}
The term $\theta\Big(\Big(\frac{\MV{p}}{Q{\,}\T_{R}}\Big)^{\!1/b} \ns - 1\Big)\, \Big\vert 1\!-\Big(\frac{Q{\,}\T_{R}}{\MV{p}}\Big)^{\!2/b}\Big\vert^{-\eta}$ in this equation couples the variables $\T_{R}$($\T_L$) and $\vt{p}$($\vt{k}$). In order to avoid dealing with coupled distributions in $\T_{R}$ and $\vt{p}$ ($\T_{L}$ and $\vt{k}$), which are difficult to interpret, we perform a change of variables. Noting that the coupled Heaviside step function is equivalent to $\theta(\MV{p}-Q\, \T_R)$ for $b>0$, we introduce the variables
\begin{align}
&v_{L} = \frac{Q \T_{L}}{k_{\perp}} \,, \quad \quad \quad  v_{R} = \frac{Q \T_{R}}{p_{\perp}} \, , \label{eq:PositiveBDefinition} 
\end{align}
where $p_{\perp} \equiv \MV{p}$ and $k_{\perp} \equiv \MV{k}$. Accordingly, for positive-$b$ angularities,
\begin{align}
\label{eq:SoftNewVariables}
{\cal S}^{(1+)}(\T_{L},\T_{R},\V{p},\V{k}) & = {\cal S}^{(1+)}(v_{L},v_{R},\V{p},\V{k})\, \bigg[\bigg\vert\frac{\df v_{L}}{\df \T_{L}}\bigg\vert \bigg\vert \frac{\df v_{R}}{\df \T_{R}}\bigg\vert\bigg]\, \nn \\ 
& = \frac{\as(\mu){\,}C_F}{\pi} \frac{e^{\e \gamma_E}}{\Gamma(1-\e)}\, w^2 \Big(\frac{\nu}{\mu}\Big)^{\eta}\, \delta(v_{L})\, \delta(\V{k})\,  \frac{\theta(1-v_{R})\, (1-v_{R}^{2/b})^{-\eta}}{b\, v_{R}^{1-\eta/b}} \nn \\ 
& \quad \times \frac{1}{\mu^2} \frac{1}{(\V{p}/\mu^2)^{1+\e+\frac{\eta}{2}}}\, \bigg[\bigg\vert\frac{\df v_{L}}{\df \T_{L}}\bigg\vert \bigg\vert \frac{\df v_{R}}{\df \T_{R}}\bigg\vert\bigg]\,  + \bigg\{ 
  \begin {array} {c} 
  v_L \leftrightarrow v_R \\ 
  \V{k} \leftrightarrow \V{p} 
  \end {array} \bigg\}\, ,
\end{align}
where 
\begin{align}
\bigg\vert\frac{\df v_{L}}{\df \T_{L}}\bigg\vert = \frac{Q}{k_{\perp}} \, , \qquad  \bigg\vert\frac{\df v_{R}}{\df \T_{R}}\bigg\vert = \frac{Q}{p_{\perp}}\, ,  \label{eq:PositiveBJacobian} 
\end{align}
which enter the Jacobian factor. Keeping track of Jacobian factors is necessary since the soft function is a distribution and its arguments represent the variables of the integration measure.  Thanks to \eq{eq:PositiveBDefinition}, the expansions in $\eta$ and $\e$ give rise to decoupled distributions in $v_{R}$ and $\vt{p}$ ($v_{L}$ and $\vt{k}$).
The distribution in $v_{R}$, upon expansion in $\eta$ leads to
\begin{equation}
\label{eq:v_exp}
\bigg[\frac{\theta(1-v_{R})\, (1-v_{R}^{2/b})^{-\eta}}{v_{R}^{1-{\eta/b}}}\bigg]_{+}^{\infty} \!\!\! = \theta(1-v_{R})\bigg( \bigg[\frac{1}{v_{R}}\bigg]_{+} \ns\, +\frac{\eta}{b} \bigg[\frac{\ln v_{R}}{v_{R}}\bigg]_{+} \ns -\eta \bigg[\frac{\ln(1-v_{R}^{2/b})}{v_{R}}\bigg]_{+} \bigg)\! + a\, \delta(v_R) + {\cal O}(\eta^2) ,
\end{equation}
where we follow the standard convention for plus distributions, i.e. $\displaystyle \!\! \int_0^1 \!\! \df x \Big[\frac{\ln^n x }{x^{1+\al}}\Big]_{+} \ns = 0 = \!\! \int_0^1 \!\! \df x \Big[\frac{\ln^n(1-x^{\al})}{x}\Big]_{+}$. To fix the coefficient $a$ of the $\delta(v_{R})$ term, we integrate the above equation over $v_{R}$ from 0 to 1, which leads to
\begin{align}
\label{eq:v_fullexp}
\bigg[\frac{\theta(1-v_{R})\, (1-v_{R}^{2/b})^{-\eta}}{v_{R}^{1-{\eta/b}}}\bigg]_{+}^{\infty} \!\!\! &= \frac{b}{\eta}\, \delta(v_{R}) + \bigg[\frac{\theta(1-v_{R})}{v_{R}}\bigg]_{+} \ns\, + \eta \bigg(\frac{\pi^2}{12}\, b\, \delta(v_R) + \frac{1}{b} \bigg[\frac{\theta(1-v_{R}) \ln v_{R}}{v_{R}}\bigg]_{+} \ns \nn \\
&\quad  - \bigg[\frac{\theta(1-v_{R}) \ln(1-v_{R}^{2/b})}{v_{R}}\bigg]_{+}  \bigg) + {\cal{O}}(\eta^{2})\, .
\end{align}
The expansion in the regulators for the one-loop bare soft function thus gives
\begin{align}
\label{eq:SoftExpanded}
& {\cal S}^{(1+)}(v_{L},v_{R},\V{p},\V{k}) = \frac{\as(\mu){\,}C_F}{\pi} \Bigg\{ \frac{w^2\,e^{\e \gamma_E}}{\eta\, \Gamma(1-\e)} \frac{1}{\mu^2}\bigg[\frac{1}{(\V{p}/\mu^2)^{1+\e}} \bigg]_{+}^{\infty} \ns\,\, \delta(v_{R}) \!-\! \frac{1}{b\, \e}\bigg[\frac{\theta(1-v_{R})}{v_{R}}\bigg]_{+}\!\! \delta(\V{p})  \nn  \\ 
& \quad + \delta(v_{R})\, \delta(\V{p}) \bigg( \frac{1}{2\e^2} \!-\! \frac{1}{\e}\, \ln\!\frac{\nu}{\mu} \bigg)\! - \frac{1}{2}\delta(v_{R})\frac{1}{\mu^2} \bigg[\frac{\ln(\V{p}/\mu^2)}{\V{p}/\mu^2}\bigg]_{+}^{\cal D_{\mu}}  \ns + \ln\!\frac{\nu}{\mu}\, \delta(v_{R})\, \frac{1}{\mu^2} \bigg[\frac{1}{\V{p}/\mu^2}\bigg]_{+}^{\cal D_{\mu}} \ns \nn \\
& \quad  + \frac{1}{b}\bigg[\frac{\theta(1-v_{R})}{v_{R}}\bigg]_{+} \frac{1}{\mu^2} \bigg[\frac{1}{\V{p}/\mu^2}\bigg]_{+}^{\cal D_{\mu}} \ns - \frac{\pi^2}{24} \delta(v_{R})\, \delta(\V{p}) + {\cal O}(\eta,\e) \Bigg\} \, \delta(v_L)\, \delta(\V{k}) \! + \bigg\{ 
  \begin {array} {c} 
  v_L \leftrightarrow v_R \\ 
  \V{k} \leftrightarrow \V{p} 
  \end {array} \bigg\}\, .  
\end{align}
The $\eta$-divergence in this equation is written with full $\epsilon$-dependence in terms of a plus distribution with infinity boundary. 
The plus distributions over a vector domain of the type $\displaystyle {\cal L}_n^\beta(\mu,\vt{p}) \equiv \frac{1}{2\pi\mu^2} \bigg[\frac{1}{(\V{p}/\mu^2)^{1+\beta}}\ln^n(\V{p}/\mu^2)\bigg]_{+}^{\cal D_{\mu}}\!\!$ used in this equation, appear also in the jet broadening case~\cite{Chiu:2012ir} as the integration measure is two-dimensional. They satisfy the condition 
\begin{equation}
\label{eq:VectorPlusDistribution}
\int_{{\cal D}_{\mu}} \frac{\df^2\vec{p}_{\perp}}{(2\pi)^2}\; {\cal L}_n^\beta(\mu,\vt{p}) = 0~,
\end{equation}
where $\cal D_{\mu}$ denotes a disc of radius $\mu$ 
around the origin in the $\vt{p}$-space. For further details on these distributions we refer to App.~\ref{appsec:VectorPlusDistributions}.


\subsection{Soft function calculation for $b<0$}
\label{subsec:SoftNLONegativeB}

The results for the tree-level and the one-loop soft function given in \eqs{eq:TreeLevelSoft}{eq:OneLoopSoft} remain the same for negative-$b$ angularities as well. However, for $b<0$, the Heaviside $\theta$-function constraint is now equivalent to $\theta(Q\, \T_R-\MV{p})$, which leads to the new set of variables
\begin{align}
u_{L} = \frac{k^2_{\perp}}{Q^2 \T^2_{L}} \,, \qquad  u_{R} = \frac{p^2_{\perp}}{Q^2 \T^2_{R}}\, ,  \label{eq:NegativeBDefinition}
\end{align}
to be used in \eq{eq:OneLoopSoft} in order to obtain decoupled distributions. This yields
\begin{align}
\label{eq:SoftNewVariables2}
{\cal S}^{(1-)}(\T_{L},\T_{R},\V{p},\V{k}) & = {\cal S}^{(1-)}(\T_{L},\T_{R},u_L,u_R)\, \bigg[\bigg\vert\frac{\df u_{L}}{\df k^2_{\perp}}\bigg\vert\, \bigg\vert\frac{\df u_{R}}{\df p^2_{\perp}}\bigg\vert\bigg]\,  \\ 
& = - \frac{\as(\mu){\,}C_F}{\pi} \frac{e^{\e \gamma_E}\, \mu^{2\e}}{\Gamma(1-\e)}\, w^2 \nu^{\eta}\, Q\, \delta(\T_{L})\, \delta(u_L)\,  \frac{1}{(Q\T_R)^{1+2\e+\eta}} \nn \\ 
& \quad \times  \frac{\theta(1-u_{R})\, (1-u_{R}^{-1/b})^{-\eta}}{b\, u_{R}^{1+\e+\eta/2+\eta/2\,b}} \, \bigg[\bigg\vert\frac{\df u_{L}}{\df k^2_{\perp}}\bigg\vert\, \bigg\vert\frac{\df u_{R}}{\df p^2_{\perp}}\bigg\vert\bigg] + \bigg\{ 
  \begin {array} {c} 
  \T_L \leftrightarrow \T_R \\ 
  u_{L} \leftrightarrow u_{R} 
  \end {array} \bigg\}\, \nn ,  
\end{align}
where
\begin{align}
\bigg\vert\frac{\df u_{L}}{\df k^2_{\perp}}\bigg\vert = \frac{1}{Q^2\T_L^2} \, , \qquad \bigg\vert\frac{\df u_{R}}{\df p^2_{\perp}}\bigg\vert = \frac{1}{Q^2\T_R^2} \, .\label{eq:NegativeBJacobian}
\end{align}
The numerator term $(1-u_{R}^{-1/b})^{-\eta}$, like before, does not contribute any singular nor finite term. 
Moreover, in the present case, all distributions in both $\tau_{L,R}$ and $u_{L,R}$ are already regulated by $\e$. Thus no rapidity divergence is contained in \eq{eq:SoftNewVariables2}. Upon expansion of this equation we obtain
\begin{align}
\label{eq:SoftExpandedNegativeB}
& {\cal S}^{(1-)}(\T_{L},\T_{R},u_{L},u_{R}) = \frac{\as(\mu){\,}C_F}{\pi}\,  \delta(\T_L)\, \delta(u_L)\, \Bigg\{\!\!\! -\!\frac{1}{2\, b\e^2}\delta(\T_{R})\, \delta(u_R)\! + \frac{1}{2\, b\, \e}\bigg[\frac{\theta(1-u_{R})}{u_{R}}\bigg]_{+}\!\! \delta(\T_R)   \nn \\ 
& \quad + \frac{1}{b\, \e} \delta(u_R) \frac{Q}{\mu} \bigg[\frac{1}{Q\, \T_R/\mu}\bigg]_{+}\ns  - \frac{2}{b}\, \delta(u_{R}) \frac{Q}{\mu} \bigg[\frac{\ln(Q\, \T_R/\mu)}{Q\, \T_R/\mu}\bigg]_{+} \ns - \frac{1}{2\, b} \bigg[\frac{\theta(1-u_R)\, \ln u_R}{u_R}\bigg]_{+} \delta(\T_R) \nn \\
& \quad   - \frac{1}{b}\bigg[\frac{\theta(1-u_{R})}{u_{R}}\bigg]_{+} \frac{Q}{\mu} \bigg[\frac{1}{Q\, \T_R/\mu}\bigg]_{+} \ns  + \frac{\pi^2}{24\, b}\,\delta(\T_{R})\, \delta(u_R) + {\cal O}(\eta,\e) \Bigg\} + \bigg\{ 
  \begin {array} {c} 
  \T_L \leftrightarrow \T_R \\ 
  u_{L} \leftrightarrow u_{R} 
  \end {array} \bigg\}\, .
\end{align}
The absence of rapidity divergences in the soft function is consistent with the fact that the jet function for $b<0$ in \eq{eq:JetResultNegativeB} is also free of rapidity divergences.


\subsection{Jet Broadening Limit of the one-loop soft function}
\label{subsec:SoftBroadLimit}

A non-trivial cross-check on our calculation is provided by the fact that we can reproduce the jet broadening soft function by taking the limit $b \rightarrow 0$. We will treat the cases $b>0$ and $b<0$ separately after making the suitable change of variable to $v_{L,R}$ and $u_{L,R}$, respectively, and {\it before} making expansion in $\eta$ and $\e$. 

\subsubsection{Limit $b \rightarrow 0^{+}$}

Since the functional dependence on $b$ is contained only through the redefined variable $v_{R}$ in \eq{eq:SoftNewVariables},  taking the limit $b \rightarrow 0^+$ gives
\begin{align}
\label{eq:BroadeningSoftFunc}
\lim_{b \rightarrow 0^{+}} \frac{\theta(1-v_{R})\, (1-v_{R}^{2/b})^{-\eta}}{b\, v_{R}^{-\eta/b}}\, \frac{1}{v_{R}} \bigg\vert\frac{\df v_{R}}{\df \T_{R}}\bigg\vert &= 2 \lim_{b \rightarrow 0^{+}}\frac{1}{2\,b} e^{-\vert \ln(v_{R}^{\eta})\vert/b} \, \frac{1}{v_{R}} \bigg\vert\frac{\df v_{R}}{\df \T_{R}}\bigg\vert \nn \\ 
&= 2\, \frac{\delta(\ln(v_{R}^{\eta}))}{v_{R}} \, \bigg\vert\frac{\df v_{R}}{\df \T_{R}}\bigg\vert \nn \\
&= 2\, Q^2\T_{R} \frac{ \delta(Q^2\, \T_R^2-\V{p})}{\eta} \, \bigg\vert\frac{\df \T^2_{R}}{\df \T_{R}}\bigg\vert\, ,  
\end{align}
In the first line of this equation, we have used the fact that $v_{R}^{2/b} \ll 1$ and $\ln(v_{R}^{\eta}) < 0$ for  $0 < v_{R} < 1$ as $b \rightarrow 0^+$.
Here the factor $\bigg\vert\frac{\df \T^2_{R}}{\df \T_{R}}\bigg\vert = 2$ is the Jacobian required for the variable change.
The second line of \eq{eq:BroadeningSoftFunc} follows from
\begin{equation}
\label{eq:DiracDeltaDefinition}
\delta(x) = \lim_{\e \rightarrow 0}\, \frac{1}{2\, \e} \,e^{-\frac{\vert x\vert}{\e}}\, ,
\end{equation}
while in the final step we have used the identity
\begin{equation}
\label{eq:DiracDeltaIdentity}
\delta(\ln(v_{R}^{\eta})) = \frac{\delta(v_{R}-1)}{\eta}\, ,
\end{equation}
which when combined with the Jacobian gives a distribution in terms of $\T_{R}$ and $p_{\perp}$. Substituting the result of \eq{eq:BroadeningSoftFunc} in the limit $b \rightarrow 0^{+}$ into \eq{eq:SoftNewVariables}, we get
\begin{align}
\label{eq:SoftBroadeningLimit}
\lim_{b \rightarrow 0^{+}}{\cal S}^{(1+)}(\T_{L},\T_{R},\V{p},\V{k}) &= \frac{2\,\as(\mu){\,}C_F}{\pi} \frac{ w^2\, e^{\e \gamma_E}}{\eta\,\Gamma(1-\e)} \Big(\frac{\mu}{Q}\Big)^{\! 2\e}\! \Big(\frac{\nu}{Q}\Big)^{\!\eta} \, \delta(\T_{L})\, \delta(\V{k})\, \frac{1}{(\T_R)^{1+2\e+\eta}} \nn \\
& \quad \times \delta(Q^2\T_R^2-\V{p}) + \bigg\{ 
  \begin {array} {c} 
  \T_L \leftrightarrow \T_R \\ 
  \V{k} \leftrightarrow \V{p} 
  \end {array} \bigg\} \, .  
\end{align}
After expanding in $\eta$ and $\e$, this gives
\begin{align}
\label{eq:BroadeningSoftExpanded}
\lim_{b \rightarrow 0^{+}}{\cal S}^{(1+)}(\T_{L},\T_{R},\V{p},\V{k}) &= \frac{\as(\mu) \, C_F}{\pi}\Bigg\{\frac{2\,w^2 e^{\e \gamma_E}}{\eta \, \Gamma(1-\e)} \Big(\frac{\mu}{Q}\Big)^{2\e} \bigg[\frac{1}{\T_{R}^{1+2\e}}\bigg]_{+}^{\infty} \ns + \delta(\T_{R}) \bigg(\frac{1}{2\e^2} - \frac{1}{\e} \ln\! \frac{\nu}{\mu}  \bigg) \nn \\
& \quad + 2 \ln\!\frac{\nu}{\mu}\, \frac{Q}{\mu}\bigg[\frac{\mu}{Q\T_{R}}\bigg]_{+} \ns  - \frac{2\,Q}{\mu}\bigg[\frac{\mu\, \ln(Q\,\T_{R}/\mu)}{Q\,\T_{R}}\bigg]_{+} \ns - \frac{\pi^2}{24}\, \delta(\T_R) \Bigg\}\delta(\T_{L})\, \delta(\V{k}) \nn \\
& \quad \delta(Q^2\T_{R}^2-\V{p}) + \bigg\{ 
  \begin {array} {c} 
  \T_L \leftrightarrow \T_R \\ 
  \V{k} \leftrightarrow \V{p} 
  \end {array} \bigg\} ,
\end{align}
which agrees with the one-loop soft function for jet broadening in Eq.~(6.45) of Ref.~\cite{Chiu:2012ir}.

\subsubsection{Limit $b \rightarrow 0^{-}$}

The $b \rightarrow 0^{-}$ limit can be obtained in an analogous way by writing the distribution in $u_{R}$ appearing in \eq{eq:SoftNewVariables2} as 
\begin{align}
\label{eq:SoftNegativeB}
\lim_{b \rightarrow 0^{-}} \frac{\theta(1-u_{R})\, (1-u_{R}^{-1/b})^{-\eta}}{(-b)\, u_{R}^{\eta/2\, b}} \frac{1}{(u_{R})^{1+\e+\frac{\eta}{2}}} \bigg\vert\frac{\df u_{R}}{\df p^2_{\perp}}\bigg\vert &= 2 \lim_{\vert b\vert \rightarrow 0^{+}}\frac{1}{2\,\vert b\vert} e^{-\vert \ln(u_{R}^{\eta/2})\vert/\vert b\vert} \frac{1}{(u_{R})^{1+\e+\frac{\eta}{2}}} \bigg\vert\frac{\df u_{R}}{\df p^2_{\perp}}\bigg\vert \nn \\
&= 2\, \frac{\delta(u_R-1)}{\eta}\, \bigg\vert\frac{\df u_{R}}{\df p^2_{\perp}}\bigg\vert\, ,
\end{align}
which leads to the same result as in \eqs{eq:SoftBroadeningLimit}{eq:BroadeningSoftExpanded}.


\section{One-loop angularity distributions from broadening-like factorization}
\label{sec:AngularityCrossSection}

In this section we will present our one-loop results for the angularity cross sections
at arbitrary values of $b$, which we obtain using the jet and soft functions calculated above. 
We will provide results for positive-$b$ and negative-$b$ angularities separately. We will show that our analytic results correctly reproduce the known one-loop results for thrust, jet broadening and large-$b$ angularities. As claimed earlier in Sec.~\ref{sec:Factorization}, we will show that, compared to thrust-like analyses of angularity distributions with $0.5 < b < 1$, our results based on broadening-like factorization contain an extra sub-leading singular correction of the type shown in \eq{eq:posbCS}.  Our treatment of the small-$b$ case shows explicitly that our results for the cross section are continuous in the limit $b \to 0$. 
In addition, we will derive the first analytic results for the recoil-sensitive angularity cross sections for negative values of $b$.


\subsection{One-loop cross section for $b>0$}
\label{subsec:AngularityNLO}

We now calculate the separate contributions  to the cross section from the one-loop jet function and the one-loop soft function, which we denote by jet$_{1+}$ and soft$_{1+}$, respectively. To obtain jet$_{1+}$, we perform a convolution of the one-loop jet function in \eq{eq:JetResult} with the tree-level soft function in \eq{eq:TreeLevelSoft} and the tree-level hard function $H^{(0)}(Q;\mu) = 1$, according to the factorization theorem in \eq{eq:BroadeningfactorizeRestated}.
Including the contribution from both 
right and left hemispheres, we have
\begin{align}
\label{eq:JetContrib}
\frac{\rm jet_{1+}}{\sigma_0}  
&= \frac{\as(\mu){\,}C_F}{\pi}\, \delta(\T_{L}) \bigg(\!\! -\! \frac{2{\,} w^2\, e^{\e \gamma_E}}{\eta{\,}(1 + b){\,}\Gamma(1-\e)} \Big(\frac{\mu}{Q} \Big)^{\! 2\e} \bigg[ \frac{1}{\T_{R}^{1 + \frac{2\e}{1+b}}} \bigg]_{+}^{\infty}\!\!\! + \frac{1}{\e}\, \ln\!\frac{\nu}{Q}\, \delta(\T_{R}) + \frac{3}{4{\,}\e}\, \delta(\T_{R}) \nn  \\ 
& \quad -   \frac{2}{1 + b}\, \ln\!\frac{\nu}{Q}\, \frac{1}{m} \bigg[ \frac{1}{\T_{R}/m}\bigg]_{+}\!\!\! - \frac{3}{2{\,}(1 + b)} \frac{1}{m} \bigg[ \frac{1}{\T_{R}/m} \bigg]_{+}\!\!\! + {\cal J}^{(+)}_0(b)\, \delta(\T_{R})\! \bigg) + \left \{ \T_{L} \leftrightarrow \T_{R} \right \} \, , 
\end{align}
where $\sigma_0$ denotes the Born cross section and
\begin{align}
\label{eq:J0Define}
{\cal J}^{(+)}_0(b) &\equiv i_a - i_b^{(+)} = \frac{2}{1+b} \Bigg[ \frac{1+8\,b}{8} +\frac{\pi^2}{6\, b} - \int_0^1 \df x\, \frac{1-x+x^2/2}{x}\, \ln\bigg(\!1+\!\Big(\!\frac{x}{1-x}\Big)^{\!\! b}\bigg) \Bigg]  \nn \\
&= \frac{2}{1+b} \Bigg[  \frac{1+8\,b}{8} + \frac{\pi^2}{12\, b} - \frac{\pi^2}{12} b - \int_0^1 \df x\, \bigg(\frac{x}{2}-1\bigg)\, \ln \bigg(\!1+\!\Big(\!\frac{x}{1-x}\Big)^{\!\! b}\bigg) \Bigg]\, .
\end{align}
The integral in the second line is finite as $b \to 0$ and will be treated numerically. 
The singularity due to $1/x$ 
in the integrand in the first line of \eq{eq:J0Define} is regulated when $b>0$.
Here we have used
\begin{align}
\label{eq:J0AnalyticFactor}
\int_0^1 \df x \, \frac{1}{x} \, \ln \bigg( 1+\!\Big(\!\frac{x}{1-x}\Big)^{\!\! b}\bigg) = \frac{\pi^2}{12\, b} + \frac{\pi^2}{12}  b \, ,
\end{align}
to obtain the expression given in the second line. The last equation can be easily checked numerically.

Similarly, soft$_{1+}$ is obtained by doing a convolution of the one-loop soft function in \eq{eq:SoftExpanded} with the hard and the jet functions at tree level, leading to
\begin{align}
\label{eq:SoftContrib}
&\frac{\rm soft_{1+}}{\sigma_0}  
= \frac{\as(\mu){\,}C_F}{\pi}\, \delta(\T_{L}) \bigg( \frac{2{\,}w^2\, e^{\e \gamma_E}}{\eta{\,}(1 + b){\,}\Gamma(1-\e)} \Big(\frac{\mu}{Q} \Big)^{\! 2\e}\! \bigg[\frac{1}{\T_{R}^{1 + \frac{2\e}{1+b}}} \bigg]_{+}^{\infty} \ns + \frac{1}{2\, \e^2}\, \delta(\T_{R}) \! - \frac{1}{\e} \ln\!\frac{\nu}{\mu}\, \delta(\T_{R}) \nn \\
&  \qquad - \frac{2}{(1 + b)^2{\,}m} \bigg[ \frac{\ln(\T_{R}/m)}{\T_{R}/m} \bigg]_{+}\!\!\! + 2 \ln\!\frac{\nu}{\mu}\, \frac{1}{(1 + b){\,}m} \bigg[\frac{1}{\T_{R}/m}\bigg]_{+}\!\!\! - \frac{\pi^2}{24}\, \delta(\T_{R}) + {\text{I}}^{(+)}_s(\T_{R})\! \bigg) + \left \{ \T_{L} \leftrightarrow \T_{R} \right \},  
\end{align}
where
\begin{align}
\label{eq:ISIntegralDefine}
{\text{I}}^{(+)}_s(\T_{R}) &= \frac{2}{b} \int_0^{Q{\,}\T_{R}^{1/(1+b)}}\bnspace {\df p_{\perp}}{\,} \theta\Big(\frac{p_{\perp}}{Q} - \T_{R} + \Big(\frac{p_{\perp}}{Q}\Big)^{\!1 + b}\Big) \frac{Q}{p_{\perp}}\, \frac{1}{\mu} \bigg[\frac{1}{p_{\perp}/\mu}\bigg]_{+} \bigg[\frac{p_{\perp}/Q}{\T_{R} -  \Big(\frac{p_{\perp}}{Q}\Big)^{\!1 + b}}\bigg]_{+} . 
\end{align}
Writing the integral in the last equation in terms of dimensionless quantities, $\rho$ and $\tau_{R}$, with
\begin{equation}
\label{eq:IsBNewVariables1}
\rho \equiv \Big(\frac{p_{\perp}}{Q}\Big)^{\!1 + b} \, ,
\end{equation}
we get
\begin{equation}
\label{eq:IsBIntegral1}
\text{I}^{(+)}_s(\T_{R}) = \frac{2}{b{\,}(1+b)}\int_{r\, \T_{R}}^{\T_{R}}\df\rho\, \frac{1}{\rho}\frac{Q}{\mu}\left[\frac{\mu/Q}{\rho^{\frac{1}{1 + b}}}\right]_{+}\left[\frac{\rho^{\frac{1}{1 + b}}}{\T_{R} - \rho}\right]_{+} .
\end{equation}
Here $r$ in the lower limit of integration, $r\, \T_{R}$, is obtained by solving the constraint equation
\begin{align}
\label{eq:r}
\frac{r}{(1-r)^{\! 1+b}} &= \T_{R}^{\! b} \, , 
\end{align}
which stems from the Heaviside $\theta$-function in \eq{eq:ISIntegralDefine} and
makes one of the plus prescriptions superfluous. The details of the calculation of the integral $\text{I}^{(+)}_s(\T_{R})$ can be found in App.~\ref{appsubsec:PositiveBSoftFactor}. Our final result amounts to
\begin{align}
\label{eq:SoftIntegralISPositive}
&\text{I}^{(+)}_s(\T_{R}) = \!-\frac{2\,b}{(1+b)^2} \bigg[\frac{\ln\T_R}{\T_R}\bigg]_{+} \!\!\! - \frac{\pi^2}{3\, b(1+b)} \delta(\T_R) - \frac{2}{1+b} \frac{\ln(1-r)}{\T_{R}} \, .
\end{align}
Note that in the small-$\T_R$ limit we have $r \sim \T_{R}^{b}$, which implies that the last term is less singular compared to $1/\tau_R$ and thus does not require a plus prescription, unless $b$ is vanishingly small. Since $b$ regulates the small-$\tau_R$ limit in the last term, for vanishingly small $b$ the last term needs to be expressed as a distribution as shown in App.~\ref{appsec:SmallBRExp}.

Finally, according to \eq{eq:Hard}, the 
remaining contribution to the one-loop normalized cross section due to the one-loop hard function is
\begin{equation}
\label{eq:OneLoopHard}
\frac{\rm hard_1}{\sigma_0} = \frac{\as(\mu)\, C_F}{2\pi}\bigg(\!\! -\frac{1}{\e^2} - \frac{2}{\e} \ln\!\frac{\mu}{Q} - \frac{3}{2\, \e} - 2 \ln^2\!\frac{\mu}{Q} - 3 \ln\!\frac{\mu}{Q} - 4 + \frac{7\pi^2}{12}\bigg) \delta(\T_{L})\, \delta(\T_{R})+\! \left \{ \T_{L}\! \leftrightarrow \!\T_{R} \right \}\, .
\end{equation}
Adding the terms jet$_{1+}$, soft$_{1+}$ and hard$_1$, we obtain the ${\cal O}(\alpha_s)$ contribution to the double-differential angularity cross section in ${\rm SCET_{II}}$, 
\begin{align}
\label{eq:NLOCS}
\bigg[\frac{1}{\sigma_0}\, \frac{\rm \df^2\sigma^{(+)}}{\df\T_{L}\, \df\T_{R}}\bigg]^{\rm {\cal O}(\alpha_s)}_{\rm SCET_{II}}\!\! &= \frac{\as(\mu){\,}C_F}{\pi}\, \delta(\T_{L}) \bigg[\delta(\T_{R}) \bigg(\!\!\!-2\! +\! \frac{\pi^2}{4}\! - \frac{\pi^2}{3\, b(1+b)} +\! {\cal J}^{(+)}_0(b)\! \bigg)\! - \frac{3}{2(1 + b)} \bigg[ \frac{1}{\T_{R}} \bigg]_{+}\ns \nn \\
&\quad  -\! \frac{2}{(1 + b)} \bigg[ \frac{\ln\T_{R}}{\T_{R}} \bigg]_{+}\ns - \frac{2}{1+b} \frac{\ln(1-r)}{\T_{R}} \bigg] +\! \left \{ \T_{L}\! \leftrightarrow \!\T_{R} \right \} \, . 
\end{align}
Notice that the rapidity divergences, identified by $1/\eta$ factors, as well as the $\nu$-dependence cancel out in the sum of the jet and soft contributions, as expected. The $\epsilon$-divergent structure in the sum of jet$_{1+}$ and soft$_{1+}$ is proportional to $\delta(\T_{L}){\;}\delta(\T_{R})$ and cancels exactly against the $\epsilon$ divergences of the hard$_1$-term, leaving the cross section independent of the choice of regulators and scales, up to a residual dependence on $\mu$ through the renormalized coupling $\alpha_s(\mu)$.

\eq{eq:NLOCS} serves as the master formula for positive-$b$ angularities and, as we are going to show in the following two subsections, it yields the known expressions for broadening, thrust and large-$b$ angularities in suitable limits.

\subsubsection{Small-$\T$ limit}

When $b$ is away from 0, the general solution of \eq{eq:r} in the small-$\T$ limit can be found as detailed in App.~\ref{appsec:SmallTauRExp}. The double-differential ${\cal O}(\alpha_s)$ cross section in the small-$\T$ 
limit is then given as
\begin{align}
\label{eq:NLOCrossSection}
&\bigg[\frac{1}{\sigma_0}\, \frac{\rm \df^2\sigma^{(+)}}{\df\T_{L}\, \df\T_{R}}\bigg]^{\rm {\cal O}(\alpha_s)}_{\rm SCET_{II}} \ns = \frac{\as(\mu){\,}C_F}{\pi}\, \delta(\T_{L}) \bigg[\delta(\T_{R}) \bigg(\!\!\!-2\! +\! \frac{\pi^2}{4}\! - \frac{\pi^2}{3\, b(1+b)} +\! {\cal J}^{(+)}_0(b)\! \bigg)\! - \frac{3}{2(1 + b)} \bigg[ \frac{1}{\T_{R}} \bigg]_{+}\ns \nn \\
&\tab[2cm] - \frac{2}{1 + b} \bigg[ \frac{\ln\T_{R}}{\T_{R}} \bigg]_{+}\ns - \frac{2}{1+b}\!\!\sum_{n=1}^{\ceil*{1/b}-1} \!\!\frac{c_n}{\T_{R}^{1-n\,b}} \bigg] + \left \{ \T_{L}\! \leftrightarrow \!\T_{R} \right \} + \text{power corrections} ,  
\end{align}
with ${\cal J}^{(+)}_0(b)$ as in \eq{eq:J0Define} and $c_{n}$ depending only on $b$. 
The first few coefficients appearing in the last equation are given by
\begin{align}
\label{eq:c_n}
&\text{c}_1 = -1, \quad \text{c}_2 = \frac{1}{2}(1+2\,b), \quad \text{c}_3 = -\frac{1}{6}(2+9\,b+9\,b^2),  \\
& \text{c}_4 = \frac{1}{12}(3+22\,b+48\,b^2+32\,b^3), \quad \text{c}_5 = -\frac{1}{120}(24+250\,b+875\,b^2+1250\,b^3+625\,b^4), \, \dots{} \; . \nn 
\end{align}
The result in \eq{eq:NLOCrossSection} has the structure presented in Sec.~\ref{sec:Factorization}.
For any given (positive) value of $b$, there exists an integer $N=\ceil*{1/b}-1$, up to which terms under the summation contribute as sub-leading singularities. 
For $n > N$, the recoil effects accounted for 
by ${\rm SCET_{II}}$ factorization amount only to power corrections. The other terms agree with the ${\rm SCET_{I}}$ factorization result. 

As discussed earlier in Sec.~\ref{sec:Factorization}, although the jet function in Sec.~\ref{sec:OneLoopCalculations} and the soft function in Sec.~\ref{sec:SoftFunc} are homogenous in the power counting, the cross section in \eq{eq:NLOCrossSection} obtained after the transverse momentum convolution contains sub-leading terms that are not, since the transverse momenta scale differently in the jet and the soft sectors. While the new terms are formally sub-leading, their numerical effects can be significant though. Tab.~\ref{tab:SizeOfCorrec} of Sec.~\ref{sec:NumericalResults} shows the numerical size of these sub-leading singular contributions relative to the leading singular ones, for several values of $b$. 

Using the result in \eq{eq:NLOCrossSection}, we obtain the singular contribution to the one-loop single-differential angularity cross section for $b>0$, by integrating over the hemisphere angularities subject to the constraint $\tau = \tau_{L} +\tau_{R}$,
\begin{align}
\label{eq:NLODiffCrossSection}
&\bigg[\frac{1}{\sigma_0}\, \frac{\rm \df\sigma^{(+)}}{\df\T}\bigg]^{\rm {\cal O}(\alpha_s)}_{\rm sing.}\!\!  = \frac{\as(\mu){\,}C_F}{\pi}\,  \bigg\{\!\!- \! \frac{3}{(1 + b)} \bigg[ \frac{1}{\T} \bigg]_{+}\ns - \frac{4}{1+b} \bigg[\frac{\ln\T}{\T}\bigg]_{+} \ns\,\, - \frac{4}{1+b}\sum_{n=1}^{\ceil*{1/b}-1} \frac{c_n}{\T^{1-n\,b}} \nn \\
&\qquad + \delta(\T) \bigg[\!\! -\frac{7}{2(1+b)}\! +\! \frac{\pi^2}{6\,b} \,\frac{b^2+3\,b-2}{1+b} - \frac{4}{1+b}  \int_0^1\!\! \df x\, \bigg(\frac{x}{2}-1\bigg) \! \ln\!\bigg(\!1+\!\Big(\!\frac{x}{1-x}\Big)^{\!\! b}\bigg) \bigg]\! \bigg\} . 
\end{align}
Here the first two terms as well as the coefficient of $\delta(\T)$-piece agree with the ${\cal O}(\alpha_s)$-singular cross section obtained from SCET$_{\rm I}$ thrust-like factorization in Ref.~\cite{Hornig:2009vb}. 
The terms in the summation 
are singular only when $0<b<1$.
When $b \geq 1$, recoil effects amount to regular terms, which can be dropped to obtain the result in~\cite{Hornig:2009vb}. 
By taking the limit $b \rightarrow 1$ in \eq{eq:NLODiffCrossSection} and neglecting 
power corrections, we obtain the ${\cal O}(\alpha_s)$-singular thrust distribution  
\begin{align}
\label{eq:ThrustResult} 
\bigg[\frac{1}{\sigma_0}\, \frac{\df\sigma^{(+)}}{\df\T}\bigg]^{\rm {\cal O}(\alpha_s)}_{\rm sing.} \!\! & = \frac{\as(\mu)\, C_F}{\pi}\bigg[ \delta(\T) \bigg( -\frac{1}{2} + \frac{\pi^2}{6} \bigg) - \frac{3}{2}\Big[\frac{\theta(\T)}{\T}\Big]_{+} \ns - 2\Big[\frac{\theta(\T)\ln\T}{\T}\Big]_{+} \bigg]\, ,  
\end{align}
which agrees with the known result in the literature (see {\it e.g.} Eq. (36.8) of~\cite{Schwartz:2013pla}).

\subsubsection{Small-$b$ and the jet broadening limit}
In the last term of the master formula in \eq{eq:NLOCS}, the limit $\T \rightarrow 0$ is regulated by the angularity exponent $b$, implying that the small-$b$ limit requires special care. While $b$ and $\T$ can be small, $\T^b$ need not be small. Thus in this case we need an expansion of $r$ in small $b$ while treating $\T^b \sim 1$. This is worked out in App.~\ref{appsec:SmallBRExp}. Using the result in \eq{appeq:RSmallB}, we get, for the last term of \eq{eq:NLOCS},
\begin{align}
\label{eq:LogRSmallB}
\frac{\ln(1-r)}{\T_R} &= \bigg(\!\!-\frac{\pi^2}{12\,b} + \frac{\ln^2 2}{2} -\frac{b}{4} \ln^2 2 \bigg) \, \delta(\T_R) - \ln 2\bigg[\frac{1}{\T_R}\bigg]_{+} \ns  -\frac{b}{2} \bigg[\frac{\ln\T_R}{\T_R}\bigg]_{+}\ns +\frac{b}{2} \ln 2\bigg[\frac{1}{\T_R}\bigg]_{+} \ns \,+ {\cal O}(b^2)\, . 
\end{align}
The double-differential ${\cal O}(\alpha_s)$ cross section, in this limit, up to order $b^2$ is thus given as
\begin{align}
\label{eq:NLOCSbSmall}
&\bigg[\frac{1}{\sigma_0}\, \frac{\rm \df^2\sigma^{(+)}}{\df\T_{L}\, \df\T_{R}}\bigg]^{\rm {\cal O}(\alpha_s)}_{\rm SCET_{II}}\!\! = \frac{\as(\mu){\,}C_F}{\pi}\, \delta(\T_{L})  \\
&\quad \times \bigg[\delta(\T_{R}) \bigg(\!\!\! - \frac{\pi^2}{6\, b}+\! \frac{5\pi^2}{12}\! -2\!  - \ln^2 2 - \frac{\pi^2\,b}{6} +\frac{3\,b}{2} \ln^2 2 +\! {\cal J}^{(+)}_0(b)\! \bigg)\!  \nn \\
&\quad - \frac{3}{2} \bigg[ \frac{1}{\T_{R}} \bigg]_{+}\ns -\! 2 \bigg[ \frac{\ln\T_{R}}{\T_{R}} \bigg]_{+}\ns+ 2\ln 2\bigg[\frac{1}{\T_R}\bigg]_{+} \ns+ b\Big(\,\frac{3}{2} \bigg[ \frac{1}{\T_{R}} \bigg]_{+}\ns + 3 \bigg[ \frac{\ln\T_{R}}{\T_{R}} \bigg]_{+}\ns  - 3 \ln 2\bigg[\frac{1}{\T_R}\bigg]_{+}\Big) \bigg]\! + {\cal O}(b^2) \! +\! \left \{ \T_{L}\! \leftrightarrow \!\T_{R} \right \} . \nn
\end{align}
The jet convolution factor ${\cal J}^{(+)}_0$ (see \eq{eq:J0Define}), in the small-$b$ limit, amounts to
\begin{align}
\label{eq:J0BroadeningLimit}
{\cal J}^{(+)}_0(b) &= \frac{\pi^2}{6\, b} - \frac{\pi^2}{6} + \frac{1}{4} + \frac{3}{2} \ln 2 - b\bigg(\frac{3}{2} \ln 2 - \frac{3}{2} \bigg)+ {\cal O}(b^2) \, .  
\end{align}
It is clear from \eqs{eq:NLOCSbSmall}{eq:J0BroadeningLimit} that the $1/b$ singularities cancel, yielding a smooth behavior for the cross section for vanishingly small positive values of $b$. To order $b^2$, we have
\begin{align}
\label{eq:NLOCSbSmallFull}
&\bigg[\frac{1}{\sigma_0}\, \frac{\rm \df^2\sigma^{(+)}}{\df\T_{L}\, \df\T_{R}}\bigg]^{\rm {\cal O}(\alpha_s)}_{\rm SCET_{II}}\!\! = \frac{\as(\mu){\,}C_F}{\pi}\, \delta(\T_{L}) \\
& \times \bigg[\delta(\T_{R}) \bigg(\!\!\!-\frac{7}{4}\! +\! \frac{\pi^2}{4}\! + \frac{3}{2} \ln 2 - \ln^2 2 +b\Big(\!\! -\frac{\pi^2}{6}  - \frac{3\ln 2}{2}  + \frac{3}{2} +\frac{3\ln^2 2}{2}\Big)\!  \bigg)\!  \nn \\
& -\! \frac{3}{2} \bigg[ \frac{1}{\T_{R}} \bigg]_{+}\ns -\! 2 \bigg[ \frac{\ln\T_{R}}{\T_{R}} \bigg]_{+}\ns+ 2\ln 2\bigg[\frac{1}{\T_R}\bigg]_{+} \ns+ b\Big(\,\frac{3}{2} \bigg[ \frac{1}{\T_{R}} \bigg]_{+}\ns + 3 \bigg[ \frac{\ln\T_{R}}{\T_{R}} \bigg]_{+}\ns  - 3 \ln 2\bigg[\frac{1}{\T_R}\bigg]_{+}\Big) \bigg]\! + {\cal O}(b^2)\! +\! \left \{ \T_{L}\! \leftrightarrow \!\T_{R} \right \} \, . \nn
\end{align}
Taking the $b \rightarrow 0^{+}$ limit of the last equation gives 
 \begin{align}
 \label{eq:BroadeningLimit}
&\bigg[\frac{1}{\sigma_0}\, \frac{\df^{\,2}\sigma^{(+)}}{\df\T_{L} \df\T_{R}}\bigg]^{\rm {\cal O}(\alpha_s)}_{\rm sing.} \!\! =  \frac{\as{\,}(\mu)C_F}{\pi}\, \bigg\{ \delta(\T_{R})\, \delta(\T_{L})\bigg(\!\!\! - \frac{7}{2} + \frac{\pi^2}{2} - 3 \ln\!\frac{\mu}{Q} - 2\ln^2\!\frac{\mu}{Q}\bigg)  \\
& \qquad  + \bigg[\delta(\T_{L}) \bigg(\!\! - \frac{3{\,}Q}{4{\,}\mu} \left[\frac{2{\,}\mu}{Q\T_{R}}\right]_{+}\ns - \frac{Q}{\mu} \left[\frac{2{\,}\mu}{Q\T_{R}}\right]_{+}\!\!\! \ln\frac{\mu}{Q} - \frac{Q}{\mu} \left[\frac{2{\,}\mu \ln(Q\T_{R}/2{\,}\mu)}{Q\T_{R}}\right]_{+}\! \bigg)\! + \left \{ \T_{L}\! \leftrightarrow \!\T_{R} \right \}\!\! \bigg] \bigg\}\, ,  \nn
 \end{align}
which agrees with the double-differential jet broadening distribution obtained in~\cite{Chiu:2012ir}.


\subsection{One-loop cross section for $b<0$}
\label{sec:b<0Angularities}
To obtain  
the contribution from the one-loop jet function to the cross section for negative-$b$ (jet$_{1-}$), we convolve our result in \eq{eq:JetResultNegativeB} with the tree-level soft function of \eq{eq:TreeLevelSoft} and the tree-level hard function $H^{(0)}(Q;\mu) = 1$,
\begin{align}
\label{eq:JetContribNegativeBCase}
\frac{\rm jet_{1-}}{\sigma_0}  
&= \frac{\as(\mu){\,}C_F}{\pi}\, \delta(\T_{L}) \bigg( \frac{1+b}{2\, b\, \e^2} \delta(\T_R) + \frac{3}{4{\,}\e}\, \delta(\T_{R}) - \frac{1}{b\, \e} \frac{1}{m} \bigg[\frac{1}{\T_{R}/m}\bigg]_{+}\ns - \frac{3}{2{\,}(1 + b)} \frac{1}{m} \bigg[ \frac{1}{\T_{R}/m} \bigg]_{+}\ns \nn  \\ 
& \quad + \frac{2}{b(1+b)} \frac{1}{m} \bigg[\frac{\ln(\T_{R}/m)}{\T_{R}/m}\bigg]_{+} \ns  -\frac{\pi^2}{24} \frac{1+b}{b} \delta(\T_R) + {\cal J}^{(-)}_0(b)\, \delta(\T_{R})\! \bigg) + \left \{ \T_{L} \leftrightarrow \T_{R} \right \} , 
\end{align}
where 
\begin{align}
\label{eq:J0primeDefine}
{\cal J}^{(-)}_0(b) &\equiv i_a - i_b^{(-)}= \frac{2}{1+b} \Bigg[  \frac{1+6\,b}{8} -\frac{\pi^2}{6} \Big(\frac{b^2+1}{b}\Big) - \!\!\int_0^1\!\! \df x\, \frac{1-x+x^2/2}{x}\, \ln\! \bigg(\!1+\!\Big(\!\frac{x}{1-x}\Big)^{\!\! -b}\bigg) \Bigg] \nn \\
&= \frac{2}{1+b} \Bigg[  \frac{1+6\,b}{8} -\frac{\pi^2}{12\, b} -\frac{\pi^2}{12}\, b - \!\!\int_0^1\!\! \df x\, \bigg(\frac{x}{2}-1\bigg)\, \ln\! \bigg(\!1+\!\Big(\!\frac{x}{1-x}\Big)^{\!\! -b}\bigg) \Bigg]\, . 
\end{align}
The integral in the first line of \eq{eq:J0primeDefine} has the same form as that given in \eq{eq:J0Define} but with $b$ replaced by $-b$. The result of \eq{eq:J0AnalyticFactor} holds also in the case of negative $b$, which we have used to obtain the second line of \eq{eq:J0primeDefine}. Here again the integral in the second line of the last equation is regular in $b$ and can be computed numerically for a given $b$.

The contribution to soft$_{1-}$ is derived by taking the convolution of the one-loop soft function of \eq{eq:SoftExpandedNegativeB} with the hard and the jet functions at tree level, 
\begin{align}
\label{eq:SoftContribNegativeBCase}
\frac{\rm soft_{1-}}{\sigma_0} & = \frac{\as(\mu){\,}C_F}{\pi}\, \delta(\T_{L}) \bigg(\!\!\! -\! \frac{1}{2\,b \e^2}\, \delta(\T_{R}) + \frac{1}{b\, \e} \frac{Q}{\mu} \bigg[\frac{1}{Q\T_R/\mu}\bigg]_{+} \ns - \frac{2}{b} \frac{Q}{\mu}\, \bigg[\frac{\ln(Q\T_R/\mu)}{Q\T_R/\mu}\bigg]_{+} \ns + \frac{\pi^2}{24\,b} \delta(\T_{R}) \nn \\
&\tab[8cm] + {\text{I}}^{(-)}_s(\T_{R})\! \bigg) +\! \left \{ \T_{L} \leftrightarrow \T_{R} \right \}. 
\end{align}
It is interesting to note that the $b$-dependent $\e$-divergences in soft$_{1-}$ cancel exactly against  jet$_{1-}$, while the $b$-independent $\e$-divergences (contained only in  jet$_{1-}$) cancel against $\rm hard_1$ of \eq{eq:OneLoopHard}. The integral ${\text{I}}^{(-)}_s(\T_{R})$ in soft$_{1-}$ is defined as
\begin{align}
\label{eq:SoftFactorNegativeB} 
{\text{I}}^{(-)}_s(\T_{R}) &= -\frac{2}{b(1+b)} \int_{s\,\T_{R}}^{\T_{R}} {\df\T^s_n}{\,} \frac{1}{(\T^s_n)^2} \frac{1}{(\T_{R}-\T^s_n)^{\frac{b-1}{1+b}}} \bigg[\frac{(\T^s_n)^2}{(\T_{R}-\T^s_n)^{\frac{2}{1+b}}}\bigg]_{+} \frac{Q}{\mu} \bigg[\frac{1}{Q\T^s_n/\mu} \bigg]_{+}\, ,
\end{align}
where $s$ in the lower limit of integration, $s\,\tau_{R}$, satisfies the constraint equation
\begin{align}
\label{eq:s}
\frac{s}{(1-s)^{\!\frac{1}{1+b}}} = \T_{R}^{\!-\frac{b}{1+b}}\, .
\end{align}
We evaluate the integral in \eq{eq:SoftFactorNegativeB} in a similar manner as discussed for the positive-$b$ case. The details of this calculation are presented in App.~\ref{appsec:SoftNegBInt}. The final result is given by
\begin{equation}
\label{eq:SoftIntegralIS}
\text{I}^{(-)}_s(\T_{R}) = \frac{2\,b}{(1+b)^2} \bigg[\frac{\ln\T_{R}}{\T_{R}}\bigg]_{+} \ns+ \frac{\pi^2}{3\,b(1+b)} \delta(\T_{R})  - \frac{2}{(1+b)^2} \frac{\ln(1-s)}{\T_{R}} \, .
\end{equation}
Interestingly, this result is similar in structure to \eq{eq:SoftIntegralISPositive} for positive-$b$, with the last term yielding sub-leading integrable singularities and/or power corrections, depending on the value of $b$.

Adding the contribution of jet$_{1-}$ and soft$_{1-}$ to hard$_{1}$ given in \eq{eq:OneLoopHard}, the double-differential ${\cal O}(\alpha_s)$ cross section for negative-$b$ in SCET$_{\rm II}$  amounts to
\begin{align}
\label{eq:NLOCSNegB}
\bigg[\frac{1}{\sigma_0}\, \frac{\rm d^2\sigma^{(-)}}{\df\T_{L}\, \df\T_{R}}\bigg]^{\rm {\cal O}(\alpha_s)}_{\rm SCET_{II}}\!\!   
&= \frac{\as(\mu){\,}C_F}{\pi}\, \delta(\T_{L}) \bigg[\delta(\T_{R}) \bigg(\!\!\!-2\! +\! \frac{\pi^2}{4}\! + \frac{\pi^2}{3\,b(1+b)} +\! {\cal J}^{(-)}_0(b)\! \bigg)\!  -\! \frac{3}{2(1 + b)} \bigg[ \frac{1}{\T_{R}} \bigg]_{+}\ns \nn \\ 
& \quad  -\! \frac{2}{(1 + b)^2} \bigg[ \frac{\ln\T_{R}}{\T_{R}} \bigg]_{+}\ns - \frac{2}{(1+b)^2} \frac{\ln(1-s)}{\T_{R}} \bigg] \!+\! \left \{ \T_{L}\! \leftrightarrow \!\T_{R} \right \}  .  
\end{align}
where ${\cal J}^{(-)}_0$ is given in \eq{eq:J0primeDefine}. The last equation represents the master formula for our analysis for $-1<b<0$ angularity distributions. This novel result of ours compares successfully against the numerical output of \Eventtwo, as shown in Sec.~\ref{sec:NumericalResults}.

\subsubsection{Small-$\T$ limit}
\label{subsec:SmallTNegB}

For $b$ not close to 0, the solution of \eq{eq:s} in the small-$\T$ limit can be expanded in powers of $\T_R^{-\frac{b}{1+b}}$, as exploited in App.~\ref{appsec:SmallTauSExp}. 
The double-differential ${\cal O}(\alpha_s)$ cross section in the small-$\T$  
limit is then given as
\begin{align}
\label{eq:NLOCrossSectionNegB}
&\bigg[\frac{1}{\sigma_0}\, \frac{\df\sigma^{(-)}}{\df\T_{L}\df\T_{R}}\bigg]^{\rm {\cal O}(\alpha_s)}_{\rm SCET_{II}} 
= \frac{\as(\mu){\,}C_F}{\pi} \,\delta(\T_{L}) \bigg[\delta(\T_{R}) \bigg(\!\!\!-2\! +\! \frac{\pi^2}{4}\! + \frac{\pi^2}{3\,b(1+b)} +\! {\cal J}^{(-)}_0(b)\! \bigg)\! -\! \frac{3}{2(1 + b)} \bigg[ \frac{1}{\T_{R}} \bigg]_{+}\ns \nn \\ 
&\tab[1.2cm] - \frac{2}{(1+b)^2} \bigg[\frac{\ln\T_{R}}{\T_{R}}\bigg]_{+}\ns - \frac{2}{(1+b)^2}\ns\,\sum_{n=1}^{\ceil*{1/|b|}-2}\ns\,\, \frac{c^{'}_n}{\T_{R}^{1+\frac{n\,b}{1+b}}} 
\bigg] +\! \left \{ \T_{L}\! \leftrightarrow \!\T_{R} \right \} +\text{power corrections}~,
\end{align}
where $c^{'}_{n}$ depend only on the value of $b$. The first few coefficients are 
\begin{align}
\label{eq:c^{'}_n}
&\text{c}^{'}_1 = -1, \quad \text{c}^{'}_2 = \frac{1-b}{2(1+b)}, \quad \text{c}^{'}_3 = -\frac{2-5\,b+2\,b^2}{6(1+b)^2},  \\
& \text{c}^{'}_4 = \frac{3-13\,b+13\,b^2-3\,b^3}{12(1+b)^3}, \quad \text{c}^{'}_5 = -\frac{24-154\,b+269\,b^2-154\,b^3+24\,b^4}{120(1+b)^4}, \, \dots{} \; . \nn 
\end{align}
The summation is truncated up to the greatest integer which is strictly less than $1/\vert b\vert -1$ and includes sub-leading singular terms for $b>-0.5$. For $b\leq -0.5$, only power corrections remain. 
Using this result, the ${\cal O}(\alpha_s)$ singular contribution to the single-differential cross section for negative-$b$ is
\begin{align}
\label{eq:NLODiffCrossSectionNegativeBCase}
&\bigg[\frac{1}{\sigma_0}\, \frac{\rm \df\sigma^{(-)}}{\df\T}\bigg]^{\rm {\cal O}(\alpha_s)}_{\rm sing.}\!\! 
 = \frac{\as(\mu){\,}C_F}{\pi}\,  \bigg\{ \!\!- \frac{3}{1 + b} \bigg[ \frac{1}{\T} \bigg]_{+}\ns -\frac{4}{(1+b)^2} \bigg[\frac{\ln\T}{\T}\bigg]_{+} \ns -\frac{4}{(1+b)^2} \ns\,\, \sum_{n=1}^{\ceil*{1/\vert b\vert}-2} \ns\,\, \frac{c^{'}_n}{\T^{1+\frac{n\,b}{1+b}}}  \nn \\
&\qquad + \delta(\T) \bigg[-\frac{7+2\,b}{2(1+b)} \! + \frac{\pi^2(b+2)}{6\,b} - \frac{4}{1+b} \int_0^1\! \df x\, \bigg(\frac{x}{2}-1\bigg)\, \ln \bigg(\!1+\!\Big(\!\frac{x}{1-x}\Big)^{\!\! -b}\bigg) \bigg] \bigg\} \, .
\end{align}
It is worth to stress that the coefficients of the $[\ln\T/\T]_{+}$ and $\delta(\T)$ terms in the equation above differ from the corresponding ones in the positive-$b$ case (see \eq{eq:NLODiffCrossSection}) while the coefficient of $[1/\T]_{+}$ remains the same.  

\subsubsection{Small-$b$ and the jet broadening limit}
As in the case of positive-$b$ angularities, the last term of our master formula in \eq{eq:NLOCSNegB}, needs to be treated with special care when $b$ and $\T$ are both small. This case is analysed in App.~\ref{appsec:SmallBSExp}. Using \eq{appeq:SBsmall}, we get for the last term of \eq{eq:NLOCSNegB}, 
\begin{align}
\label{eq:LogSSmallB}
\frac{\ln(1-s)}{\T_R} 
&= \bigg(\frac{\pi^2}{12\,b} + \frac{\pi^2}{12} + \frac{\ln^2 2}{2} + \frac{b}{4}\ln^2 2\bigg)\, \delta(\T_R) - \ln 2\bigg[\frac{1}{\T_R}\bigg]_{+} \ns +\frac{b}{2} \bigg[\frac{\ln\T_R}{\T_R}\bigg]_{+}\ns -\frac{b}{2} \ln 2\bigg[\frac{1}{\T_R}\bigg]_{+} \ns \,+ {\cal O}(b^2)\,.
\end{align}
For the double-differential cross section for negative, small $b$, we obtain to order $b^2$
\begin{align}
\label{eq:NLOCSbSmallNegB}
&\bigg[\frac{1}{\sigma_0}\, \frac{\rm \df^2\sigma^{(-)}}{\df\T_{L}\, \df\T_{R}}\bigg]^{\rm {\cal O}(\alpha_s)}_{\rm SCET_{II}}\!\! = \frac{\as(\mu){\,}C_F}{\pi}\, \delta(\T_{L})  \\
&\quad \times \bigg[\delta(\T_{R}) \bigg(\frac{\pi^2}{6\,b} +\! \frac{\pi^2}{12} \!-2 - \ln^2 2 + \frac{\pi^2\,b}{6} +\frac{3\,b}{2} \ln^2 2 +\! {\cal J}^{(-)}_0(b)\! \bigg)\!   \nn \\ 
& \quad  -\! \frac{3}{2} \bigg[ \frac{1}{\T_{R}} \bigg]_{+}\ns -\! 2 \bigg[ \frac{\ln\T_{R}}{\T_{R}} \bigg]_{+}\ns  + 2 \ln 2\bigg[\frac{1}{\T_R}\bigg]_{+} \ns +b \Big(\, \frac{3}{2} \bigg[ \frac{1}{\T_{R}} \bigg]_{+}\ns +\! 3 \bigg[ \frac{\ln\T_{R}}{\T_{R}} \bigg]_{+}\ns - 3 \ln 2\bigg[\frac{1}{\T_R}\bigg]_{+}\Big) \bigg] \! + {\cal O}(b^2) \!+\! \left \{ \T_{L}\! \leftrightarrow \!\T_{R} \right \} , \nn
\end{align}
where jet convolution integral ${\cal J}^{(-)}_0$ in \eq{eq:J0primeDefine}, up to order $b^2$, is
\begin{align}
\label{eq:J0PrimeBroadeningLimit}
{\cal J}^{(-)}_0(b) 
& = -\frac{\pi^2}{6\, b} + \frac{\pi^2}{6} + \frac{1}{4} + \frac{3}{2} \ln 2 + b\bigg(\frac{3}{2}-\frac{\pi^2}{3} -\frac{3}{2} \ln 2 \bigg) + {\cal O}(b^2) \, .
\end{align}
As in the positive-$b$ case, both jet and soft terms contain $1/b$ singularities which cancel out in the cross section, leading to 
\begin{align}
\label{eq:NLOCSbSmallNegBFull}
\bigg[\frac{1}{\sigma_0}\, &\frac{\rm \df^2\sigma^{(-)}}{\df\T_{L}\, \df\T_{R}}\bigg]^{\rm {\cal O}(\alpha_s)}_{\rm SCET_{II}}\!\! = \frac{\as(\mu){\,}C_F}{\pi}\, \delta(\T_{L}) \\
& \times \bigg[\delta(\T_{R}) \bigg(\!\!\!-\frac{7}{4}\! +\! \frac{\pi^2}{4}\! -\! \ln^2 2 + \frac{3 \ln 2}{2} \!+ \!b\Big(\frac{3}{2} \!-\! \frac{\pi^2}{3} \!-\!\frac{3 \ln 2}{2} +\frac{3\ln^2 2}{2}  \Big)\! \bigg)\!   \nn \\ 
& -\! \frac{3}{2} \bigg[ \frac{1}{\T_{R}} \bigg]_{+}\ns -\! 2 \bigg[ \frac{\ln\T_{R}}{\T_{R}} \bigg]_{+}\ns  + 2 \ln 2\bigg[\frac{1}{\T_R}\bigg]_{+} \ns +b \Big(\, \frac{3}{2} \bigg[ \frac{1}{\T_{R}} \bigg]_{+}\ns +\! 3 \bigg[ \frac{\ln\T_{R}}{\T_{R}} \bigg]_{+}\ns - 3 \ln 2\bigg[\frac{1}{\T_R}\bigg]_{+}\Big) \bigg]+ {\cal O}(b^2) \!+\! \left \{ \T_{L}\! \leftrightarrow \!\T_{R} \right \} \nn .
\end{align}
Taking the $b \rightarrow 0^{-}$ limit of this result gives 
 \begin{align}
 \label{eq:BroadeningLimit}
&\bigg[\frac{1}{\sigma_0}\, \frac{\df^{\,2}\sigma^{(-)}}{\df\T_{L} \df\T_{R}}\bigg]^{\rm {\cal O}(\alpha_s)}_{\rm sing.} \!\! =  \frac{\as{\,}(\mu)C_F}{\pi}\, \bigg\{ \delta(\T_{R})\, \delta(\T_{L})\bigg(\!\!\! - \frac{7}{2} + \frac{\pi^2}{2} - 3 \ln\!\frac{\mu}{Q} - 2\ln^2\!\frac{\mu}{Q}\bigg)  \\
& \qquad  + \bigg[\delta(\T_{L}) \bigg(\!\! - \frac{3{\,}Q}{4{\,}\mu} \left[\frac{2{\,}\mu}{Q\T_{R}}\right]_{+}\ns - \frac{Q}{\mu} \left[\frac{2{\,}\mu}{Q\T_{R}}\right]_{+}\!\!\! \ln\frac{\mu}{Q} - \frac{Q}{\mu} \left[\frac{2{\,}\mu \ln(Q\T_{R}/2{\,}\mu)}{Q\T_{R}}\right]_{+}\! \bigg)\! + \left \{ \T_{L}\! \leftrightarrow \!\T_{R} \right \}\!\! \bigg] \bigg\}\, ,  \nn
 \end{align}
which agrees with the double-differential jet broadening distribution obtained in~\cite{Chiu:2012ir}.

Since its left and the right limits ($b\rightarrow 0^{\pm}$) are the same, our cross section is a continuous function of the angularity parameter
at $b=0$. However, it is interesting to note that the soft and the jet functions independently, upon expansion in the regulators, are not smooth functions of $b$ at $b=0$, as in the case of angularities measured with respect to a recoil-insensitive axis~\cite{Larkoski:2014uqa}.

%
%

 \begin{figure}[t!]
  \centering
\includegraphics[height=0.35\textwidth]{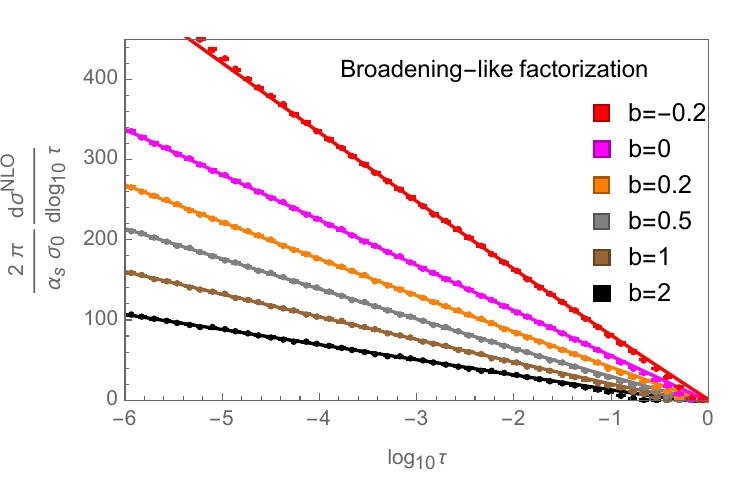}
   \caption{
Normalized one-loop angularity cross sections from \Eventtwo against our analytic one-loop results based on broadening-like factorization for a variety of exponents.}
  \label{fig:comb}
  \end{figure}

 \section{Numerical analysis and comparison against {\textsc{Event2}\xspace}}
 \label{sec:NumericalResults}
 
We compare both the broadening-like and the thrust-like normalized single-differential angularity distributions against numerical output from the \Eventtwo generator~\cite{Catani:1996vz}, for six different angularity exponents. In particular, for $b=\{2, 1, 0.2\}$, we used  $10^{12}$ events from the runs in~\cite{Procura:2018zpn} with $n_f=5$, an infrared cutoff $\rho = 10^{-10}$ and, in addition, we generated $10^{11}$ events with values of $\rho$ as low as $10^{-18}$ for $b=\{0.5, 0, -0.2\}$.\footnote{The parameter $\rho$ is defined by the fact that \Eventtwo regulates infrared divergences by cutting on the invariant mass of pairs of partons, $(p_i + p_j)^2> \rho\, Q^2$.} 

\Fig{comb} shows the normalized $\df \sigma/ \df \log_{10}{\tau}$ distribution from \Eventtwo against our analytic one-loop expressions from \eq{eq:NLOCS} and \eq{eq:NLOCSNegB}, which include both sub-leading singular terms and power corrections due to recoil. Here we use our untruncated results since our power corrections can be large and can significantly contribute to the agreement with \Eventtwo. 

\begin{figure}[t]
  \centering
   \includegraphics[width=0.49\textwidth]{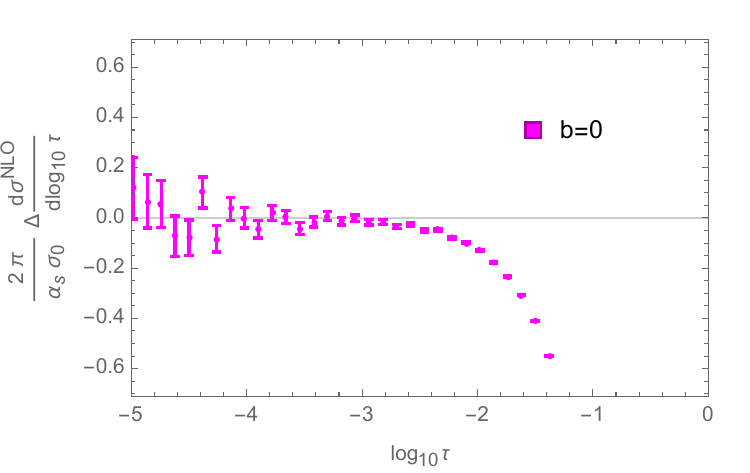}
   \includegraphics[width=0.49\textwidth]{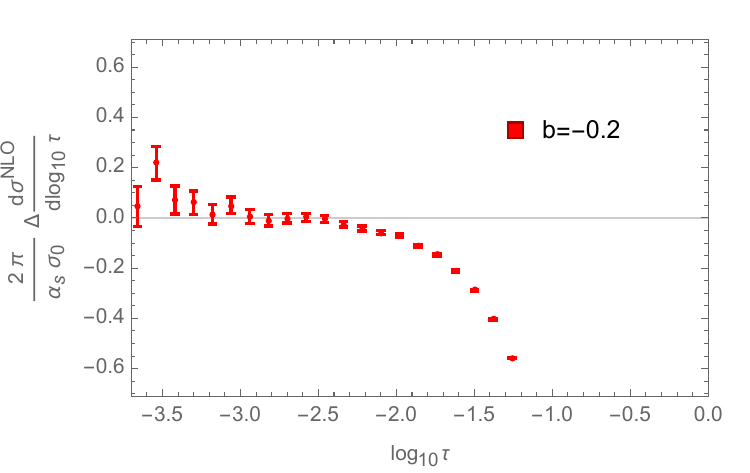}
   \caption{Differences between \Eventtwo and our one-loop results from SCET$_{\rm II}$  factorization for $\df \sigma/ \df \log_{10}{\tau}$ in the cases of $b=0$ (jet broadening) and $b=-0.2$.}
   \label{fig:logtauplots}
  \end{figure}

\begin{figure}[t]
  \centering
   \includegraphics[width=0.49\textwidth]{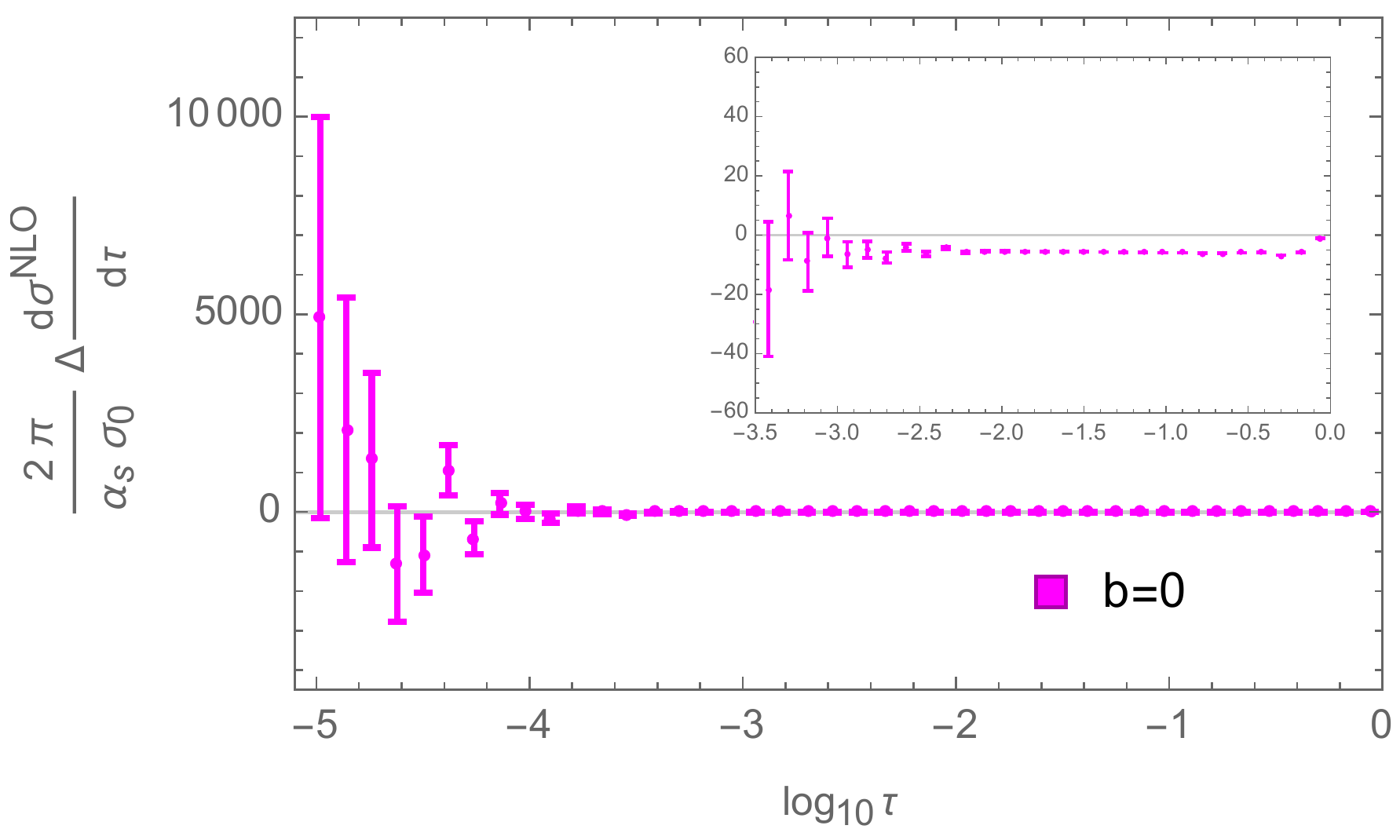}
   \includegraphics[width=0.49\textwidth]{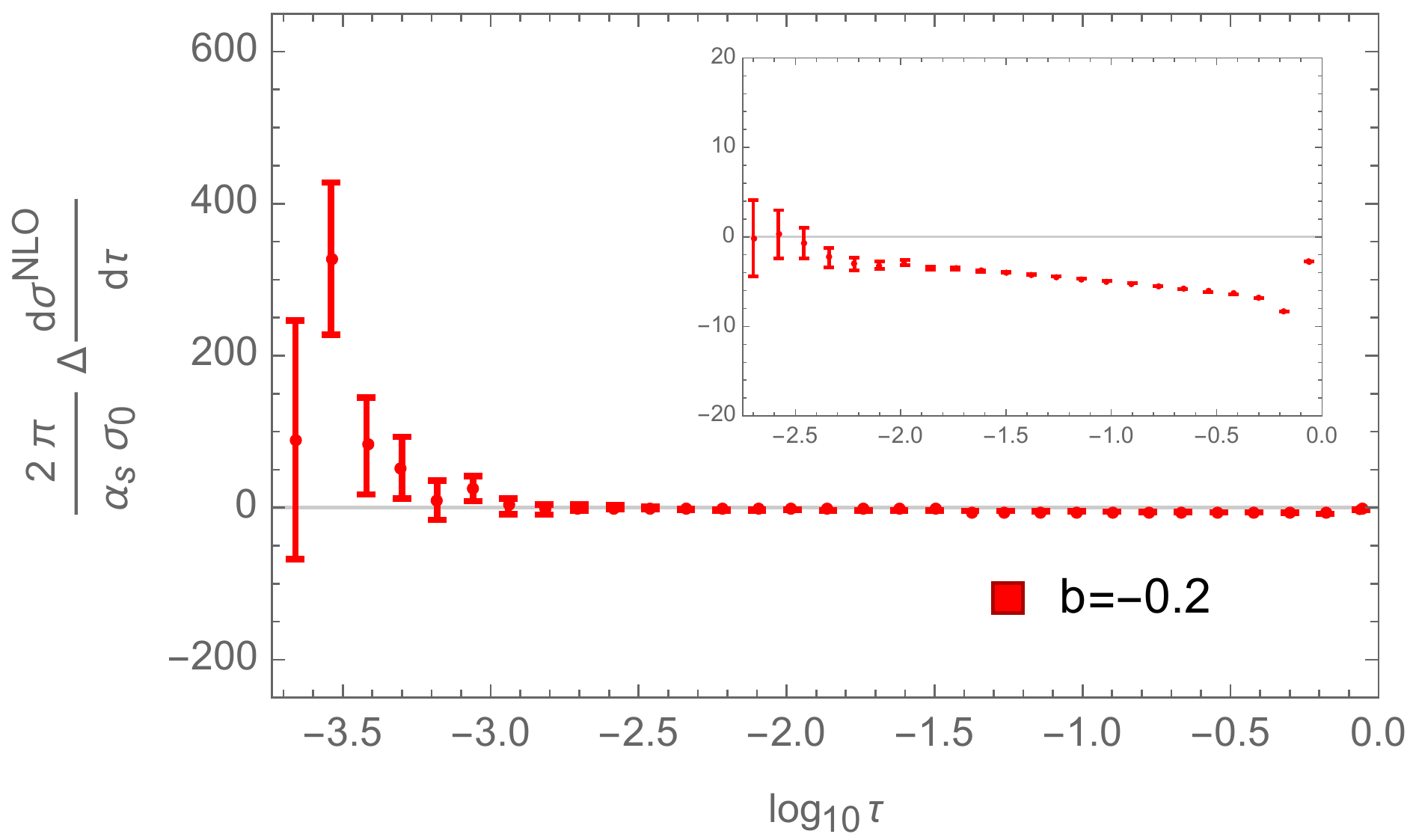}
   \caption{Differences between \Eventtwo and our one-loop predictions from SCET$_{\rm II}$   factorization for $\df \sigma/ \df{\tau}$ for $b=0$ and $b=-0.2$. The insets magnify the region of larger values of $\tau$.}
   \label{fig:tauplots}
  \end{figure}

We found agreement within numerical uncertainties between \Eventtwo and broadening-like SCET$_{\rm II}$  factorization for sufficiently small values of $\tau$ for all values of $b$ we tested. As an example, the differences between the \Eventtwo output and our expressions for $\df \sigma/ \df \log_{10}{\tau}$ for $b=0$ (jet broadening case) and $b=-0.2$ are shown in \Fig{logtauplots} for ranges of $\tau$ where no visible cutoff effects are present. The comparison is successful also for the differential cross section $\df \sigma/ \df{\tau} = 1/(\log_{10}\tau)\,\df \sigma/ \df \log_{10}\tau$, as illustrated in \Fig{tauplots} again for $b=0$ and $b=-0.2$.

\begin{figure}[t]
  \centering
   \includegraphics[width=0.49\textwidth]{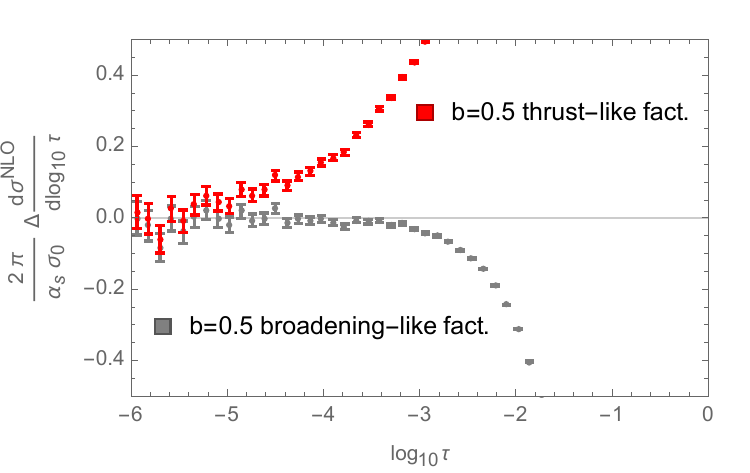}
   \includegraphics[width=0.49\textwidth]{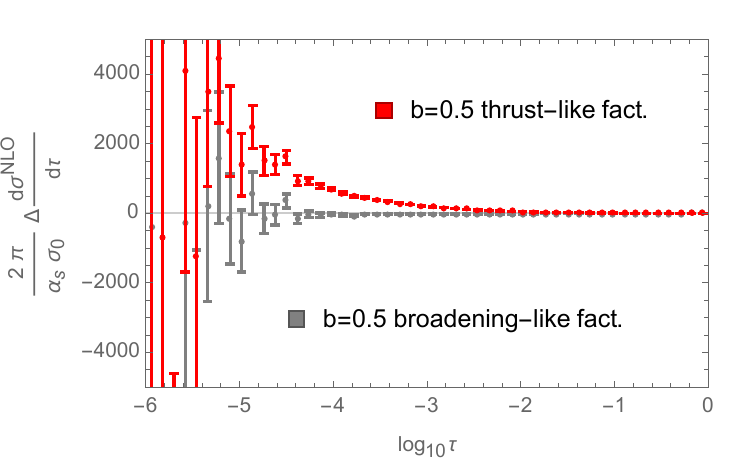}
   \caption{Differences between \Eventtwo and thrust-like (SCET$_{\rm I}$ )/broadening-like (SCET$_{\rm II}$) one-loop results for $\df \sigma/ \df \log_{10}{\tau}$ (left panel) and $\df \sigma/ \df {\tau}$ (right panel) for $b=0.5$.}
   \label{fig:b05plots}
  \end{figure}

For $b \geq 0.5$, we found agreement within error bars between \Eventtwo and both thrust-like and broadening-like factorization for sufficiently small values of $\tau$.\footnote{The consistency between \Eventtwo and thrust-like factorization up to NNLO for $b\geq0.5$ has been recently discussed and exploited in~\cite{Bell:2018gce}.} \Fig{b05plots} illustrates the case of $b=0.5$, both for $\df \sigma/ \df \log_{10}{\tau}$ and $\df \sigma/ \df {\tau}$. As shown in these plots, the extra terms provided by the SCET$_{\rm II}$ factorization theorem clearly improve the agreement with \Eventtwo in the region of intermediate values of $\tau$. 

The relative size of the sum of the sub-leading singular terms compared to the leading singular contribution is shown Tab.~\ref{tab:SizeOfCorrec}. Since these contributions turn out to be non-negligible in the peak region also for $0.5<b<1$, it is possible that the resummation of these terms plays an important role in precision calculations and extractions of the strong coupling. This will be addressed in a future publication~\cite{Budhraja:2018}.

\begin{table}
\centering
\begin{tabular}{|c|c|c|c|}
     \hline 
$b$ &  $N$ = max($n$)  & \% correction for $\T_b = 0.05$ & \% correction for $\T_b = 0.1$ \\
\hline
1 & 0 & $ 0 $ & 0 \\
\hline
0.8 &1 &  4 & 10 \\ 
\hline
0.5 & 1 & 10 & 20 \\
\hline
0.25 & 3 & 17 & 30 \\
\hline
0 & $\infty$ & 31 & 45 \\
\hline
-0.2 & 3 & 16 & 27 \\
\hline
-0.3 & 2 & 8 & 14 \\
\hline
-0.5 & 0 & 0 & 0 \\
\hline
\end{tabular}
\caption{Relative size of the sum of the sub-leading singular terms compared to the leading singular contribution in the peak region for the $\T_b$ distribution, for various values of $b$, see \eq{eq:NLOCrossSection} and \eq{eq:NLOCrossSectionNegB}. For a given $b$, $N$ denotes the maximum value of the summation index $n$ up to which the sub-leading terms are singular.}
\label{tab:SizeOfCorrec}
 \end{table}


\section{Conclusions}
\label{sec:Conclusions}

In this paper, we have started to investigate how a theoretical framework based on SCET$_{\rm II}$ factorization allows us to compute angularity distributions measured with respect to the thrust axis for any value of the exponent $b$. Working out the analytic expressions of these cross sections is a rich and complex problem due to the presence of two kinematic scales ($Q \tau$ and transverse momentum), two types of divergences and renormalization scales, and a continuous dependence of the mode scalings on $b$, which interpolates between different regimes. As a first step, we have focused on fixed-order calculations, and determined origin and nature of the various divergent structures appearing in the bare angularity jet and soft functions for arbitrary $b$, which get cancelled in the computation of the cross section.
We have produced novel one-loop results for the range $b<1$, which open up the possibility to use our formalism to extend present analyses of high-precision LEP data. We have computed and numerically quantified previously unknown recoil effects in the perturbative part of the spectrum for all values of $b$ and showed how the known limits of thrust and broadening are reproduced in our framework. All our one-loop distributions are found to be in agreement with \Eventtwo and our SCET$_{\rm II}$ formalism is shown to capture important sub-leading effects that clearly improve the agreement with \Eventtwo in the region of intermediate values of $\tau$.

Depending on the value of $b$, recoil effects amount to sub-leading singular terms and/or power corrections in the differential angularity distributions. These singular terms are integrable fractional powers of $\tau_b$ and their contribution at fixed-order turns out to be non-negligible even for $b>0.5$. It will thus be interesting to assess how the resummation of these terms is going to affect precision calculations and the extraction of the strong coupling from data.


\begin{acknowledgments}
We thank Andr\'e Hoang and Wouter Waalewijn for fruitful discussions and for feedback on the manuscript. 
We thank Wouter Waalewijn for high-statistics \Eventtwo runs on the computer cluster at NIKHEF.
We are also grateful to Jui-Yu Chiu, Wouter Waalewijn and Duff Neill for collaboration at an early stage of this work. AJ acknowledges support by a Ramanujan Fellowship of SERB DST. We also thank the Erwin-Schrödinger International Institute for Mathematics and Physics in Vienna for partial support. 
\end{acknowledgments}



\appendix
\section{Plus Distributions and their Identities}

Here we collect definitions and properties of plus distributions relevant for our calculations.

\subsection{Plus Distributions over scalar domains}
\label{appsec:PlusDistributions}
The standard plus distribution with boundary at $x = x_0$ for a function $f(x)$ which is less singular than $1/x^2$ as $x \to 0$, is defined as
\begin{equation}
\label{appeq:PlusDefine}
[\theta(x)\, f(x)]_{+}^{x=x_0} = \lim_{\e \rightarrow 0}[\theta(x-\e)\, f(x) + \delta(x-\e)\, F(x,x_0)]\, ,
\end{equation}
with
\begin{equation}
\label{appeq:BoundaryFunctionDefine}
F(x,x_0) = \int_{x_0}^x\! \df x'\, f(x') \, .
\end{equation}
In addition, the distribution satisfies the boundary condition
\begin{equation}
\label{eq:BoundaryConditionDefine}
\int_{0}^{x_0}\! \df x\, [\theta(x)\, f(x)]_{+}^{x=x_0} = 0\, .
\end{equation}
Plus distributions with two different boundaries can be related to each other. For boundaries $x = x_0$ and $x = +1$,
\begin{equation}
\label{appeq:BoundaryChangeRelation}
[\theta(x)\, f(x)]_{+}^{x=x_0} = [\theta(x)\, f(x)]_{+} + \delta(x) \int_{x_0}^1\! \df x'\, f(x')\, .
\end{equation}
In particular, we define the distributions relevant for this work as 
\begin{equation}
\label{appeq:PlusNewNotation}
{\cal L}_n^{\beta}(x ; x_0) \equiv \left[\frac{\theta(x)}{x^{1+\beta}}\, \ln^n x \right]_{+}^{x_0}\, ,
\end{equation}
so that, 
\begin{align}
\label{appeq:PlusNewNotationCases}
{\cal L}_0^{\beta}(x ; 1) = {\cal L}^{\beta}(x) &\equiv \left[\frac{\theta(x)}{x^{1+\beta}}\right]_{+}\, , \nn \\
\quad {\cal L}_n^{0}(x ; 1) = {\cal L}_n(x) &\equiv \left[\frac{\theta(x)}{x}\, \ln^n x \right]_{+}\, .
\end{align}
Note that from the definition above, ${\cal L}^{0} = {\cal L}_{0}$. 
The identity in \eq{appeq:BoundaryChangeRelation} for ${\cal L}^{\beta}$ and ${\cal L}_1^{\beta}$ then gives
\begin{align}
{\cal L}^{\beta}(x;\infty) = -\frac{1}{\beta}\, \delta(x) + {\cal L}^{0}(x) - \beta\, {\cal L}_{1}(x) + {\cal O}(\beta^2)\; , \label{appeq:PlusIdentity} \\
\quad {\cal L}_1^{\beta}(x;\infty) = -\frac{1}{\beta^{\, 2}}\, \delta(x) + {\cal L}_{1}(x) - \beta \, {\cal L}_{2}(x) + {\cal O}(\beta^2)\; . \label{appeq:PlusIdentityLog}
\end{align}
Another useful property is the rescaling identity (for $\zeta > 0$) for the two special cases above,
\begin{align}
\label{appeq:PlusFunction}
{\cal L}^{\beta = 0}(x/\zeta) \equiv \frac{1}{\zeta}\left[\frac{\theta(x)}{x/\zeta}\right]_{+}\!\! &=  {\cal L}^{0}(x) - \ln \zeta\, \delta(x)\, , \\
{\cal L}_{n=1}(x/\zeta) \equiv \frac{1}{\zeta}\left[\frac{\theta(x)\ln(x/\zeta)}{x/\zeta}\right]_{+}\!\!\! &= {\cal L}_{1}(x) - \ln\zeta\,\, {\cal L}^{0}(x) + \frac{1}{2} \ln^2\! \zeta\,\, \delta(x)\, .
\end{align}
\subsection{Plus Distributions over vector domains}
\label{appsec:VectorPlusDistributions}
For the class of vector distributions of the type~\cite{Chiu:2012ir}
\begin{equation}
\label{appeq:VectorDist}
{\cal L}_n^\beta(\mu,\vt{p}) \equiv \frac{1}{2\pi\mu^2}\bigg[\frac{1}{(\V{p}/\mu^2)^{1+\beta}}\ln^n(\V{p}/\mu^2)\bigg]_{+}^{\cal D_{\mu}}\, ,
\end{equation}
used in \eq{eq:SoftExpanded}, we define two limiting cases:
\begin{align}
\label{appeq:LimitingCases}
{\cal L}_0^\beta(\mu,\vt{p}) = {\cal L}^\beta(\mu,\vt{p}) \equiv \frac{1}{2\pi\mu^2} \bigg[\frac{1}{(\V{p}/\mu^2)^{1+\beta}}\bigg]_{+}^{\cal D_{\mu}}\, , \\
{\cal L}_n^0(\mu,\vt{p}) = {\cal L}_n(\mu,\vt{p}) \equiv \frac{1}{2\pi\mu^2} \bigg[\frac{\ln^n(\V{p}/\mu^2)}{(\V{p}/\mu^2)}\bigg]_{+}^{\cal D_{\mu}} \, . 
\end{align}
Here $\cal D_{\mu}$ denotes a disc of radius $\mu$ 
about the origin in the $\vt{p}$-space. The plus distribution  ${\cal L}_n^\beta(\mu,\vt{p})$ over a vector function is then defined as 
\begin{align}
\label{appeq:VectorPlusDistribution}
\int \frac{\df^2\vt{p}}{(2\pi)^2}\, g(\vt{p}) {\cal L}_n^\beta(\mu,\vt{p}) = \lim_{\e \rightarrow 0^+} \frac{1}{2\pi}\! \int\! \frac{\df^{2+2\e}\vt{p}}{(2\pi)^{2+2\e}}\, g(\vt{p})  \frac{1}{\mu^{2+2\e}} \frac{\ln^n(\V{p}/\mu^2)}{(\V{p}/\mu^2)^{1+\beta}} - \frac{g(0)}{2\pi} B_\e[f;\mu] \, ,
\end{align}
where, $g(\vt{p})$ is a suitably chosen test function that has no singularities at the origin. 
The boundary term, $B_\e$, defined as
\begin{align}
\label{appeq:BoundaryFuncDefinition}
B_\e[{\cal L}_n^\beta;\mu] = \frac{1}{2\pi} \int_{{\cal D}_{\mu}} \frac{\df^{2+2\e}\vt{q}}{(2\pi)^{2+2\e}} \frac{1}{\mu^{2+2\e}} \frac{\ln^n(\V{q}/\mu^2)}{(\V{q}/\mu^2)^{1+\beta}}  \, ,
\end{align}
is such that the condition
\begin{align}
\label{appeq:BoundaryCondition}
\int_{{\cal D}_{\mu}} \frac{\df^{2}\vt{q}}{(2\pi)^{2}}\, {\cal L}_n^\beta(\mu,\vt{q}) = 0 \, 
\end{align}
is satisfied. The explicit expression for the boundary term, in dimensional regularization, for the distributions ${\cal L}^\beta$ and ${\cal L}_{n}$, is given as
\begin{align}
\label{appeq:BoundaryTermsLimitCases}
B_\e[{\cal L}^\beta;\mu] &= \frac{1}{(4\pi)^{\!1+\e}\, \Gamma(1+\e)\,(\e-\beta)}\, , \\
B_\e[{\cal L}_n;\mu] &= \frac{(-1)^{n}\, \Gamma(1+n)}{(4\pi)^{\!1+\e}\, \Gamma(1+\e)\,\e^{1+n}}\, .  
\end{align}
For completeness, we also mention the rescaling identities ($\rho>0$),
\begin{align}
\label{appeq:VectorDistRescaling}
{\cal L}^{\beta}\Big(\rho\mu,\vt{p}\Big) &= \rho^{2\beta} {\cal L}^{\beta}\Big(\mu,\vt{p}\Big) - \frac{\rho^{2\beta}-1}{2\,\beta}\, \delta^{(2)}(\vt{p}) \, , \\
{\cal L}_n\Big(\rho\mu,\vt{p}\Big) &= \sum_{m=0}^{n} \,_nC_m\, (-1)^{n-m} \ln^m(\rho^2) {\cal L}_{n-m}\Big(\mu,\vt{p}\Big) - \frac{\ln^{n+1}(\rho^2)}{2(n+1)}\, \delta^{(2)}(\vt{p}) \, ,
\end{align}
with $\,_nC_m = \Gamma(1+n)/[\Gamma(1+m)\Gamma(1+n-m)]$.
\section{Evaluation of the one-loop jet function}
\label{appsec:JetIntegrals}

In this appendix, we present the details of the calculation for the jet function diagrams and the relevant zero-bin subtractions for all $b>-1$ angularities. 
\subsection{Jet Diagram (a)}
\label{appsubsec:JetA}
Since the jet diagram (a) in \Fig{JetDiagrams} contains no rapidity divergence, the $\eta$-regulator can be dropped. The diagram in Feynman gauge is then given as 
\begin{align}
\label{appeq:JetNaiveAIntegral}
{\cal J}^{(1)}_{\rm{a}}(\T_{R}, 0) &= \left(\frac{e^{\gamma_{E}} \mu^2}{4\pi}\right)^{\!\!\e} \,\frac{(2\pi)^{3-2\e}}{N_c} \!\int\! \frac{\df^d l}{(2\pi)^{d-1}}\, \delta(l^2)\, \theta(l^0) \int\! \frac{\df^d p}{(2\pi)^{d-1}}\, \delta(p^2)\, \theta(p^0)\, \delta(p^-\!+l^-\!-Q) \nn\\ 
& \quad \times \delta^{d - 2}(\vt{l} + \vt{p})\,\, \delta\bigg(\!\T_{R} - \frac{\MV{p}}{Q}\Big(\frac{p^+}{p^-}\Big)^{\!\frac{b}{2}} \ns\;- \frac{\MV{l}}{Q}\Big(\frac{l^+}{l^-}\Big)^{\!\frac{b}{2}}\bigg)\, {\rm{tr}}\bigg[\frac{\bar{\slashed{n}}}{2}\, \frac{i (\slashed{l} + \slashed{p})}{(l+p)^2}\, (i g{\,} \gamma^{\mu} T^a)\,   \nn \\
&\quad \times  u_s(p)\, {\bar{u}}_s(p)\, (- g_{\mu \nu})\, (i g{\,} \gamma^{\nu} T^a)\, \frac{i (\slashed{l} + \slashed{p})}{(l+p)^2} \bigg]  \nn \\ 
&=-\frac{\as(\mu)\,C_F}{2\pi^{2-\e}} e^{\e\gamma_{E}} \mu^{\!2\e} \!\int\! \df^d l\, \delta(l^2)\, \theta(l^0) \int\! \df^d p\, \delta(p^2)\, \theta(p^0)\, \delta(p^-\!+l^-\!-Q)\, \delta^{(d - 2)}(\vt{l} + \vt{p}) \nn\\ 
& \quad \times \delta\bigg(\!\T_{R} - \frac{\MV{p}}{Q}\Big(\frac{p^+}{p^-}\Big)^{\!\frac{b}{2}} \ns\;- \frac{\MV{l}}{Q}\Big(\frac{l^+}{l^-}\Big)^{\!\frac{b}{2}}\bigg) {\rm{tr}}\bigg[\frac{\bar{\slashed{n}}}{2}\, \frac{(\slashed{l} + \slashed{p})}{(l+p)^2}\, \gamma^{\mu}\, \slashed{p}\, \gamma_{\mu} \, \frac{(\slashed{l} + \slashed{p})}{(l+p)^2} \bigg] \nn \\ 
& = \frac{\as(\mu) C_F}{\pi}\, \frac{e^{\e \gamma_E}}{\Gamma(1-\e)} \Big(\frac{\mu}{Q} \Big)^{\!\!2\e}\! \frac{1}{\T_{R}^{1+\frac{2\e}{1+b}}}\,\,  \frac{1-\e}{1+b} \int_{0}^{1}\! \df x\, x\, ((1 - x)^{-b} + x^{- b})^{\!\frac{2\e}{1+b}}\, .
\end{align}
Here $x = l^{-}/Q$ is the light-cone momentum fraction of the emitted gluon. The trace in the second line of the integral simplifies as follows 
\begin{align}
\label{appeq:TraceA}
{\rm{tr}}\bigg[\frac{\bar{\slashed{n}}}{2}\, \frac{(\slashed{l} + \slashed{p})}{(l+p)^2}\, \gamma^{\mu}\, \slashed{p}\, \gamma_{\mu} \, \frac{(\slashed{l} + \slashed{p})}{(l+p)^2} \bigg] &= (2-d) {\rm{tr}}\bigg[\frac{\bar{\slashed{n}}}{2}\, (\slashed{l} + \slashed{p})\, \slashed{p}\, (\slashed{l} + \slashed{p}) \bigg]\times \frac{1}{[(l+p)^2+i\e]^2} \nn \\
&= (2-d)\,p^{+}\,\frac{(l^{-}+p^{-})^{\!2}}{[(l+p)^2+i\e]^2}\times {\rm tr}\bigg[\frac{\bar{\slashed{n}}}{2}\, \frac{\slashed{n}}{2}\, \frac{\bar{\slashed{n}}}{2}\, \frac{\slashed{n}}{2}\bigg] \nn \\
& = -2(1-\e)\,\frac{p^{+}}{\,(l^{+}+p^{+})^{\!2}}\,\times 2 \, ,
\end{align}
and the final result is obtained by performing the delta function integrals. 
The analytical expression of the integral in the last line of \eq{appeq:JetNaiveAIntegral} is not known. 
However, the divergence structure of interest
can be obtained by expanding the result of \eq{appeq:JetNaiveAIntegral} in powers of $\e$. The result, upon expansion, is given as
\begin{eqnarray}
\label{appeq:JetDiagramA}
{\cal J}^{(1)}_{\rm{a}}(\T_{R}, 0) = \frac{\as(\mu) C_F}{\pi} \bigg[\!\!-\! \frac{1}{4\, \e}\, \delta(\T_{R})\! +\! \frac{1}{2(1+b)}\, {\cal L}^{0}\Big(\frac{\T_{R}}{m}\Big) + i_a \, \delta(\T_{R})  + \mathcal{O}(\e) \bigg] , 
\end{eqnarray}
where
\begin{equation}
\label{appeq:IntegralADefine}
i_a = \frac{1}{4(1+b)} - \frac{1}{1+b}\int_{0}^{1}\! \df x\, x\, \ln\!\Big(1+\Big(\frac{x}{1-x} \Big)^{\! b} \Big)\, ,
\end{equation} 
and $m = (\mu/Q)^{\! 1+b}$, is a scale factor. The same result holds for negative-$b$ angularities as well.
\subsection{Jet Diagram (b)}
\label{appsubsec:JetB}
The jet diagram (b), in general, contains rapidity divergences for both the unsubtracted jet integral and its corresponding zero-bin contribution. However, as discussed in the text, the net result of the jet diagram (b), i.e. ${\cal J}_{\rm b,unsub} - {\cal J}_{b,0}$, contains no rapidity divergence for $-1<b<0$. Here we will present the calculation of the unsubtracted diagram separately for positive-$b$ and negative-$b$ cases. The zero-bin contribution is evaluated in App.~\ref{appsubsec:ZeroBin}.

\subsubsection{Positive-$b$ angularities}
\label{appsubsubsec:JetBPositive}
The Feynman integral for the unsubtracted jet diagram (b) and its mirror image, is given as
\begin{align}
\label{appeq:JetNaiveBIntegral}
{\cal J}^{(1+)}_{\rm{b, unsub}}(\T_{R},0) &= 2 \left(\frac{e^{\gamma_{E}} \mu^2}{4\pi}\right)^{\!\!\e} \,\frac{(2\pi)^{3-2\e}}{N_c} \!\int\! \frac{\df^d l}{(2\pi)^{d-1}}\, \delta(l^2)\, \theta(l^0) \int\!\! \frac{\df^d p}{(2\pi)^{d-1}}\, \delta(p^2)\, \theta(p^0)\, \delta(p^-\!+ l^-\!- Q) \nn \\ 
& \quad \times \delta\bigg(\!\T_{R} - \frac{\MV{p}}{Q}\Big(\frac{p^+}{p^-}\Big)^{\!\frac{b}{2}} \ns \,\,- \frac{\MV{l}}{Q}\Big(\frac{l^+}{l^-}\Big)^{\!\frac{b}{2}}\bigg)\, \delta^{d - 2}(\vt{l} + \vt{p})\, {\rm{tr}}\bigg[\frac{\bar{\slashed{n}}}{2}\, \frac{i (\slashed{l} + \slashed{p})}{(l+p)^2}\, (i g{\,} \gamma^{\mu} T^a) \nn \\ 
& \quad \times u_s(p)\, \bar{u}_s(p) \, (- g_{\mu \nu}) \frac{(g{\,} \bn^{\nu} T^a\, {\nu}^{\eta}w^2)}{(\bn \cdot l)^{1+\eta}}\bigg] \nn \\    
& = \frac{2}{1+b}\, \frac{\as(\mu) C_F}{\pi}\, \frac{e^{\e \gamma_E}w^2}{\Gamma(1-\e)}\, \Big(\frac{\mu}{Q} \Big)^{\!2\e}\! \Big(\frac{\nu}{Q} \Big)^{\!\eta}\! \frac{1}{\T_{R}^{1+\frac{2\e}{1+b}}}\! \int_0^1\! \df x\, ((1 - x)^{- b} + x^{-b})^{\!\frac{2\e}{1+b}}\, \frac{(1-x)}{x^{1+\eta}} \nn \\
&= \frac{2}{1+b}\, \frac{\as(\mu) C_F}{\pi}\, \frac{e^{\e \gamma_E}w^2}{\Gamma(1-\e)}\, \Big(\frac{\mu}{Q} \Big)^{\!2\e}\! \Big(\frac{\nu}{Q} \Big)^{\!\eta}\! \frac{1}{\T_{R}^{1+\frac{2\e}{1+b}}}\! \int_0^1\! \df x\, \bigg(\!1+\!\Big(\frac{x}{1-x}\Big)^{\! b} \bigg)^{\!\!\frac{\!2\e}{1+b}} \! \frac{(1-x)}{x^{1+\eta+\frac{2\e\,b}{1+b}}} .
\end{align}
For positive-$b$ angularities, the factor $((1 - x)^{- b} + x^{-b})^{\!\frac{2\e}{1+b}}$ yields a singular contribution as $x\rightarrow 0$, which we have factored out in the final step of \eq{appeq:JetNaiveBIntegral}. 
This implies that for $b>0$ the $\eta$-regulator is not required since the divergence due to $1/x^{1+\eta+\frac{2\e\,b}{1+b}}$ as $x\rightarrow 0$ is already taken care of by the $\e$-regulator. Dropping the $\eta$-regulator, we expand the result for the unsubtracted jet diagram (b), for positive-$b$ values, in powers of $\e$. The result, upon expansion, is given as
\begin{align}
\label{appeq:JetNaivePositiveB}
{\cal J}^{(1+)}_{b,\rm{unsub}}(\T_R,0) &= \frac{\al_s(\mu)\, C_F}{\pi}  \bigg[\frac{1+b}{2\,b\,\e^2}\, \delta(\T_R) + \frac{1}{\e} \delta(\T_R) - \frac{1}{b\,\e}\, {\cal L}^{0}\Big(\frac{\T_{R}}{m}\Big)  + \frac{2}{b(1+b)}\, {\cal L}_{1}\Big(\frac{\T_{R}}{m}\Big) \nn \\
&\quad  -\frac{2}{1+b}\, {\cal L}^{0}\Big(\frac{\T_{R}}{m}\Big)  \! -\! \frac{2}{1+b}\,\delta(\T_R)\!\! \int_0^1 \df x\, \frac{(1-x)}{x} \ln\Big(1+\Big(\frac{x}{1-x}\Big)^{\! b}\Big)\!\! +\frac{2\,b}{1+b}\, \delta(\T_R) \nn \\
&\quad - \frac{\pi^2}{24}\, \frac{1+b}{b}\, \delta(\T_R) \!  + \mathcal{O}(\e) \bigg] \, .
\end{align}
This expression contains $b$-dependent $\e$-divergences which will cancel upon taking the appropriate zero-bin contribution into account.

\subsubsection{Negative-$b$ angularities}
\label{appsubsec:JetBNegative}
As the test function, $((1 - x)^{- b} + x^{-b})^{\!\frac{2\e}{1+b}}$, is well defined for $b<0$, we rewrite the second line in \eq{appeq:JetNaiveBIntegral} for negative-$b$ angularities as
\begin{align}
\label{appeq:ReWriteIntegralNegativeB}
{\cal J}^{(1-)}_{\rm{b, unsub}}(\T_{R},0) &= \frac{2}{1+b} \frac{\as(\mu) C_F}{\pi}\, \frac{e^{\e \gamma_E}w^2}{\Gamma(1-\e)}\, \Big(\frac{\mu}{Q} \Big)^{\!2\e}\! \Big(\frac{\nu}{Q} \Big)^{\!\eta}\! \frac{1}{\T_{R}^{1+\frac{2\e}{1+b}}}\! \int_0^1\! \df x\, \bigg(\!1+\!\Big(\frac{x}{1-x}\Big)^{\!-b} \bigg)^{\!\frac{2\e}{1+b}}\, \nn \\
&\tab[9cm] \times \frac{(1-x)^{1-\frac{2\e\,b}{1+b}}}{x^{1+\eta}}\, ,
\end{align}
to obtain the same form for the integral as in the positive-$b$ case.

The result of the unsubtracted jet diagram (b), now requires the explicit use of the $\eta$-regulator to regularize the divergence as $x \rightarrow 0$. We expand the integrand in \eq{appeq:ReWriteIntegralNegativeB} in $\eta$ and $\e$ ($\eta \ll \e$) and perform the integral to obtain 
\begin{align}
\label{appeq:JetNaiveNegativeB}
{\cal J}^{(1-)}_{\rm{b, naive}}(\T_R,0) &= \frac{\al_s(\mu)\, C_F}{\pi} \bigg[\!\!-\! \frac{2{\,}e^{\e \gamma_E} w^2}{\eta{\,}(1+b){\,}\Gamma(1-\e)}\, \Big(\frac{\mu}{Q}\Big)^{\!2\e} {\cal L}^{\frac{2\e}{1+b}}(\T_R;\infty) +\frac{1}{\e}\, \delta(\T_R) + \frac{1}{\e} \ln\!\frac{\nu}{Q}\, \delta(\T_R) \nn \\
&\quad -\frac{2}{1+b}\,{\cal L}^{0}\Big(\frac{\T_{R}}{m}\Big)+\frac{2\,b}{1+b}\,\delta(\T_R) - \frac{\pi^2\,b}{3(1+b)}\, \delta(\T_R) - \frac{2}{1+b} \ln\!\frac{\nu}{Q}\,{\cal L}^{0}\Big(\frac{\T_{R}}{m}\Big) \nn \\
&\quad -\! \frac{2}{1+b}\,\delta(\T_R) \!\!\int_0^1\!\! \df x\, \frac{(1-x)}{x} \ln\!\Big(1+\!\Big(\frac{x}{1-x}\Big)^{\! -b}\Big) \!+ \mathcal{O}(\eta,\e)\bigg]\, , 
\end{align}
where the full $\e$-dependence has been retained in the $1/\eta$--term according to the rapidity renormalization prescription~\cite{Chiu:2012ir}. 
 
\subsection{Zero-Bin Contribution}
\label{appsubsec:ZeroBin}
The zero-bin subtraction to the jet diagram (b) can be evaluated by taking the gluon momentum to scale as $l^{\mu} \sim Q(\lambda^{\!1+b},\lambda^{\!1+b},\lambda^{\!1+b})$. This results to
\begin{align}
{\cal J}^{(1\pm)}_{\rm{b, 0}}(\T_{R},0) &=  2 \bigg(\frac{e^{\gamma_E} \mu^2}{4\pi}\bigg)^{\!\!\e}\, \frac{(2\pi)^{3-2\e}}{N_c} \!\!\int\!\! \frac{\df^d l}{(2\pi)^{d-1}}\delta(l^2)\, \theta(l^0) \!\!\int\!\! \frac{\df^d p}{(2\pi)^{d-1}} \delta(p^2)\, \theta(p^0)\, \delta(p^{-}\!\!- Q) \nn \\ 
& \quad  \times \delta\bigg(\!\T_{R} - \frac{\MV{p}}{Q}\Big(\frac{p^+}{p^-}\Big)^{\!\frac{b}{2}} \ns - \frac{\MV{l}}{Q}\Big(\frac{l^+}{l^-}\Big)^{\!\frac{b}{2}}\bigg)\, \delta^{d-2}(\vt{l} + \vt{p})\, {\rm{tr}}\bigg[\frac{\bar{\slashed{n}}}{2}\, \frac{i (\slashed{l} + \slashed{p})}{p^- l^+}\, (i g{\,}\gamma^{\mu} T^a) \nn \\ 
& \quad \times  u_s(p){\,}\bar{u}_s(p)\, (- g_{\mu \nu}) \, \frac{g{\,}\bn^{\nu} T^a}{\bn \cdot l} \bigg] \nn \\
&= \frac{2 \as(\mu) C_F}{\pi} \frac{e^{\e \gamma_E}}{\Gamma(1-\e)} w^2\, \Big(\frac{\nu}{Q}\Big)^{\!\eta}\, \Big(\frac{\mu}{Q}\Big)^{\!2\e}\! \frac{1}{1+b} \frac{1}{\T_{R}^{1+\frac{1\e}{1+b}}}\! \int_0^{\infty}\!\! \df x\, \frac{(1+x^{-b})^{\frac{2\e}{1+b}}}{x^{1+\eta}}\, \nn \\ 
&= {\rm sgn}(b)\frac{2 \as(\mu) C_F}{\pi} \frac{e^{\e \gamma_E}}{\Gamma(1-\e)} w^2\, \Big(\frac{\nu}{Q}\Big)^{\!\eta}\, \Big(\frac{\mu}{Q}\Big)^{\!2\e}\! \frac{1}{1+b} \frac{1}{\T_{R}^{1+\frac{2\e}{1+b}}} \frac{\Gamma(\frac{\eta}{b})\, \Gamma(- \frac{2\e}{1+b} - \frac{\eta}{b})}{b{\,}\Gamma(-\frac{2\e}{1+b})}\, , \label{appeq:ZeroBinIntegral}
\end{align}
which holds for all $b>-1$.
It is worth mentioning that the integral in the second line of the equation changes sign for negative-$b$ angularities and hence we write the outcome as given in \eq{appeq:ZeroBinIntegral}. The final result, upon expansion in the regulators $\eta \ll \e$ yields
\begin{align} 
\label{appeq:ZerBin}
&{\cal J}_{b,0}^{(1\pm)}(\T_{R}, 0) = \frac{\as(\mu)\, C_F}{\pi} {\rm sgn}(b) \bigg[\frac{2\, w^2\, e^{\e\, \gamma_E}}{\eta\, (1+b)\, \Gamma(1-\e)}\, \Big(\frac{\mu}{Q}\Big)^{\! 2\e} {\cal L}^{\frac{2\e}{1+b}}(\T_R;\infty)\!+\! \frac{1+b}{2\,b\,\e^2}\, \delta(\T_R) \!-\! \frac{1}{b\,\e}\, {\cal L}^{0}\Big(\frac{\T_{R}}{m}\Big)   \nn \\
&\quad -\frac{1}{\e} \ln\!\frac{\nu}{Q}\, \delta(\T_R) + \frac{2}{b(1+b)}\, {\cal L}_{1}\Big(\frac{\T_{R}}{m}\Big)+\frac{2}{1+b}\, \ln\!\frac{\nu}{Q}\,{\cal L}^{0}\Big(\frac{\T_{R}}{m}\Big) -\frac{\pi^2}{24\,b}\, \frac{b^2+2\,b+9}{1+b}\,  \delta(\T_R) + \mathcal{O}(\eta,\e) \bigg] \, .
\end{align} 
We stress that the zero-bin integral contains an $\eta$-divergence for all angularities. It is only the unsubtracted jet diagram (b) which does not require a rapidity regulator for $b>0$ angularities.
\section{Evaluation of the one-loop soft function}
\label{appsec:SoftIntegrals}
Here we present the calculation of the soft function diagrams for all angularities ($b>-1$). 
The one-loop contribution to the soft function due to diagrams (a) and (b) of \Fig{SoftDiagrams}, which give the same result due to symmetry, amounts to
\begin{align}
\label{appeq:OneLoopSoft}
{\cal S}^{(1)}(\T_{L},\T_{R},\V{p},\V{k}) &=  2 \bigg(\frac{e^{\gamma_E} \mu^2}{4\pi}\bigg)^{\!\!\e}\, \frac{\pi^{2-2\e}\vt{k}^{-2\e}\, \vt{p}^{\,-2\e}}{N_c\, \Gamma^2(1-\e)} \nu^{\eta}\, w^2 \!\!\int\!\! \frac{\df^d l}{(2\pi)^{d-1}}\delta(l^2)\, \theta(l^0) \vert2\,l^3\vert^{-\eta} \bigg\{ \theta(l^3)\, \delta(\T_L) \nn \\ 
& \quad \delta\bigg(\!\T_{R} \!- \!\frac{\MV{l}}{Q}\Big(\frac{l^+}{l^-}\Big)^{\!\frac{b}{2}}\bigg) \delta^{2-2\e}(\vt{l}-\vt{p})\, \delta^{2-2\e}(\vt{k}) +  \theta(-l^3)\, \delta\bigg(\!\T_{L} \!-\! \frac{\MV{l}}{Q}\Big(\frac{l^-}{l^+}\Big)^{\!\frac{b}{2}}\bigg) \nn \\
&\quad \delta(\T_{R})\, \delta^{2-2\e}(\vt{l}-\vt{k})\, \delta^{2-2\e}(\vt{p}) \bigg\} {\rm tr}\bigg[\frac{g\,n^{\mu} T^a}{l^+}\, \frac{g\,\bar{n}^{\nu} T^a}{-\,l^-}\bigg] \sum_{\rm pol} \e_{\mu}^r(l)\, \e_{\nu}^{*r}(l)  \nn \\ 
&= \frac{2\,\as(\mu) C_F}{\pi} e^{\e\gamma_E} \mu^{2\e} \frac{\vt{p}^{\,-2\e}}{\Gamma(1-\e)} \nu^{\eta}\, w^2\, \delta(\V{k})\, \delta(\T_L) \!\!\int\!\! \df^d l\,\, \delta(l^2)\, \theta(l^0) \vert2\,l^3\vert^{-\eta}  \nn \\
& \quad \Bigg\{ \theta(l^3)\, \delta\bigg(\!\T_{R} \!-\! \frac{\MV{l}}{Q}\Big(\frac{l^+}{l^-}\Big)^{\!\frac{b}{2}}\bigg) \delta^{2-2\e}(\vt{l}-\vt{p}) + \bigg\{ 
  \begin {array} {c} 
  \T_L \leftrightarrow \T_R \\ 
  \V{k} \leftrightarrow \V{p} 
  \end {array} \bigg\} \Bigg\} \frac{1}{l^{+}}\, \frac{1}{l^{-}} \nn \\
&= \frac{\as(\mu){\,}C_F}{\pi} \frac{\mu^{2\e}e^{\e \gamma_E}}{\Gamma(1-\e)} \nu^{\eta}\, w^2\, Q\, \delta(\V{k})\, \delta(\T_{L})\, \theta\Big(\Big(\frac{\MV{p}}{Q{\,}\T_{R}}\Big)^{\!1/b} \ns - 1\Big)  \nn \\ 
& \quad \times (p^2_{\perp})^{-1-\e-\frac{\eta+1}{2}} \frac{\Big\vert 1\!-\Big(\frac{Q{\,}\T_{R}}{\MV{p}}\Big)^{\!2/b}\Big\vert^{-\eta}}{\big\vert b\big\vert{\,}\Big(\frac{Q{\,}\T_{R}}{\MV{p}}\Big)^{\!1-\eta/b}} + \bigg\{ 
  \begin {array} {c} 
  \T_L \leftrightarrow \T_R \\ 
  \V{k} \leftrightarrow \V{p} 
  \end {array} \bigg\}\, . 
\end{align}
As noted in the text, this expression contains coupled distribution in the variables $\T_{R}$ and $p_{\perp}$ (or $\T_{L}$ and $k_{\perp}$) which we decouple by making appropriate change of variables as shown in \eqs{eq:PositiveBDefinition}{eq:NegativeBDefinition}, respectively. 


\section{Evaluation of the soft convolution integral}
In this appendix, we calculate the soft convolution integrals $\text{I}^{(+)}_s(\T_{R})$ and $\text{I}^{(-)}_s(\T_{R})$ entering the cross sections for positive and negative values of $b$, respectively. 

\subsection{Soft convolution integral $\text{I}^{(+)}_s(\T_{R})$}
\label{appsubsec:PositiveBSoftFactor}
The integral $\text{I}^{(+)}_s(\T_{R})$ as defined in \eq{eq:ISIntegralDefine}, is given by
\begin{equation}
\label{appeq:SoftConvolutionIntegralDefine}
{\text{I}}^{(+)}_s(\T_{R}) \equiv \frac{2}{b} \int_0^{Q{\,}\T_{R}^{1/(1+b)}}\bnspace {dp_{\perp}}{\,} \theta\Big(\frac{p_{\perp}}{Q} - \T_{R} + \Big(\frac{p_{\perp}}{Q}\Big)^{\!1 + b}\Big) \frac{Q}{p_{\perp}}\, \frac{1}{\mu} \bigg[\frac{1}{p_{\perp}/\mu}\bigg]_{+} \bigg[\frac{p_{\perp}/Q}{\T_{R} - \Big(\frac{p_{\perp}}{Q}\Big)^{\!1 + b}}\bigg]_{+} .
\end{equation}
First of all, we perform a change of variables as
\begin{equation}
\label{appeq:SoftConvolutionNewVariables}
\Big(\frac{p_{\perp}}{Q}\Big)^{\!1 + b} = \rho\, \;\;\;\;{\rm implying}\;\;\;\; dp_{\perp} = \frac{Q}{1+b}\, \rho^{\frac{-b}{1+b}}\, d\rho \, .
\end{equation}
Writing the integral in terms of dimensionless quantities $\rho$ and $\T_{R}$ gives
\begin{align}
\label{appeq:SoftConvolutionIntegral}
\text{I}^{(+)}_s(\T_{R}) &= \frac{2}{b{\,}(1+b)}\int_{0}^{\T_{R}}d\rho\, \theta(\rho+\rho^{\frac{1}{1+b}}-\T_{R}) \frac{1}{\rho}\frac{Q}{\mu}\left[\frac{\mu/Q}{\rho^{\frac{1}{1 + b}}}\right]_{+}\left[\frac{\rho^{\frac{1}{1 + b}}}{\T_{R} - \rho}\right]_{+} \nn \\
&= \frac{2}{b{\,}(1+b)}\int_{0}^{\T_{R}}d\rho\, \theta(\rho+\rho^{\frac{1}{1+b}}-\T_{R}) \frac{1}{\rho} \bigg(\left[\frac{1}{\rho^{\frac{1}{1 + b}}}\right]_{+}\ns -(1+b)\ln\frac{\mu}{Q}\, \rho^{\frac{b}{1+b}}\delta(\rho)\bigg) \left[\frac{\rho^{\frac{1}{1 + b}}}{\T_{R} - \rho}\right]_{+} \nn \\
&=\frac{2}{b{\,}(1+b)}\int_{0}^{\T_{R}}d\rho\, \theta(\rho+\rho^{\frac{1}{1+b}}-\T_{R}) \frac{1}{\rho}\left[\frac{1}{\rho^{\frac{1}{1 + b}}}\right]_{+} \left[\frac{\rho^{\frac{1}{1 + b}}}{\T_{R} - \rho}\right]_{+}\, .
\end{align}
In the second line of the above equation, the $\delta(\rho)$ term does not contribute. Using the following identities for the plus distributions appearing above,
\begin{align}
\label{appeq:identities}
\bigg[\frac{\rho^{\frac{1}{1+b}}}{\T-\rho}\bigg]_{+} &= -\frac{\rho^{\frac{1}{1+b}}}{1+b} \ln\rho\, \delta(\rho-\T) + \rho^{\frac{1}{1+b}} \bigg[\frac{1}{\T-\rho}\bigg]_{+} , \\
 \frac{\rho^{\frac{1}{1+b}}}{\rho} \bigg[\frac{1}{\rho^{\frac{1}{1+b}}}\bigg]_{+} &= \bigg[\frac{1}{\rho}\bigg]_{+} , \\
 \frac{\rho^{\frac{1}{1+b}} \ln\rho}{\rho} \bigg[\frac{1}{\rho^{\frac{1}{1+b}}}\bigg]_{+} &= \bigg[\frac{\ln\rho}{\rho}\bigg]_{+}\, ,
\end{align}
\eq{appeq:SoftConvolutionIntegral} can be rewritten in the form 
\begin{align}
\label{appeq:RewriteIsPlus}
\text{I}^{(+)}_s(\T_{R}) &= \frac{2}{b(1+b)} \int_{0}^{\T_{R}}d\rho\, \theta(\rho+\rho^{\frac{1}{1+b}}-\T_{R}) \frac{1}{\rho} \left[\frac{1}{\rho^{\frac{1}{1 + b}}}\right]_{+}\bigg(\!\!-\frac{\rho^{\frac{1}{1+b}}}{1+b} \ln\rho\, \delta(\rho-\T_R) + \rho^{\frac{1}{1+b}} \bigg[\frac{1}{\T_R-\rho}\bigg]_{+}\bigg) \nn \\
&= \frac{2}{b(1+b)} \bigg(\!\!-\frac{1}{1+b} \bigg[\frac{\ln\T_R}{\T_R}\bigg]_{+} + \int_{r\, \T_{R}}^{\T_{R}}d\rho\, \left[\frac{1}{\rho}\right]_{+} \bigg[\frac{1}{\T_R-\rho}\bigg]_{+}\bigg)\, .
\end{align}
Here the lower limit of integration corresponding to the Heaviside $\theta$-function is determined by the solution of the equation
\begin{align}
\label{appeq:SoftFactorRhoEquation}
\frac{r}{(1-r)^{\! 1+b}} = \T_{R}^{\! b} \, .
\end{align}
\eq{appeq:RewriteIsPlus} has the familiar form of a convolution of two plus distributions except for the fact that the lower limit is now regulated by $r\, \T_R$. This lower cut-off regulates the singularity due to $1/\rho$ as $\rho \rightarrow 0$, thereby allowing us to drop the first plus prescription in the last line of \eq{appeq:RewriteIsPlus}. The other singular limit for $\rho \rightarrow \T_{R}$ can be regulated explicitly by raising the second term in the convolution to the power of $1+\al$. The final result, which is obtained by extracting the $\al^0$-terms of this regulated integral, gives
\begin{equation}
\label{eq:SoftIntegralISPositive}
\text{I}^{(+)}_s(\T_{R}) = \!-\frac{2\,b}{(1+b)^2} \bigg[\frac{\ln\T_R}{\T_R}\bigg]_{+} \!\!\! - \frac{\pi^2}{3\, b(1+b)} \delta(\T_R) - \frac{2}{1+b} \frac{\ln(1-r)}{\T_{R}}\, . 
\end{equation}
\subsection{Soft convolution integral $\text{I}^{(-)}_s(\T_{R})$}
\label{appsec:SoftNegBInt}
The integral $\text{I}^{(-)}_s(\T_{R})$ as defined in \eq{eq:SoftFactorNegativeB}, is given by
\begin{align}
\label{appeq:SoftConvolutionFactorNegativeB} 
{\text{I}}^{(-)}_s(\T_{R}) &\equiv -\frac{2}{b(1+b)} \int_{s\,\T_R}^{\T_{R}} {\df\T^s_n}{\,} \frac{1}{(\T^s_n)^2} \frac{1}{(\T_{R}-\T^s_n)^{\frac{b-1}{1+b}}} \bigg[\frac{(\T^s_n)^2}{(\T_{R}-\T^s_n)^{\frac{2}{1+b}}}\bigg]_{+} \frac{Q}{\mu} \bigg[\frac{1}{Q\T^s_n/\mu} \bigg]_{+} \nn \\
&= -\frac{2}{b(1+b)} \bigg(-(1+b) \bigg[\frac{\ln\T_R}{\T_R}\bigg]_{+}\ns\, + \int_{s\, \T_R}^{\T_R} \df\T_n^s \bigg[\frac{1}{\T_R-\T_n^s}\bigg]_{+} \bigg[\frac{1}{\T_R}\bigg]_{+}\bigg)\, .
\end{align}
Here again the lower limit of integration is given by the solution of a transcendental equation,
\begin{align}
\label{appeq:SoftFactorSEqn}
\frac{s}{(1-s)^{\frac{1}{1+b}}} = \T_{R}^{\!-\frac{b}{1+b}} \, .
\end{align}
\eq{appeq:SoftConvolutionFactorNegativeB} has the form of a convolution of plus distributions but with a modified lower limit. This convolution integral can be calculated in a similar way as outlined for the positive-$b$ case, leading to
\begin{equation}
\label{eq:SoftIntegralISNegative}
\text{I}^{(-)}_s(\T_{R}) = \frac{2\,b}{(1+b)^2} \bigg[\frac{\ln \T_R}{\T_{R}}\bigg]_{+}\ns + \frac{\pi^2}{3\, b(1+b)} \delta(\T_R) - \frac{2}{(1+b)^2} \frac{\ln(1-s)}{\T_{R}}\, . 
\end{equation}
\section{Expansion of the constraint equations in various limits}
Here we discuss the appropriate expansions for the constraint \eqs{appeq:SoftFactorRhoEquation}{appeq:SoftFactorSEqn} in various limits. 
\subsection{Solution to the constraint equation in $r$ ($b>0$)}
We start with positive values of the angularity exponent, and treat separately the case of vanishingly small $b$. 
\subsubsection{Small-$\T$ expansion for $r$}
\label{appsec:SmallTauRExp}
In the small-$\T$ limit, the lowest-order solution to \eq{appeq:SoftFactorRhoEquation} goes like $\T_R^b$. The general solution can then be written as a series in $\T_R^b$, 
\begin{align}
\label{appeq:RSolution}
r = a_1 \T_{R}^{\! b} + a_2 \T_{R}^{\! 2\,b} + a_3 \T_{R}^{\! 3\,b} + a_4 \T_{R}^{\! 4\,b} + a_5 \T_{R}^{\! 5\,b} +  \dots{} \, ,
\end{align}
where 
\begin{align}
\label{appeq:a_n}
&a_1 = 1, \quad a_2 = -(1+b), \quad a_3 = \frac{3\,b^2}{2}+\frac{5\,b}{2}+1, \quad a_4 = -\frac{8\,b^3}{3}-6\,b^2
-\frac{13\,b}{3}-1 \nn \\
&a_5 = \frac{125\,b^4}{24}+\frac{175\,b^3}{12}+\frac{355\,b^2}{24}+\frac{77\,b}{12}+1, \quad \dots{} \; .
\end{align}
The $\ln(1-r)/\T_{R}$ term in \eq{eq:SoftIntegralISPositive} is well-defined for $r \sim \T_R^b$ and does not require any plus prescription since a finite value of $b$ regulates the limit $\T_{R} \rightarrow 0$. \eq{eq:SoftIntegralISPositive}, in this limit, can then be written as
\begin{align}
\label{appeq:IsSolThrustLike}
\text{I}^{(+)}_s(\T_{R}) &= \!-\frac{2\,b}{(1+b)^2} \bigg[\frac{\ln\T_R}{\T_R}\bigg]_{+} \!\!\! - \frac{\pi^2}{3\, b(1+b)} \delta(\T_R) - \frac{2}{1+b} \frac{\ln(1-a_1\T_R^b-a_2\T_R^{2\,b}-a_3\T_R^{3\,b}+\dots{})}{\T_{R}}\nn \\
&= \!-\frac{2\,b}{(1+b)^2} \bigg[\frac{\ln\T_R}{\T_R}\bigg]_{+} \!\!\! - \frac{\pi^2}{3\, b(1+b)} \delta(\T_R) - \frac{2}{1+b} \sum_{n=1}^{\infty}\frac{c_n}{\T_{R}^{1-n\,b}} \, ,
\end{align}
with coefficients $c_n$ given in \eq{eq:c_n}. The last term accounts for sub-leading singular contributions and/or power corrections depending on the value of $b$.

\subsubsection{Small-$b$ expansion for $r$}
\label{appsec:SmallBRExp}

Any truncation of the power series expansion in \eq{appeq:RSolution} ceases to provide the correct solution to \eq{appeq:SoftFactorRhoEquation} when $\T$ and $b$ are both close to zero such that $\T_R^b \sim 1$.
In this case, the general solution can instead be represented as an expansion in powers of $b$,
\begin{align}
\label{appeq:RSmallB}
r = r_0 + b\, r_1 + b^2\, r_2 +  \dots{} \, ,
\end{align}
where the leading-order solution $r_0$ is determined by
\begin{align}
\label{appeq:R0}
\frac{r_0}{1-r_0} = \T_{R}^{\! b} \, .
\end{align}
Using the last equation and the power series expansion  in \eq{appeq:RSmallB}, the coefficients $r_1$, $r_2$, \dots{} can be obtained from \eq{appeq:SoftFactorRhoEquation} as
\begin{align}
\label{appeq:r_n}
&r_0 = \frac{\T_R^b}{1+\T_R^b}, \quad r_1 = -\frac{\T_R^b \ln(1+\T_R^b)}{(1+\T_R^b)^2}, \nn \\
&r_2 = 
-\frac{\T_R^b\Big((\T_R^b-1) \ln(1+\T_R^b) -2\, \T_R^b\Big)\ln(1+\T_R^b)}{2\,(1+\T_R^b)^3}\, \quad \dots{} \; .
\end{align}
Note that the leading-order solution $r_0$ in this case has a well-defined limit, i.e. as $b$ become vanishingly small, $r_0$ converges to 1/2 as expected.

Finally, to obtain the small-$b$ limit of the soft convolution integral of \eq{eq:SoftIntegralISPositive}, we need to take the small-$b$ limit of $\ln(1-r)/\T_R$. Since $b$ regulates small-$\tau_R$ limit, $\ln(1-r)/\T_R$ becomes a distribution in $\T_R$. We get,
\begin{equation}
\label{appeq:LogRSmallB}
\frac{\ln(1-r)}{\T_R}  
= \bigg(\!\!-\frac{\pi^2}{12\,b} + \frac{\ln^2 2}{2} -\frac{b \ln^2 2}{4} \bigg) \, \delta(\T_R) - \ln 2\bigg[\frac{1}{\T_R}\bigg]_{+} \ns  -\frac{b}{2} \bigg[\frac{\ln\T_R}{\T_R}\bigg]_{+}\ns +\frac{b\ln 2}{2} \bigg[\frac{1}{\T_R}\bigg]_{+} \ns \,+ {\cal O}(b^2) \, .
\end{equation}
This finally gives
\begin{equation}
\label{appeq:IsSolBroadLike}
\text{I}^{(+)}_s(\T_{R}) =  \delta(\T_R) \bigg(\!\!- \frac{\pi^2}{6\, b} + \frac{\pi^2}{6} - \frac{\pi^2\, b}{6} - \ln^2 2 +\frac{3\,b}{2}\, \ln^2 2\! \bigg)\! + 2\ln 2\bigg[\frac{1}{\T_R}\bigg]_{+} \ns - 3\,b\ln 2\bigg[\frac{1}{\T_R}\bigg]_{+}\ns-b \bigg[\frac{\ln\T_R}{\T_R}\bigg]_{+} \ns \,\,+ {\cal O}(b^2) \, .
\end{equation}

\subsection{Solution to the constraint equation in $s$ ($b<0$)}

Analogously, we now treat the case of negative angularity exponents. 

\subsubsection{Small-$\T$ expansion for $s$}
\label{appsec:SmallTauSExp}
In the small-$\T$ limit, \eq{appeq:SoftFactorSEqn} admits a solution of the form
\begin{align}
\label{appeq:SSol}
s = a^{'}_1 \T_{R}^{\!-\frac{b}{1+b}} + a^{'}_2 \T_{R}^{\!-\frac{2\,b}{1+b}} + a^{'}_3 \T_{R}^{\!-\frac{3\,b}{1+b}} + a^{'}_4 \T_{R}^{\!-\frac{4\,b}{1+b}} + a^{'}_5 \T_{R}^{\!-\frac{5\,b}{1+b}} +  \dots{} \, ,
\end{align}
where 
the first few coefficients $a^{'}_n$ are given by
\begin{align}
\label{appeq:a^{'}_n}
&a^{'}_1 = 1, \quad a^{'}_2 = -\frac{1}{1+b}, \quad a^{'}_3 = -\frac{b-2}{2(1+b)^2}, \quad a^{'}_4 = -\frac{b^2-4\,b+3}{3(1+b)^3} \nn \\
&a^{'}_5 = -\frac{6\,b^3-37\,b^2+58\,b-24}{24(1+b)^4}, \quad \dots{} \; .
\end{align}
The last term in \eq{eq:SoftIntegralISNegative} is well-defined in this limit and does not require any plus prescription, as earlier. The soft convolution integral in \eq{eq:SoftIntegralISNegative} can be written as
\begin{align}
\label{appeq:IsSolThrustLikeNegB}
\text{I}^{(-)}_s(\T_{R}) &= \frac{2\,b}{(1+b)^2} \bigg[\frac{\ln\T_R}{\T_R}\bigg]_{+} \ns\,\, + \frac{\pi^2}{3\, b(1+b)} \delta(\T_R) - \frac{2}{(1+b)^2} \frac{\ln(1-a^{'}_1\T_R^{-\frac{b}{1+b}}-a^{'}_2\T_R^{-\frac{2\,b}{1+b}}+\dots{})}{\T_{R}}\nn \\
&= \frac{2\,b}{(1+b)^2} \bigg[\frac{\ln\T_R}{\T_R}\bigg]_{+} \!\!\! + \frac{\pi^2}{3\, b(1+b)} \delta(\T_R) - \frac{2}{(1+b)^2} \sum_{n=1}^{\infty}\frac{c^{'}_n}{\T_{R}^{1+\frac{n\,b}{1+b}}}\, .
\end{align}
with coefficients $c^{'}_n$ given in \eq{eq:c^{'}_n}. Notice that for a given negative value of $b$, only the terms with $N = max(n) < \ceil*{1/\vert b\vert}-1$ in the summation are singular. 
Thus for $-1<b<-0.5$, the extra terms amount only to power corrections and there is no sub-leading singular contribution from recoil. 

\subsubsection{Small-$b$ expansion for $s$}
\label{appsec:SmallBSExp}

The solution in \eq{appeq:SSol} cannot be applied when $b$ and $\T$ are both vanishingly small. In this limit, a general solution to the constraint equation can be written, similar to \eq{appeq:RSmallB}, as
\begin{align}
\label{appeq:SBsmall}
s = s_0 + b\, s_1 + b^2\, s_2 +  \dots{} \, ,
\end{align}
with the leading-order solution $s_0$ determined by 
\begin{align}
\label{appeq:S0}
\frac{s_0}{1-s_0} = \T_{R}^{\!-b} \, .
\end{align}
The first few coefficients of the series expansion in \eq{appeq:SBsmall} are 
\begin{align}
\label{appeq:x_n}
&s_0 = \frac{\T_R^{-b}}{1+\T_R^{-b}}, \quad s_1 = -\frac{\T_R^{-b} \ln\Big(\frac{\T_R^{-b}}{1+\T_R^{-b}}\Big)}{(1+\T_R^{-b})^2}, \nn \\
&s_2 = 
\frac{\T_R^{-b} \ln\Big(\frac{\T_R^{-b}}{1+\T_R^{-b}}\Big) \bigg(2-(\T_R^{-b}-1) \ln\Big(\frac{\T_R^{-b}}{1+\T_R^{-b}}\Big) \bigg)}{2\,(1+\T_R^{-b})^3}\, \quad \dots{} \; .
\end{align}
Upon expansion in $b$, the term $\ln(1-s)/\T_R$ in the soft convolution integral becomes a distribution in $\T$, as follows  
\begin{equation}
\label{appeq:LogSSmallB}
\frac{\ln(1-s)}{\T_R} 
= \bigg(\frac{\pi^2}{12\,b} + \frac{\pi^2}{12} + \frac{\ln^2 2}{2} + \frac{b}{4}\ln^2 2\bigg)\, \delta(\T_R) - \ln 2\bigg[\frac{1}{\T_R}\bigg]_{+} \ns +\frac{b}{2} \bigg[\frac{\ln\T_R}{\T_R}\bigg]_{+}\ns -\frac{b}{2} \ln 2\bigg[\frac{1}{\T_R}\bigg]_{+} \ns \,+ {\cal O}(b^2)\, ,
\end{equation}
which leads to the following result in the small-$\T$ and small-$b$ limit, 
\begin{equation}
\label{appeq:IsSolBroadLikeNegB}
\text{I}^{(-)}_s(\T_{R}) 
=  \delta(\T_R) \bigg(\frac{\pi^2}{6\, b} - \frac{\pi^2}{6} + \frac{\pi^2\,b}{6} - \ln^2 2 + \frac{3\,b}{2}  \ln^2 2 \!\bigg)\! + 2\ln 2 \bigg[\frac{1}{\T_R}\bigg]_{+} \ns -3\,b\ln 2\bigg[\frac{1}{\T_R}\bigg]_{+} \ns + b \bigg[\frac{\ln\T_R}{\T_R}\bigg]_{+} \ns \,\,+ {\cal O}(b^2) \, .
\end{equation}


\bibliographystyle{jhep}
\bibliography{JHEP_paper_v1}
\end{document}